\documentclass[aps,prc,superscriptaddress,twocolumn,nofootinbib,amsmath,amssymb,showpacs,showkeys]{revtex4-1}
\usepackage{pifont}
\usepackage{graphicx}
\usepackage{color}
\usepackage{amsmath}
\usepackage{amssymb}
\usepackage[caption=false]{subfig}
\usepackage{multirow}

\usepackage[colorlinks=true, pdfstartview=FitV, linkcolor=blue,citecolor=blue, urlcolor=blue]{hyperref}

\newcommand{\half}{\ensuremath{\frac{1}{2}}}
\renewcommand{\d}{\ensuremath{\mathrm{d}}}

\newcommand{\proton}{\ensuremath{p}}
\newcommand{\neutron}{\ensuremath{n}}
\newcommand{\pip}{\ensuremath{\pi^{+}}}
\newcommand{\pim}{\ensuremath{\pi^{-}}}
\newcommand{\kp}{\ensuremath{K^{+}}}
\newcommand{\km}{\ensuremath{K^{-}}}
\newcommand{\kbar}{\ensuremath{\bar{K}}}

\newcommand{\kshort}{\ensuremath{{K_{S}^{0}}}}

\newcommand{\pizero}{\ensuremath{\pi^{0}}}
\newcommand{\SigmaPlus}{\ensuremath{\Sigma^{+}}}
\newcommand{\SigmaZero}{\ensuremath{\Sigma^{0}}}
\newcommand{\SigmaMinus}{\ensuremath{\Sigma^{-}}}
\newcommand{\LambdaOne}{\ensuremath{\Lambda(1405)}}
\newcommand{\LambdaTwo}{\ensuremath{\Lambda(1520)}}

\newcommand{\Kstar}{\ensuremath{K^{\ast}}}
\newcommand{\KstarZero}{\ensuremath{K^{\ast 0}}}
\newcommand{\KstarPlus}{\ensuremath{K^{\ast +}}}

\newcommand{\chLambda}{\ensuremath{\Lambda \pizero}}
\newcommand{\chSigmaPlusP}{\ensuremath{\Sigma^{+}_{p} \pim}}
\newcommand{\chSigmaPlusN}{\ensuremath{\Sigma^{+}_{n} \pim}}
\newcommand{\chSigmaZero}{\ensuremath{\SigmaZero \pizero}}
\newcommand{\chSigmaMinus}{\ensuremath{\SigmaMinus \pip}} 
\newcommand{\chSigmapmPimp}{\ensuremath{\Sigma^\pm\pi^\mp}}

\newcommand{\GeV}{\ensuremath{\mathrm{GeV}}}    % GeV
\newcommand{\gevc}{\ensuremath{\mathrm{GeV}/c}} % GeV/c
\newcommand{\mevcc}{\ensuremath{\mathrm{MeV}/c^{2}}} % MeV/c2
\newcommand{\gevcc}{\ensuremath{\mathrm{GeV}/c^{2}}} % GeV/c2

\newcommand{\deltaTOF}{\ensuremath{\Delta \mathrm{TOF}}}

\newcommand{\costhetakp}{\ensuremath{\cos \theta_{\kp}^{\mathrm{c.m.}}}}

\newcommand{\etal}{\textit{et al.}}

\newcommand{\ie}{\textit{i.e.}}

\begin{document}

% To get line numbers using lineo package
%\linenumbers


\preprint{CMU/102-2012}

%%%%%%%%%%%%%%%%% date and requesting user %%%%%%%%%%%%%%%%% 
% 
% requestor reinhard, 190  
% date 2013-01-11 or Fri Jan 11 17:36:00 2013 
%%%%%%%%%%%%%%%%% time windows and condtions %%%%%%%%%%%%%%%%% 
% 
% current window TODAY-TOMORROW   
% window 1 "2004-05-01" to "2004-06-30"   
% window 2 "1987-01-01" to "1987-01-01"   
% window 3 "1987-01-01" to "1987-01-01"   
% 
%%%%%%%%%%%%%%%%%%%%%%%%%%%%%%%%%%%%%%%%%%%%%%% 
% 
% 
% The journal format chosen is RevTek 
% 
% 
%%%%%%%%%%%%%%%%%%%%%%%%%%%%%%%%%%%%%%%%%%%%%%% 

%%%%%%%%%%%%%%% Latex Macros for institute addresses  %%%%%%%%%%%%%%%%%%%%%%%%% 

\newcommand*{\CMU}{Carnegie Mellon University, Pittsburgh, Pennsylvania 15213}
\affiliation{\CMU}
\newcommand*{\ANL}{Argonne National Laboratory, Argonne, Illinois 60439}
\affiliation{\ANL}
\newcommand*{\ASU}{Arizona State University, Tempe, Arizona 85287-1504}
\affiliation{\ASU}
\newcommand*{\CSUDH}{California State University, Dominguez Hills, Carson, California 90757, USA}
\affiliation{\CSUDH}
\newcommand*{\CUA}{Catholic University of America, Washington, D.C. 20064}
\affiliation{\CUA}
\newcommand*{\SACLAY}{CEA, Centre de Saclay, Irfu/Service de Physique Nucl\'eaire, 91191 Gif-sur-Yvette, France}
\affiliation{\SACLAY}
\newcommand*{\CNU}{Christopher Newport University, Newport News, Virginia 23606}
\affiliation{\CNU}
\newcommand*{\UCONN}{University of Connecticut, Storrs, Connecticut 06269}
\affiliation{\UCONN}
\newcommand*{\EDINBURGH}{Edinburgh University, Edinburgh EH9 3JZ, United Kingdom}
\affiliation{\EDINBURGH}
\newcommand*{\FU}{Fairfield University, Fairfield CT 06824}
\affiliation{\FU}
\newcommand*{\FIU}{Florida International University, Miami, Florida 33199}
\affiliation{\FIU}
\newcommand*{\FSU}{Florida State University, Tallahassee, Florida 32306}
\affiliation{\FSU}
\newcommand*{\Genova}{Universit$\grave{a}$ di Genova, 16146 Genova, Italy}
\affiliation{\Genova}
\newcommand*{\GWUI}{The George Washington University, Washington, DC 20052}
\affiliation{\GWUI}
\newcommand*{\ISU}{Idaho State University, Pocatello, Idaho 83209}
\affiliation{\ISU}
\newcommand*{\INFNFE}{INFN, Sezione di Ferrara, 44100 Ferrara, Italy}
\affiliation{\INFNFE}
\newcommand*{\INFNFR}{INFN, Laboratori Nazionali di Frascati, 00044 Frascati, Italy}
\affiliation{\INFNFR}
\newcommand*{\INFNGE}{INFN, Sezione di Genova, 16146 Genova, Italy}
\affiliation{\INFNGE}
\newcommand*{\INFNRO}{INFN, Sezione di Roma Tor Vergata, 00133 Rome, Italy}
\affiliation{\INFNRO}
\newcommand*{\ORSAY}{Institut de Physique Nucl\'eaire ORSAY, Orsay, France}
\affiliation{\ORSAY}
\newcommand*{\ITEP}{Institute of Theoretical and Experimental Physics, Moscow, 117259, Russia}
\affiliation{\ITEP}
\newcommand*{\JMU}{James Madison University, Harrisonburg, Virginia 22807}
\affiliation{\JMU}
\newcommand*{\KNU}{Kyungpook National University, Daegu 702-701, Republic of Korea}
\affiliation{\KNU}
\newcommand*{\LPSC}{LPSC, Universite Joseph Fourier, CNRS/IN2P3, INPG, Grenoble, France}
\affiliation{\LPSC}
\newcommand*{\UNH}{University of New Hampshire, Durham, New Hampshire 03824-3568}
\affiliation{\UNH}
\newcommand*{\NSU}{Norfolk State University, Norfolk, Virginia 23504}
\affiliation{\NSU}
\newcommand*{\OHIOU}{Ohio University, Athens, Ohio  45701}
\affiliation{\OHIOU}
\newcommand*{\ODU}{Old Dominion University, Norfolk, Virginia 23529}
\affiliation{\ODU}
\newcommand*{\RPI}{Rensselaer Polytechnic Institute, Troy, New York 12180-3590}
\affiliation{\RPI}
\newcommand*{\URICH}{University of Richmond, Richmond, Virginia 23173}
\affiliation{\URICH}
\newcommand*{\ROMAII}{Universita' di Roma Tor Vergata, 00133 Rome Italy}
\affiliation{\ROMAII}
\newcommand*{\MSU}{Skobeltsyn Nuclear Physics Institute, 119899 Moscow, Russia}
\affiliation{\MSU}
\newcommand*{\SCAROLINA}{University of South Carolina, Columbia, South Carolina 29208}
\affiliation{\SCAROLINA}
\newcommand*{\JLAB}{Thomas Jefferson National Accelerator Facility, Newport News, Virginia 23606}
\affiliation{\JLAB}
\newcommand*{\UTFSM}{Universidad T\'{e}cnica Federico Santa Mar\'{i}a, Casilla 110-V Valpara\'{i}so, Chile}
\affiliation{\UTFSM}
\newcommand*{\GLASGOW}{University of Glasgow, Glasgow G12 8QQ, United Kingdom}
\affiliation{\GLASGOW}
\newcommand*{\VIRGINIA}{University of Virginia, Charlottesville, Virginia 22901}
\affiliation{\VIRGINIA}
\newcommand*{\WM}{College of William and Mary, Williamsburg, Virginia 23187-8795}
\affiliation{\WM}
\newcommand*{\YEREVAN}{Yerevan Physics Institute, 375036 Yerevan, Armenia}
\affiliation{\YEREVAN}

\newcommand*{\NOWINFNGE}{INFN, Sezione di Genova, 16146 Genova, Italy}
\newcommand*{\NOWCNU}{Christopher Newport University, Newport News, Virginia 23606}
\newcommand*{\NOWMSU}{Skobeltsyn Nuclear Physics Institute, 119899 Moscow, Russia}
\newcommand*{\NOWORSAY}{Institut de Physique Nucl\'eaire ORSAY, Orsay, France}
\newcommand*{\NOWROMAII}{Universita' di Roma Tor Vergata, 00133 Rome Italy}
\newcommand*{\NOWINDIANA}{Indiana University, Bloomington, Indiana 47405}
\newcommand*{\NOWSIENA}{Siena College, Loudonville, NY 12211}
 %%%%%%%%%%%%%%% END OF Latex Macros for institute addresses  %%%%%%%%%%%%%%%%%%%%%%%%% 

\author {K.~Moriya} 
\altaffiliation[Current address:]{\NOWINDIANA}
\affiliation{\CMU}
\author {R.A.~Schumacher} 
\email[Contact: ]{schumacher@cmu.edu}
\affiliation{\CMU}
\author {K.P. ~Adhikari} 
\affiliation{\ODU}
\author {D.~Adikaram} 
\affiliation{\ODU}
\author {M.~Aghasyan} 
\affiliation{\INFNFR}
\author {M.D.~Anderson} 
\affiliation{\GLASGOW}
\author {S. ~Anefalos~Pereira} 
\affiliation{\INFNFR}
\author {J.~Ball} 
\affiliation{\SACLAY}
\author {N.A.~Baltzell} 
\affiliation{\ANL}
\affiliation{\SCAROLINA}
\author {M.~Battaglieri} 
\affiliation{\INFNGE}
% ...

\collaboration{The CLAS Collaboration}
\noaffiliation

%%%%%%%%%%%%%%%%%%%%%%%%%%%%%%%%%%%%%%%%%%%%%%%%%%%%%%%%%%%%%%%
%
%The Southeastern Universities Research Association (SURA) operates the 
%Thomas Jefferson National Accelerator Facility for the United States 
%Department of Energy under contractDE-AC05-84ER40150. 
% 
% 
%%%%%%%%%%%%%%%%%%%%%%%%%%%%%%%%%%% 
%The following paragraph should serve as a starting point for acknowledging 
% the support CLAS has received. 
% 
% 
% This work was supported in part by 
% the Chilean Comisi\'on Nacional de Investigaci\'on Cient\'ifica y Tecnol\'ogica (CONICYT),
% the Italian Istituto Nazionale di Fisica Nucleare,
% the French Centre National de la Recherche Scientifique,
% the French Commissariat \`{a} l'Energie Atomique,
% the U.S. Department of Energy,
% the National Science Foundation,
% the Scottish Universities Physics Alliance (SUPA),
% the United Kingdom's Science and Technology Facilities Council,
% and the National Research Foundation of Korea.
%
% The Southeastern Universities Research Association (SURA) operates the
% Thomas Jefferson National Accelerator Facility for the United States
% Department of Energy under contract DE-AC05-84ER40150.
%
%%%%%%%%%%%%%%%%%%%%%%%%%%%%%%%%%%%%%%%%%%%%%%%%%%%%%%%%%%%%%%%%%%%%%%%%%

\title{Measurement of the $\mathbf{\Sigma\pi}$ photoproduction line shapes near the
  $\mathbf{\Lambda(1405)}$}


%\author{Kei Moriya}
%\author{Reinhard Schumacher}
%\email[contact: ]{schumacher@cmu.edu}
%\affiliation{Department of Physics, Carnegie Mellon University, Pittsburgh, PA 15213, USA}
%%\homepage[]{Your web page}
%%\thanks{}
%%\altaffiliation{}

%\collaboration{CLAS Collaboration}%\noaffiliation


\date{\today}

%\begin{center}
%DRAFT - NOT FOR DISTRIBUTION - DRAFT
%\end{center}

\begin{abstract}
The reaction $\gamma + p \to K^+ + \Sigma + \pi$ was used to determine
the invariant mass distributions or ``line shapes'' of the
$\Sigma^+\pi^-$, $\Sigma^-\pi^+$ and $\Sigma^0\pi^0$ final states,
from threshold at $1328$ \mevcc{} through the mass range of the
$\Lambda(1405)$ and the $\Lambda(1520)$.  The measurements were made
with the CLAS system at Jefferson Lab using tagged real photons, for
center-of-mass energies $1.95 < W < 2.85$ GeV.  The three mass
distributions differ strongly in the vicinity of the $I=0$
$\Lambda(1405)$, indicating the presence of substantial $I=1$ strength
in the reaction.  Background contributions to the data from the
$\Sigma^0(1385)$ and from $K^*\Sigma$ production were studied and
shown to have negligible influence.  To separate the isospin
amplitudes, Breit-Wigner model fits were made that included
channel-coupling distortions due to the $N\bar{K}$ threshold.  A best
fit to all the data was obtained after including a phenomenological
$I=1$, $J^P = 1/2^-$ amplitude with a centroid at $1394\pm20$ \mevcc{}
and a second $I=1$ amplitude at $1413\pm10$ \mevcc.  The centroid of
the $I=0$ $\Lambda(1405)$ strength was found at the $\Sigma\pi$
threshold, with the observed shape determined largely by
channel-coupling, leading to an apparent overall peak near $1405$
\mevcc.
\end{abstract}

%\pacs{PACS PACS PACS}
\pacs{
      {13.30.Eg}
      {13.60.Rj}
      {14.20.Gk}
      {25.20.Lj}
     } % end of PACS codes
%\pacs{
%      {13.30.-a}{ Decays of baryons},
%      {13.30.Eg}{ Hadronic decays},
%      {13.40.-f}{ Electromagnetic processes and properties},
%      {13.60.-r}{ Photon and charged lepton interactions with hadrons},
%      {13.60.Le}{ Meson  photo production},
%      {13.60.Rj}{ Baryon photo production},
%      {14.20.Gk}{ Baryon resonances with S=0},
%      {14.20.Ju}{ Hyperon properties},
%      {25.20.Lj}{ Photoproduction reactions}
%     } % end of PACS codes
\maketitle

%\keywords{}

\maketitle

%*************************************************************************************
%                                                                                    *
%                                   Introduction                                     *
%                                                                                    *
%*************************************************************************************
% Revisions:
% 09-18-12 RS small wording changes.
% 11-20-12 RS Changes after Ad Hoc comments

\section{Introduction}
\label{section:Introduction}

The \LambdaOne, situated just below $N\bar{K}$ threshold, has been an
enigmatic state in the spectrum of strange baryons for decades. First
seen in bubble chamber experiments in the 1960's~\cite{Alston_1405},
there have been remarkably few measurements of this state to date. The
most prominent feature of the state is that its invariant mass
spectrum, which we call the ``line shape'', has always been seen to be
distorted from a Breit-Wigner form, indicating that there are strong
dynamics at work that are not seen in more typical resonances. Almost
all theories agree that this is due to the state's strong coupling to
$N\bar{K}$, but the exact nature of this coupling is as yet
unknown. Due to its mass being below the $N\bar{K}$ threshold, it is
not possible to produce it directly in kaon beam experiments, so
accessing this state experimentally has been a challenge compared to
other strange baryon resonances.  Precise measurements of the line
shape should yield information on what dynamics play a significant
role in the \LambdaOne, and lead to a deeper understanding of the
additional amplitudes that may exist in this mass region.

\subsection{Theories of the \LambdaOne{} }
\label{subsection:introduction:theory}

Explaining the mass of the \LambdaOne{} has also proved to be a
problem.  The state does not fit well within the constituent quark
model that has otherwise worked remarkably well for understanding the
masses of low-lying baryon
resonances~\cite{Isgur-Karl_PRD18,*Capstick-Isgur,*Capstick:2000qj}.
Theoretical investigations into the nature of the \LambdaOne{} were
discussed from the days of its prediction by Dalitz and
others~\cite{Dalitz_Tuan:AnnPhys8,*Dalitz_Tuan:PRL,*Dalitz_Tuan:AnnPhys10},
and there has been a surge of interest in recent years. Chiral unitary
theory~\cite{Oset-Ramos,Oller:2000fj,Hyodo:2011ur} combines chiral
dynamics with unitarity constraints based on effective meson-baryon
interactions.  In this class of models, the \LambdaOne{} is
dynamically generated as a rescattering of all pseudo-scalar meson and
octet baryon states that couple to it.  Definite predictions have been
made of what the line shape of the \LambdaOne{} should be for
photoproduction near threshold~\cite{Nacher:1998mi}.  In this model
the interference between a dominant isospin $I=0$ amplitude and a
smaller $I=1$ amplitude causes the line shapes for each $\Sigma \pi$
channel to be different.

Further developments of the chiral unitary approach have shown that
the \LambdaOne{} may be composed of two $I=0$ poles, whose couplings to
various particle final states and whose initial-state populations
differ according to the reaction under
investigation~\cite{Kaiser-Siegel-Weise,Jido:2003cb,Borasoy}.  The
$\Lambda(1405)$ plays a special role in these theories as the
archetype of a dynamically generated rescattering state, but the
models also impinge on the nature of non-strange nucleon resonances
such as the $N(1535)S_{11}$ and $N(1440)P_{11}$.

In another approach~\cite{Lutz:2004sg} the pair of states
$\Sigma(1385)$ and \LambdaOne{} are treated together in a kaon
double-pole model with an explicit assumption of photon dissociation
into a real $K^+$ and a virtual $K^-$ as illustrated in
Fig.~\ref{fig:cartoon}.  This model also made specific predictions for
the mass distributions of $\Sigma\pi$ and $\Lambda\pi^0$ final states
that we will mention later.  In a meson exchange
model~\cite{Haidenbauer:2010ch} the \LambdaOne{} is generated
dynamically by the coherent addition of $\omega, \rho$ and
scalar-meson exchanges.  Here, too, it appears as a two-pole structure
in the $N \bar{K}$ $S_{01}$ partial wave.

Other theories see the \LambdaOne{} as a bound state of
$N\bar{K}$~\cite{Akaishi,*Akaishi:2010wt} alone, and the dynamics that
create it are expected to have significant repercussions on whether
bound $\km \proton \proton$ states exist.  In other views, the
\LambdaOne{} is pictured as a true 3-quark state~\cite{Wohl}, or as a
negative-parity~\cite{Kisslinger:2009dr} or
positive-parity~\cite{Kittel-Farrar1,*Kittel-Farrar2} hybrid state.

The spectrum of $I=1$ $\Sigma$ excited states is also predicted to be
quite different in different models.  As discussed recently in
Ref.~\cite{Gao:2012zh}, standard so-called quenched quark models put
the $J^P = \frac{1}{2}^-$ $\Sigma$ at about $1650$ \mevcc, while an
unquenched model that allows $[qq][qq]\bar{q}$ $S$-wave configurations
expects it near $1380$ \mevcc.  Some evidence for a light
negative-parity $\Sigma$ has been discussed for several
years~\cite{Zou:2010tc, Wu:2009tu, Wu:2009nw, Gao:2010hy}.  In a
different vein, meson-baryon dynamical models place a $\Sigma^{\ast}$
$(1/2)^-$ near $1430$~\mevcc~\cite{Khemchandari:PhysRevD.84.094018},
$1475$~\mevcc~\cite{Oh:PhysRevD.75.074002}, or $1620$
\mevcc~\cite{Oset:2001cn}.  It is therefore of interest to look for
evidence of isospin one strength in the $\Sigma\pi$ system in the same
neighborhood as the isospin zero $\Lambda(1405)$.  Discovery of a
resonant structure in that same mass range could be decisive in
picking among models of baryonic excitations.

\subsection{Experiments on the $\Lambda(1405)$}
\label{subsection:introduction:past_experiments}

While there has been continual theoretical interest in the \LambdaOne,
there have been remarkably few measurements made of this state, known
for more than half a century and given a $4$-star rating by the
PDG~\cite{Beringer:1900zz}. Bubble chamber experiments using hadronic
beams at Brookhaven~\cite{Thomas} and CERN~\cite{Hemingway} have long
been the only experiments to identify the line shape, and that with
barely adequate statistics. The mass and width of the \LambdaOne{}
cited by the PDG are based primarily on these measurements.

In recent years, with the development of higher-statistics
experimental capabilities, there has been a renewed interest in
measuring the \LambdaOne. These include measurement of the $\SigmaZero
\pizero$ line shape in proton-proton collisions at COSY~\cite{Zychor},
measurement of the $\Sigma^{\pm} \pi^{\mp}$ line shapes in
proton-proton collisions at HADES~\cite{Hades}, and again the
measurement of the $\Sigma^{\pm} \pi^{\mp}$ line shapes using
photoproduction by LEPS~\cite{Ahn,Niiyama}. The LEPS measurement has
some overlap with the energy used in our measurement, but due to
limited statistics, their results depended on broad averaging over
kinematics and line shape comparisons with existing theoretical
curves, rather than on new and decisive fits to the data.

In this paper, we report results of a measurement with large
statistics accumulated with the CLAS system in Hall B of Jefferson
Lab.  With good mass resolution for the \LambdaOne{}, we will show for
the first time a measurement of all three $\Sigma \pi$ line
shapes. The center-of-mass energies ($W$) in this experiment covered a
wide range from near production threshold of the \LambdaOne{} up to
$2.85$ GeV, which allowed us to measure the energy-dependence of the
line shapes. The results are shown after summing the line shapes over
all kaon production angles for each energy; another paper that is in
preparation will show the differential cross sections for each
energy~\cite{crosssectionpaper}.

\begin{figure}[htpb]
\resizebox{0.4\textwidth}{!}{\includegraphics[angle=0.0]{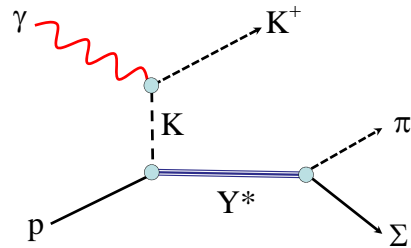}}
\caption{ (Color online) Creation of the three-body $K^+$ $\Sigma$
$\pi$ final state via an intermediate hyperon in the reaction $\gamma
+ p \to K^+ + \Sigma + \pi$.  In this particular example, a
$t$-channel exchange enables an off-shell kaon to create a
$\Lambda(1405)$ that is sub-threshold for on-shell $N\bar{K}$
reactions.}
\label{fig:cartoon}       % Give a unique label
\end{figure}

The line shapes are differential in the $\Sigma \pi$ invariant mass,
$m$, and extracted for nine bins in the initial state $\gamma p$
invariant energy $W$.  We anticipate that the $W$ dependence of the
$I=0$ and $I=1$ contributions varies slowly.  This is because the data
stem from an associated production experiment accompanied by a kaon,
as opposed to a direct formation experiment, so that the connection of
energy $W$ to the properties of the excited strange resonance(s) is
indirect.  This is indicated in Fig.~\ref{fig:cartoon}, for example,
where the intermediate hyperon is construed to be created via an
off-shell kaon as part of a $t$-channel interaction between the
incoming photon and the target proton.
% In this example
% the intermediate hyperon can have $I=0$ or $I=1$.  Via other diagrams
% it could conceivably have $I=2$, but we shall ignore this possibility.

%%%%%%%%%%%%%%%%%%%%%%%%%Fig 1%%%%%%%%%%%%%%%%%%%%%%%%%%%%%%%%%%%%
%%%%%%%%%%%%%%%%%%%%%%%%%Fig 1%%%%%%%%%%%%%%%%%%%%%%%%%%%%%%%%%%%%

The outline of this paper is as follows. 
Sections~\ref{section:ExperimentalSetup}--\ref{section:AcceptanceAndNormalization}
present the details of the setup of the experiment, the event
selection used to extract the yield of the various final states, the
acceptance corrections and the data normalization.
Section~\ref{section:YieldExtraction} describes our method of
extracting the yield of signal events from the data,
Section~\ref{section:LineshapeResults} presents the line shape results
for the three $\Sigma \pi$ final states, and
Section~\ref{section:SystematicUncertainties} discusses the systematic
uncertainties and the methods used to test the reliability of the
measurements.
Section~\ref{section:ModelFitting} explains the method used to
fit the line shapes we obtained.
In Section~\ref{section:isospin} the outcome of fitting the mass
distributions with this model is given, and we conclude with
Section~\ref{section:Conclusions}.

%*************************************************************************************
%                                                                                    *
%                                Experimental Setup                                  *
%                                                                                    *
%*************************************************************************************
\section{Experimental Setup}
\label{section:ExperimentalSetup}

The data for this experiment were obtained during May and June of 2004
with the CLAS detector, located in Hall B at the Thomas Jefferson
National Accelerator Facility The run, known as g11a, used a 40 cm
unpolarized liquid hydrogen (LH2) target and an incoming unpolarized
real-photon beam. Bremsstrahlung photons with an endpoint energy of
$4.019$ GeV were created via the CEBAF accelerator electron beam and a
$10^{-4}$ radiation length gold foil.  Electrons that radiated a
photon were identified with the CLAS tagger~\cite{Sober} to obtain
energy and timing information between 20\% and 95\% of the endpoint
energy.

Details of the CLAS detector can be found in
Ref.~\cite{CLAS-NIM}. Here we give a very brief description of the
main components used in our analysis. CLAS was equipped with a
superconducting toroidal magnet with six identical sectors surrounding
the beam line.  The field was selected to bend positive particles away
from the beam line.  A 34-layer drift chamber system in each sector
provided charge and momentum information for charged particles.
Momentum resolution $\delta p / p$ was $\approx 0.5\%$. The target was
surrounded by a 24 element plastic scintillator Start Counter used in
the trigger to select charged tracks leaving the target.  Finally, a
system of 342 time of flight (TOF) scintillators was used in the
trigger which also determined the duration of flight of each charged
particle. For the g11a run period, the trigger required a hit in the
tagger system in coincidence with Start Counter and TOF hits in at
least two of the six sectors.  A sector trigger required hits in a
Start Counter paddle and a TOF paddle within $150$ ns of each
other. With this setup, the g11a run accumulated over $20$ billion
events, including a large sample of excited hyperon states.  More
details of the setup and analysis can be found in
Ref.~\cite{Moriya-thesis}.

%*************************************************************************************
%                                                                                    *
%                                 Event Selection                                    *
%(write in past tense)                                                               *
%*************************************************************************************
% Modifications:
% 08-28-12 RS lots of wording, tense, formatting changes starting in subsection A
% 09-6-12 RS minor edits
% 09-14-12 RS minor edits
% 09-19-12 RS minor edits
% 11-16-12 RS after Ad Hoc comments
% 01-11-13 RS after CLAS comments

\section{Event selection}
\label{section:EventSelection}

\begin{figure*}[bhtp]
  \centering
  \vspace{-1.0cm}
  \includegraphics[width=0.70\textwidth]{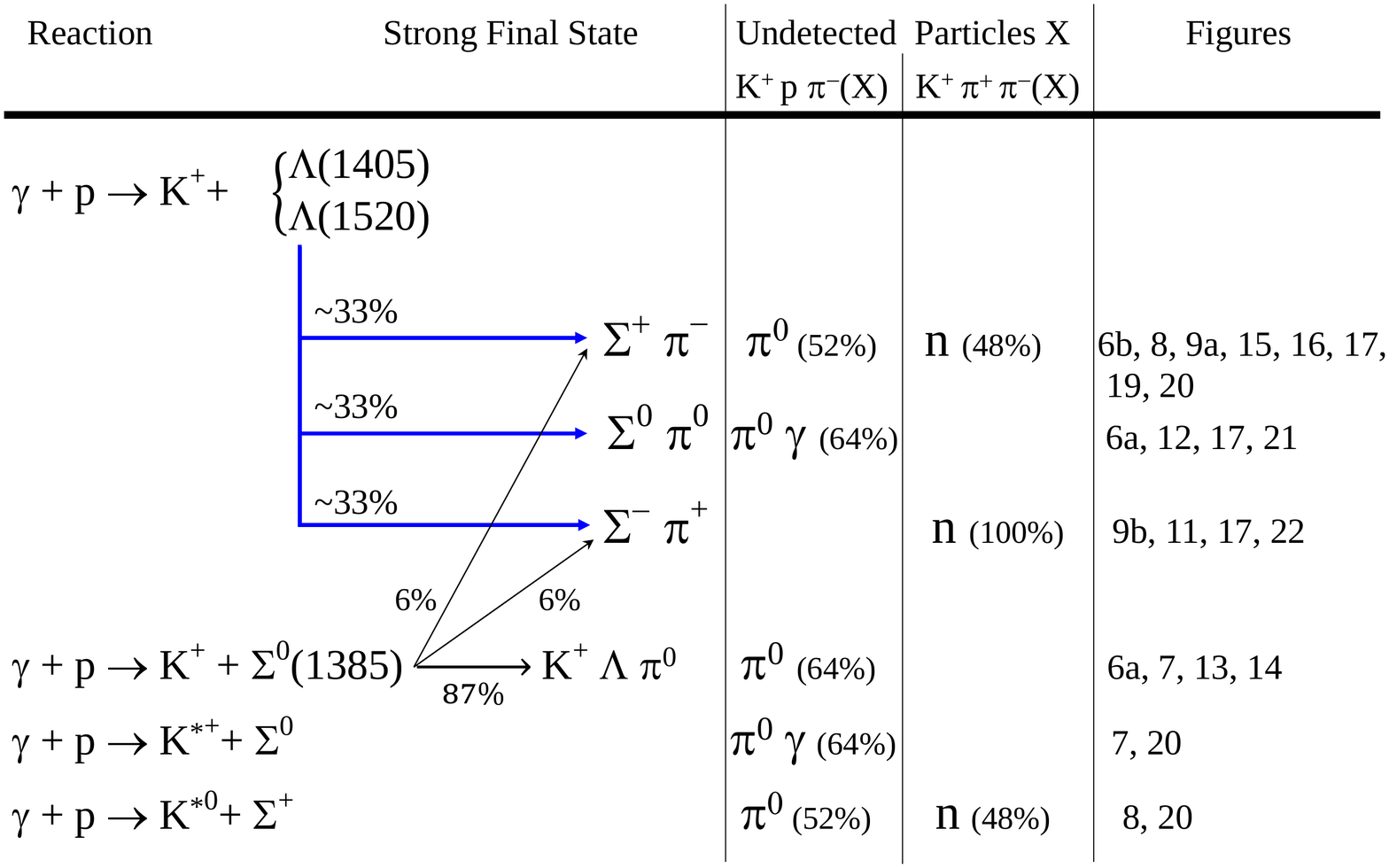}
  \vspace{-1.0cm}
  \caption{(Color online) Representation of reaction and decay
    channels used in this analysis.  Particles X given in the columns
    were reconstructed via kinematic fitting or the missing mass
    determination.  Branching fractions are indicated as percentages.
    The right-most column lists the figures related to the given
    channel.
    \label{fig:BigChart}}
\end{figure*}

For reference, we include Fig.~\ref{fig:BigChart} to illustrate the
various data-handling paths in this analysis.  The main interest
lies in the $\Sigma\pi$ final states of the hyperon decays,
particularly $\LambdaOne \to \Sigma \pi$.  The analysis channels can
be divided into two main categories depending on the final charged
particles detected. The first case is when a \kp, $p$, and \pim{}
are detected, while the second is when a \kp, \pip, and \pim{} are
detected.  The main strong final states of interest are then $\kp
\SigmaPlus \pim$, $\kp \SigmaZero \pizero$, $\kp \SigmaMinus \pip$,
and $\kp \Lambda \pizero$.  The latter is mainly due to $\Sigma(1385)
\to \Lambda \pizero$, which is one of the significant backgrounds for
isolating the \LambdaOne.  Since the $\Lambda$ and $\Sigma$ hyperons decay via the
weak force, we will call the final states including a hyperon the
strong final states.  During analysis the hyperons were reconstructed
through their weak decay products.  The ground-state hyperon decays
$\SigmaPlus \to \proton \pizero$, $\SigmaPlus \to \neutron \pip$,
$\SigmaZero \to \gamma \Lambda \to \gamma \proton \pim$, $\SigmaMinus
\to \neutron \pim$, and $\Lambda \to \proton \pim$ are detected in
our analysis.

%%%%%%%%%%%%%%%%%%%%%%%%%%%%%%%% FIGURE 2 %%%%%%%%%%%%%%%%%%%%%%%%%%%%%%%%%%%%%%%%%%%%%%
%%%%%%%%%%%%%%%%%%%%%%%%%%%%%%%% END FIGURE %%%%%%%%%%%%%%%%%%%%%%%%%%%%%%%%%%%%%%%%%%%%%%

The analysis procedure was as follows.  The raw set of reconstructed
and calibrated events was the same as used in several previous CLAS
publications~\cite{Williams1,Williams2,McCracken:2009ra,Dey}.  We
selected events with all of the required charged particles for each
channel of interest. After some cuts to reduce backgrounds, described
below, a kinematic fit was applied when all final state particles but
one were detected. Otherwise, the missing mass squared was computed
for the case of the \chSigmaZero{} channel. From this, the ground state
hyperons of $\Lambda$ or $\Sigma$ were reconstructed and selected for
analysis in each channel. For brevity, we will label the channels of
interest by their ground state hyperon--pion combination, \ie,
$\chSigmaPlusP$, $\chSigmaPlusN$, $\chSigmaZero$, $\chSigmaMinus$, and
$\chLambda$, where the two $\SigmaPlus$ decay channels are
distinguished with a subscript denoting the final-state
baryon. Analyzing both $\Sigma^+$ channels served as a cross check of
uncertainties introduced by our analysis methods.

\subsection{Initial selection of particles}
\label{subsection:EventSelection:InitialSelection}

The effects of selection cuts discussed below are summarized in
Table~\ref{tab:num_events}.  In this analysis all channels of interest
have a final state \kp. For the entire data set, the masses of charged
particles were calculated from the momenta given by the drift chamber
tracks, and the timing given by the RF-corrected tagger timing and TOF
scintillator timing. A loose mass cut was made to select events with a
kaon candidate, and these were retained for analysis.  Events were
then required to have all charged particles reconstructed in the
fiducial region of the detector, and a few malfunctioning timing
detectors were identified and removed.  The fiducial region of good
acceptance and good Monte Carlo matching was the same as used in
previously-published analyses~\cite{Williams1,Williams2,
McCracken:2009ra, Dey} of the same data set.  Copious non-strangeness
events were removed by testing the hypothesis that a candidate \kp{} was
actually a $\pi^+$ or a proton.  The leading backgrounds were $\gamma
p \to p \pi^+ \pi^-$ with nothing missing, $\gamma p \to p \pi^+
\pi^- (\pi^0)$ with the $\pi^0$ missing, and $\gamma p \to \pi^+ \pi^+
\pi^- (n)$ with the neutron missing.

Previously-established corrections to the reconstructed tracks were
applied, such as momentum corrections for small imperfections in the
magnetic field map, and energy losses due to the charged particles
traveling through the target and detector material.  The incident
photon energy was corrected for the known mechanical sagging of the
tagger hodoscope.  In all cases a timing cut was applied to remove
events where a \pip{} was misidentified as the \kp (see below). A
primary event vertex cut along the beam direction selected events very
cleanly from the LH2 target and rejected events from foils.  Minimum
momentum cuts based on the identity of the particle ($0.3~\gevc$ for
protons and \kp, $0.1~\gevc$ for $\pi^{\pm}$) were applied.

%%%%%%%%%%%%%%%%%%%%%%%%%%%%%%%%%Table 1 %%%%%%%%%%%%%%%%%%%%%%%%%%%%%%%%%%%%%%%%%%%%%%%
\begin{table*}
  \caption{\label{tab:num_events} The number of events remaining
    after each selection cut, in 1000's.}
%\resizebox{10cm}{!}{
%  \begin{ruledtabular}
    \begin{tabular}{p{5cm}| c c c c c}
      \hline 
      \hline
      \multirow{2}{*}{Selection} &  \multicolumn{5}{c}{Channel} \\
      & \chSigmaPlusP, & \chLambda, & \chSigmaZero & \chSigmaPlusN, & \chSigmaMinus \\ \hline
      Detected particles     &                   \multicolumn{3}{c|}{$K^+ p \pi^-$} & \multicolumn{2}{c}{$K^+ \pi^+ \pi^-$} \\ \hline
      Initial kaon selection &                          \multicolumn{3}{c|}{64,026} & \multicolumn{2}{c}{35,627} \\ \hline
      Fiducial cuts          &                          \multicolumn{3}{c|}{31,486} & \multicolumn{2}{c}{16,662} \\ \hline
      Remove false $K^+$ due to $\pi^+$ or $p$ &        \multicolumn{3}{c|}{4,852}  & \multicolumn{2}{c}{10,045} \\ \hline
      Loose \deltaTOF{} cuts &                          \multicolumn{3}{c|}{3,093}  & \multicolumn{2}{c}{6,576} \\ \hline
      Vertex $z$ cut &                                  \multicolumn{3}{c|}{3,066}  & \multicolumn{2}{c}{6,464} \\ \hline
      Minimum $|\vec{p}|$ requirements &                \multicolumn{3}{c|}{3,047}  & \multicolumn{2}{c}{6,233} \\ \hline
      Precise \deltaTOF{} cuts &                        \multicolumn{3}{c|}{2,415}  & \multicolumn{2}{c}{3,912} \\ \hline
      Kinematic fit or $\mathrm{MM}^{2}$ cut &          \multicolumn{2}{c|}{818} & 233 & \multicolumn{2}{|c}{1,052} \\ \hline
      Selection on ground state hyperon &               \multicolumn{1}{c|}{440} & \multicolumn{1}{c|}{238}
                                             &          \multicolumn{1}{c|}{76}  & \multicolumn{1}{c|}{316} & \multicolumn{1}{c}{338} \\ \hline
      %% ORIGINAL TABLE WITH ACCURATE NUMBERS
      %% & \chSigmaPlusP & \chLambda & \chSigmaZero & \chSigmaPlusN & \chSigmaMinus \\ \hline
      %% initial data kaon skim & \multicolumn{3}{c|}{64,026,872}  & \multicolumn{2}{c}{35,626,904} \\ \hline
      %% fiducial cuts          & \multicolumn{3}{c|}{31,486,439}  & \multicolumn{2}{c}{16,661,813} \\ \hline
      %% removal of fake $\pi$ events & \multicolumn{3}{c|}{4,852,020}  & \multicolumn{2}{c}{10,044,889} \\ \hline
      %% loose \deltaTOF{} cuts & \multicolumn{3}{c|}{3,092,659}  & \multicolumn{2}{c}{6,576,382} \\ \hline
      %% vertex $z$ cut & \multicolumn{3}{c|}{3,066,470}  & \multicolumn{2}{c}{6,464,190} \\ \hline
      %% minimum $|\vec{p}|$ requirements & \multicolumn{3}{c|}{3,047,089}  & \multicolumn{2}{c}{6,232,609} \\ \hline
      %% \deltaTOF{} cuts & \multicolumn{3}{c|}{2,414,626}  & \multicolumn{2}{c}{3,912,408} \\ \hline
      %% kinematic fit or $\mathrm{MM}^{2}$ cut & \multicolumn{2}{c|}{817,738} & 233,077 & \multicolumn{2}{|c}{1,052,038} \\ \hline
      %% selection on ground state hyperon & \multicolumn{1}{c|}{440,099} & \multicolumn{1}{c|}{238,359}
      %% & \multicolumn{1}{c|}{75,781} & \multicolumn{1}{c|}{315,750} & \multicolumn{1}{c}{338,143} \\ \hline
    \end{tabular}
%  \end{ruledtabular}
%}
\end{table*}
%
%%%%%%%%%%%%%%%%%%%%%%%%%%%%%%%%%Table 1 %%%%%%%%%%%%%%%%%%%%%%%%%%%%%%%%%%%%%%%%%%%%%%%

\begin{figure}[t!b!hp!]
  \centering
  \includegraphics[width=0.50\textwidth]{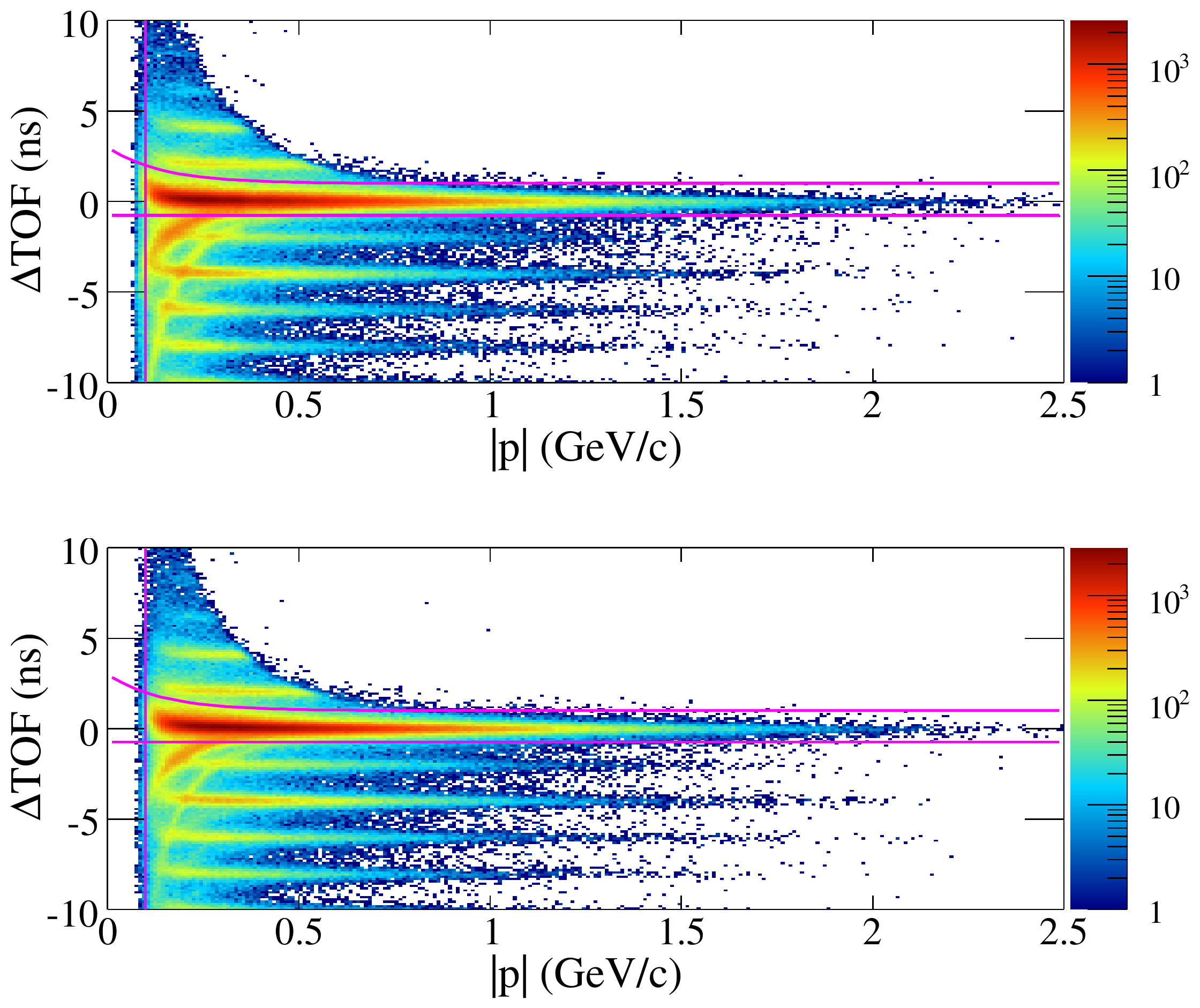}
  \caption{(Color online) Difference in particle time-of-flight,
    $\deltaTOF$, versus the measured magnitude of
    momentum for the \pip{} (top) and \pim{} (bottom) for a given
    energy bin of $2.35<W<2.45$ \GeV. The different horizontal bands
    correspond to the $2$ ns time structure of the CEBAF beam. The
    magenta lines show where the cuts were applied to select each
    particle.
    \label{fig:TOFcuts_pi}}
\end{figure}

\begin{figure}[t!b!hp!]
  \begin{center}
    \includegraphics[width=0.50\textwidth]{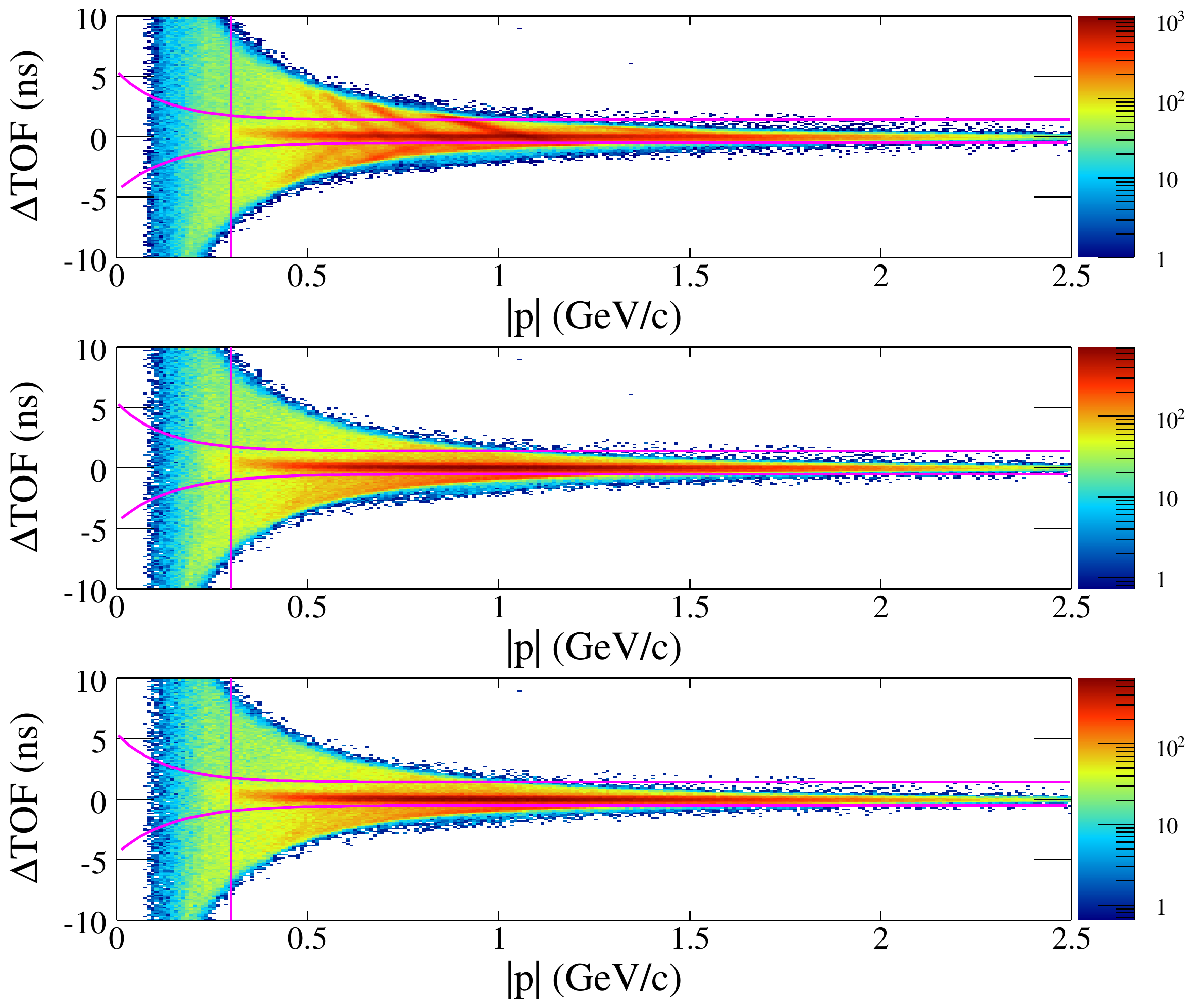}
  \end{center}
  \caption[]{(Color online) $\deltaTOF$ versus the measured magnitude
    of momentum for the
    \kp{} candidates at several analysis stages for a given energy bin
    of $2.35<W<2.45$ \GeV. From top to bottom, the plots correspond
    to: after the \kp{} misidentification rejection cut; after
    selecting the detected in-time \pim{} as shown in
    Fig.~\ref{fig:TOFcuts_pi}; and selection on both the detected
    \pim{} and \pip. The magenta lines represent the selection cut on
    the \kp. The last spectrum is seen to be much cleaner after the
    \deltaTOF{} selection cuts on the pions.  }
  \label{fig:TOFcuts_kp}
\end{figure}

As most of the background in these channels came from strangeness-free
events, cuts on the timing of particles were crucial to correctly
select kaons.  In the CLAS detector, the distance a charged particle
travels through the drift chambers ($l$), the accelerator RF- and
vertex-corrected event start time ($t_{0}$), and the time that the
particle hit the TOF paddles ($t_{1}$) were recorded along with the
particle's magnitude of momentum ($p$). From this information, the
measured travel time was calculated as
\begin{align}
  t_{\text{meas}} &= t_{1} - t_{0}.
\end{align}
Alternatively, we assumed a mass hypothesis for the particle,
$m_0$, and used the measured momentum to calculate the
velocity of the particle as
\begin{align}
  \beta_{\text{calc}} &= \frac{p}{\sqrt{p^{2} + m_0^{2}}},
\end{align}
and together with the reconstructed flight distance $l$ determined the
calculated flight time of the particle as
\begin{align}
  t_{\text{calc}} &= \frac{l}{\beta_{\text{calc}} c}.
\end{align}
Taking the difference between these two timing measures gives
\begin{align}
  \deltaTOF &= t_{\text{meas}} - t_{\text{calc}},
\end{align}
and cuts were applied on this quantity as a function of particle
momentum. Figures~\ref{fig:TOFcuts_pi} and \ref{fig:TOFcuts_kp} show the
momentum-dependent cuts applied to select the \pim, \pip, and \kp.

%%%%%%%%%%%%%%%%%%%%%%%%%%%%%%%% FIGURE 3 %%%%%%%%%%%%%%%%%%%%%%%%%%%%%%%%%%%%%%%%%%%%%%
%%%%%%%%%%%%%%%%%%%%%%%%%%%%%%%% END FIGURE %%%%%%%%%%%%%%%%%%%%%%%%%%%%%%%%%%%%%%%%%%%%%%

%%%%%%%%%%%%%%%%%%%%%%%%%%%%%%%% FIGURE 4 %%%%%%%%%%%%%%%%%%%%%%%%%%%%%%%%%%%%%%%%%%%%%%
%%%%%%%%%%%%%%%%%%%%%%%%%%%%%%%% END FIGURE %%%%%%%%%%%%%%%%%%%%%%%%%%%%%%%%%%%%%%%%%%%%%%

\subsection{Selection of events for analysis}
\label{subsection:EventSelection:EventsForAnalysis}

In all channels, the data were divided into $10$ bins of energy
spanning $100$ MeV in the center-of-mass energy $W$~$(=\sqrt{s})$, and
$20$ angle bins in the center-of-mass kaon angle. All selection cuts
and fits to the data were done independently for each bin.  For
channels with the \chSigmapmPimp combinations, kinematic fits were
applied with fixed mass of the undetected neutron or $\pi^0$
(one-constraint or 1C fits).  This optimized the information based on
the measured momenta while balancing the energy and momentum of the
reaction.

\subsubsection{Event selection for \chLambda{} and \chSigmaPlusP}
\label{subsubsection:EventSelection:EventsForAnalysis:case2pi0}

In these channels we reconstructed the particles \kp, \proton, and
\pim, with a missing \pizero. The 1-C kinematic fit was applied to the
selected particles, and those events with a confidence level (CL) of
greater than $1\%$ were retained for further analysis.  The covariance
matrix for these fits was optimized in a previous
study~\cite{Williams1}, and the confidence-level distributions were
checked for the present kinematics and found to be very flat.  The
possible combinations that yield a hyperon in the strong final state
are $\Lambda \to \proton \pim$ and $\SigmaPlus \to \proton \pizero$.

\begin{figure*}[t!b!hp!]
  \begin{center}
    \includegraphics[width=0.90\textwidth]{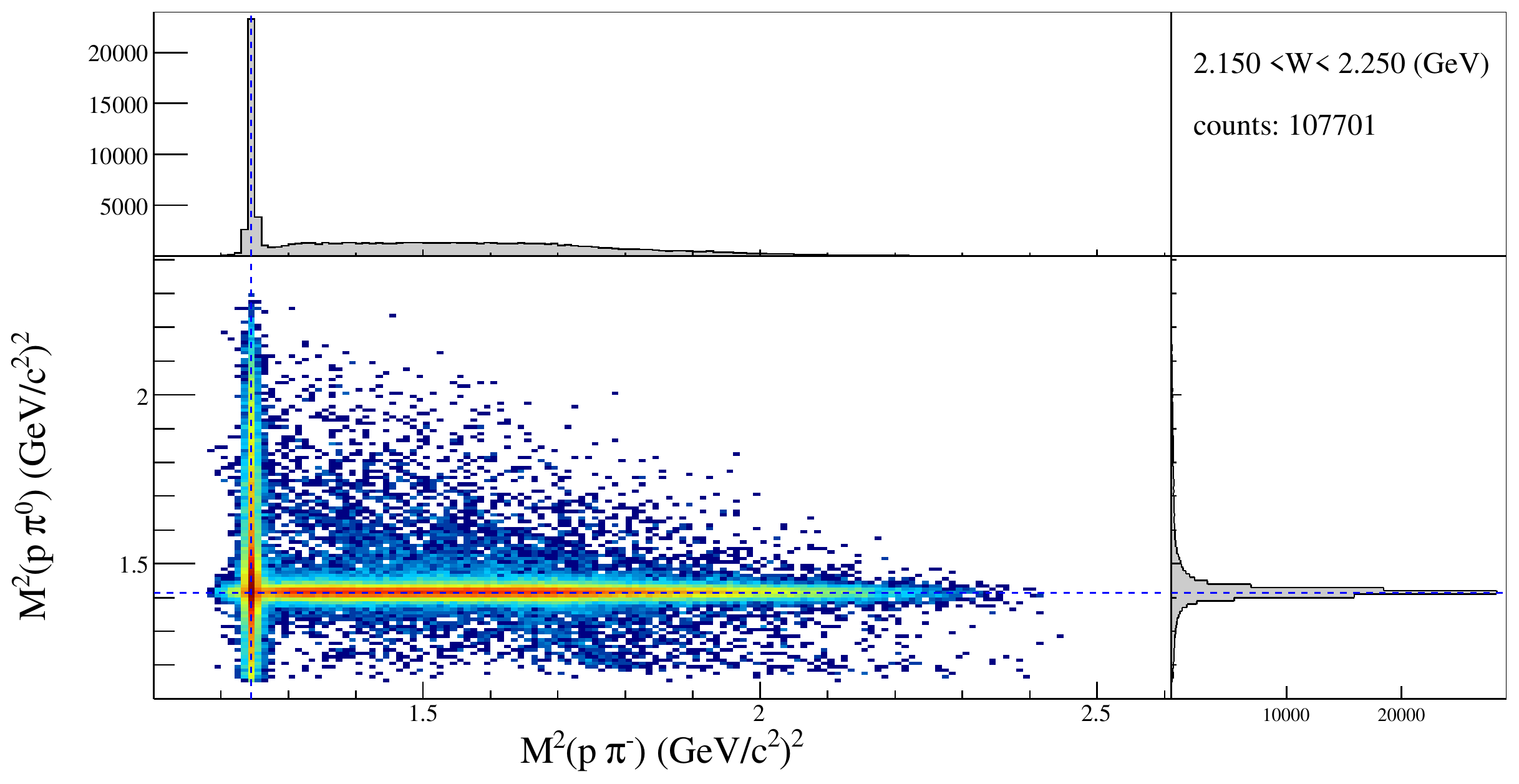}
  \end{center}
  \caption[]{(Color online) $\mathrm{M}^{2}(\proton \pizero)$ versus
    $\mathrm{M}^{2}(\proton \pim)$ for a given energy bin, with
    projections. The bands corresponding to the $\Lambda$ (vertical)
    and \SigmaPlus{} (horizontal) are clearly seen.  }
  \label{fig:case2cross}
\end{figure*}

\begin{figure}[t!b!h!p!]
  \begin{center} \subfloat[Fit to $\Lambda$.]
    {\label{fig:case2crossfits:Lambda}\includegraphics[width=0.45\textwidth]{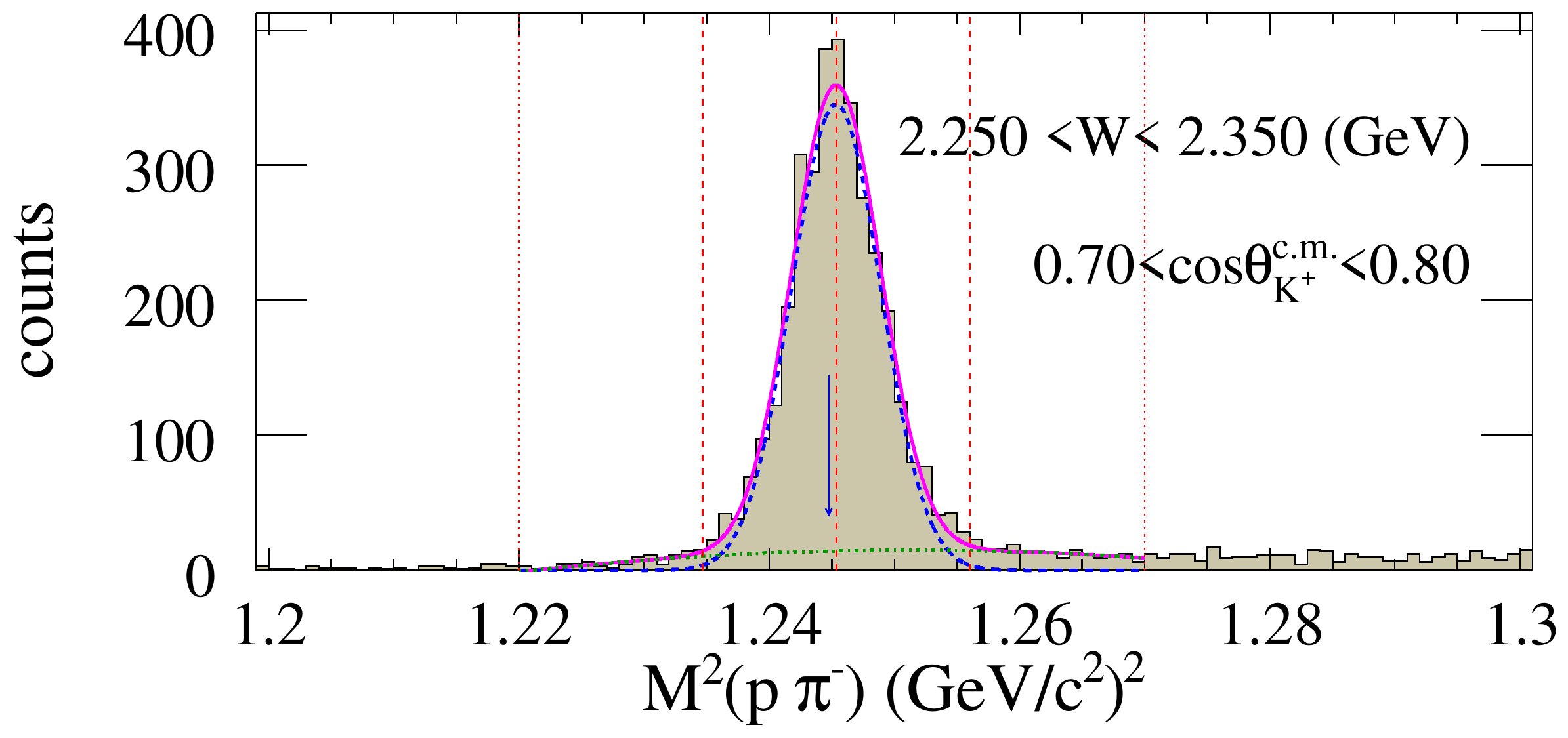}}
    \hfill \subfloat[Fit to \SigmaPlus.]
    {\label{fig:case2crossfits:SigmaPlus}\includegraphics[width=0.45\textwidth]{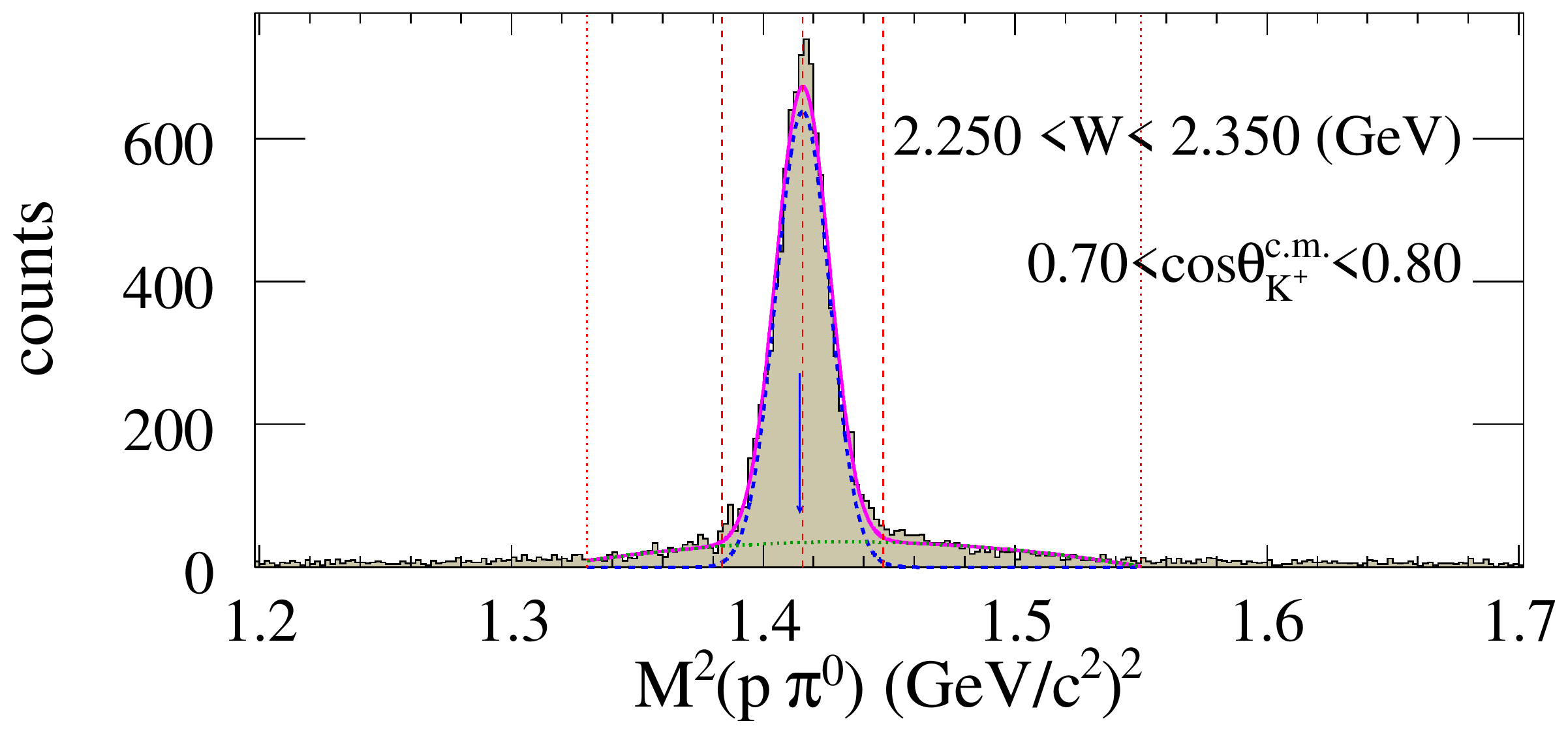}}
    \end{center} \caption[]{(Color online)
    Fits to the ground state hyperons
    \subref{fig:case2crossfits:Lambda} $\Lambda$ and
    \subref{fig:case2crossfits:SigmaPlus} \SigmaPlus ~for a single bin
    in energy and angle.  The data of each invariant mass
    squared are shown as the histograms, and the fits are shown as the
    solid curve (total), dashed curve (Gaussian), and dot-dashed curve
    (background). The outer dotted vertical lines show the range of
    the fits, the inner lines show $\pm 3 \sigma$ around the peaks,
    which is used to define the signal events.}
    \label{fig:case2crossfits}
\end{figure}

Figure~\ref{fig:case2cross} shows the invariant mass squared
distributions $\mathrm{M}^{2}(\proton \pizero)$ against
$\mathrm{M}^{2}(\proton \pim)$, summed over all kaon angles in one
particular energy bin.  For each bin in energy and angle, fits were
done to the projections of $\mathrm{M}^{2}(\proton \pizero)$ and
$\mathrm{M}^{2}(\proton \pim)$ with Gaussians and a second order
polynomial background. Figure~\ref{fig:case2crossfits} shows a
representative example of the fits to the $\Lambda$ and \SigmaPlus{}
peaks in a single energy and angle bin.  After projecting and fitting
the $\Lambda$ and \SigmaPlus{} peaks, a region of $\pm 3 \sigma$ was
chosen around each peak as the signal region. For further analysis of
events with a \SigmaPlus, the overlap region with the $\Lambda$ was
excluded so that there was no $\Lambda$ distribution underneath the
\SigmaPlus{} events. On the other hand, because the $\Lambda$ peak is
very narrow (approximately $1.3$ \mevcc{} across all bins when
converted to width around the $\Lambda$ peak, compared to
approximately $6.3$ \mevcc{} for the \SigmaPlus{} peak), the
\SigmaPlus{} region was not excluded from the $\Lambda$ signal, since
most of the $\Lambda$ signal was within this overlap region.  The
remaining backgrounds were removed as part of the later bin-by-bin
yield fits.

%%%%%%%%%%%%%%%%%%%%%%%%%%%%%%%% FIGURE 5 %%%%%%%%%%%%%%%%%%%%%%%%%%%%%%%%%%%%%%%%%%%%%%
%%%%%%%%%%%%%%%%%%%%%%%%%%%%%%%% END FIGURE %%%%%%%%%%%%%%%%%%%%%%%%%%%%%%%%%%%%%%%%%%%%%%

%%%%%%%%%%%%%%%%%%%%%%%%%%%%%%%% FIGURE 6 %%%%%%%%%%%%%%%%%%%%%%%%%%%%%%%%%%%%%%%%%%%%%%
%%%%%%%%%%%%%%%%%%%%%%%%%%%%%%%% END FIGURE %%%%%%%%%%%%%%%%%%%%%%%%%%%%%%%%%%%%%%%%%%%%%%

For these channels, a small contamination is seen in the projection of
the invariant mass squared of the \pim{} and \pizero, which comes from
the decay $K^{-} \to \pim \pizero$. Since our main channel of
interest, the \LambdaOne{}, is below the $K^{-} \proton$ threshold, we
did not cut away this contamination, but removed it later by
background subtraction.

\begin{figure*}[p!t!b!h!]
  \begin{center}
    \subfloat[]
    {\label{fig:Dalitz:case2Lambda:W1}\includegraphics[width=0.45\textwidth]{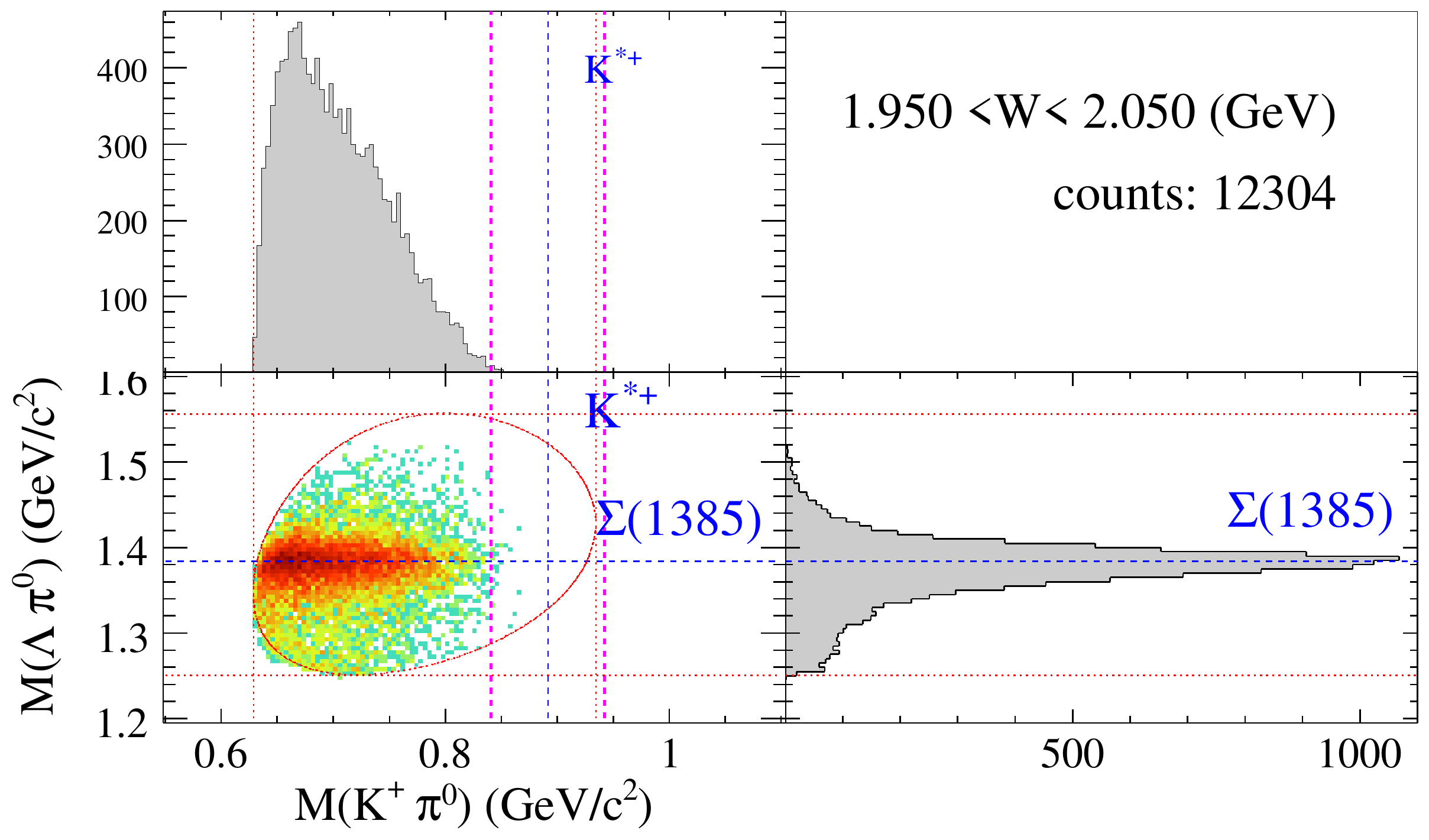}}
    \hfill
    %\subfloat[]
    %{\label{fig:Dalitz:case2Lambda:W2}\includegraphics[width=0.45\textwidth]{4___W2_Lambda-crop.pdf}}
    %\hfill
    %
    \subfloat[]
    {\label{fig:Dalitz:case2Lambda:W3}\includegraphics[width=0.45\textwidth]{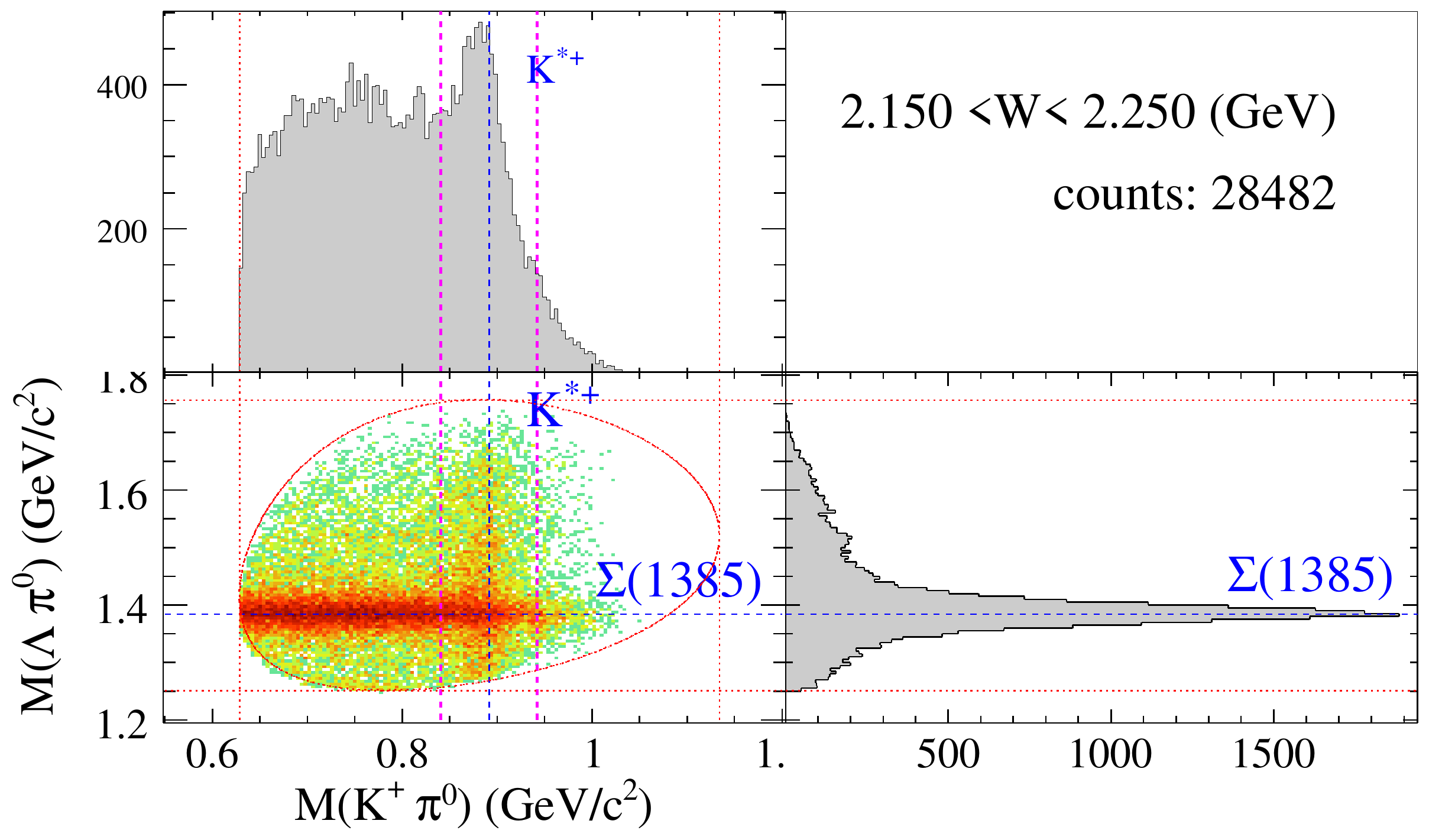}}
    \hfill
    %\subfloat[]
    %{\label{fig:Dalitz:case2Lambda:W4}\includegraphics[width=0.45\textwidth]{4___W4_Lambda-crop.pdf}}
    %\hfill
    %
    %\subfloat[]
    %{\label{fig:Dalitz:case2Lambda:W5}\includegraphics[width=0.45\textwidth]{4___W5_Lambda-crop.pdf}}
    %\hfill
    %\subfloat[]
    %{\label{fig:Dalitz:case2Lambda:W6}\includegraphics[width=0.45\textwidth]{4___W6_Lambda-crop.pdf}}
    %\hfill
    %
    \subfloat[]
    {\label{fig:Dalitz:case2Lambda:W7}\includegraphics[width=0.45\textwidth]{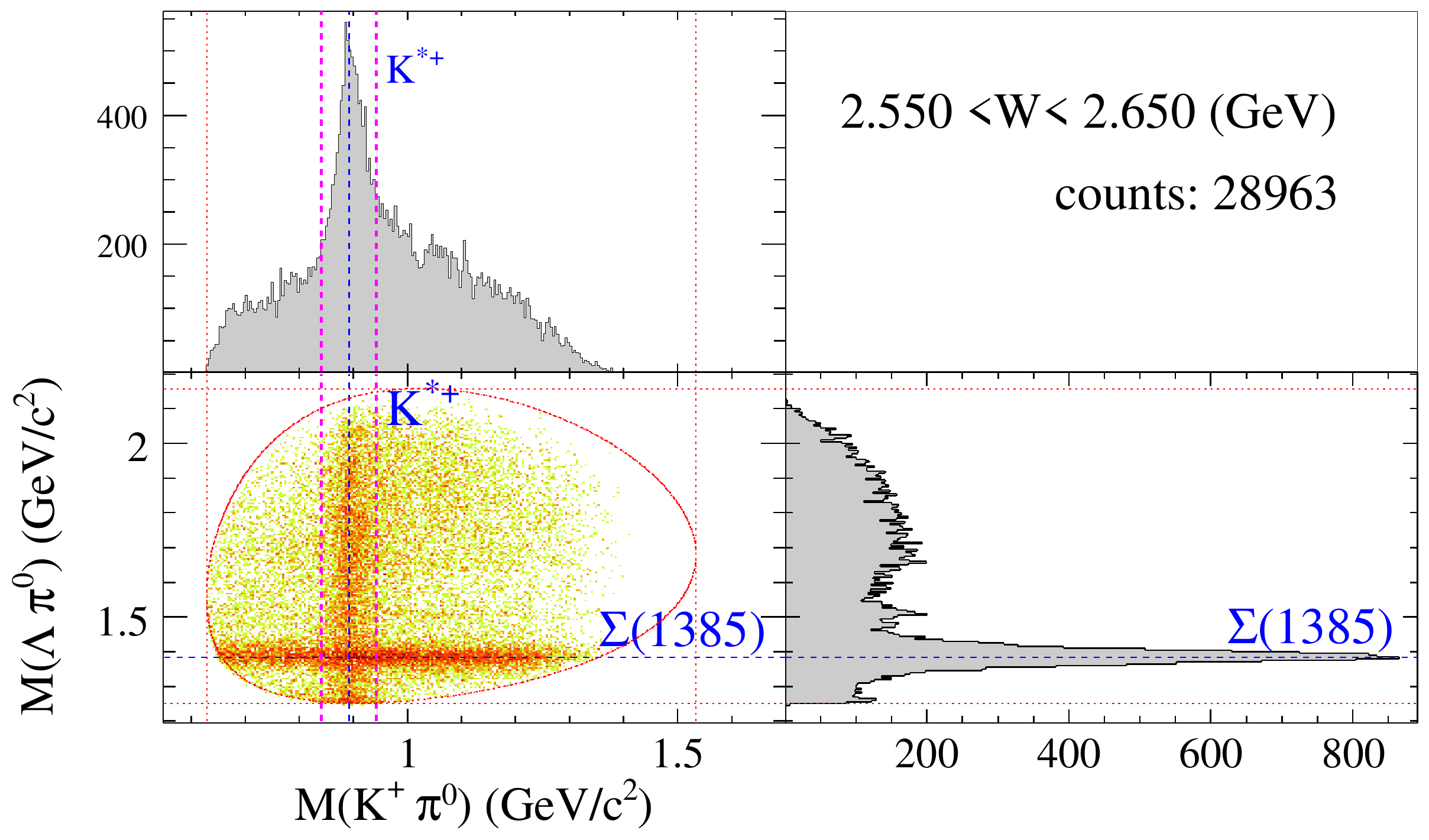}}
    \hfill
    %\subfloat[]
    %{\label{fig:Dalitz:case2Lambda:W8}\includegraphics[width=0.45\textwidth]{4___W8_Lambda-crop.pdf}}
    %\hfill
    %
    \subfloat[]
    {\label{fig:Dalitz:case2Lambda:W9}\includegraphics[width=0.45\textwidth]{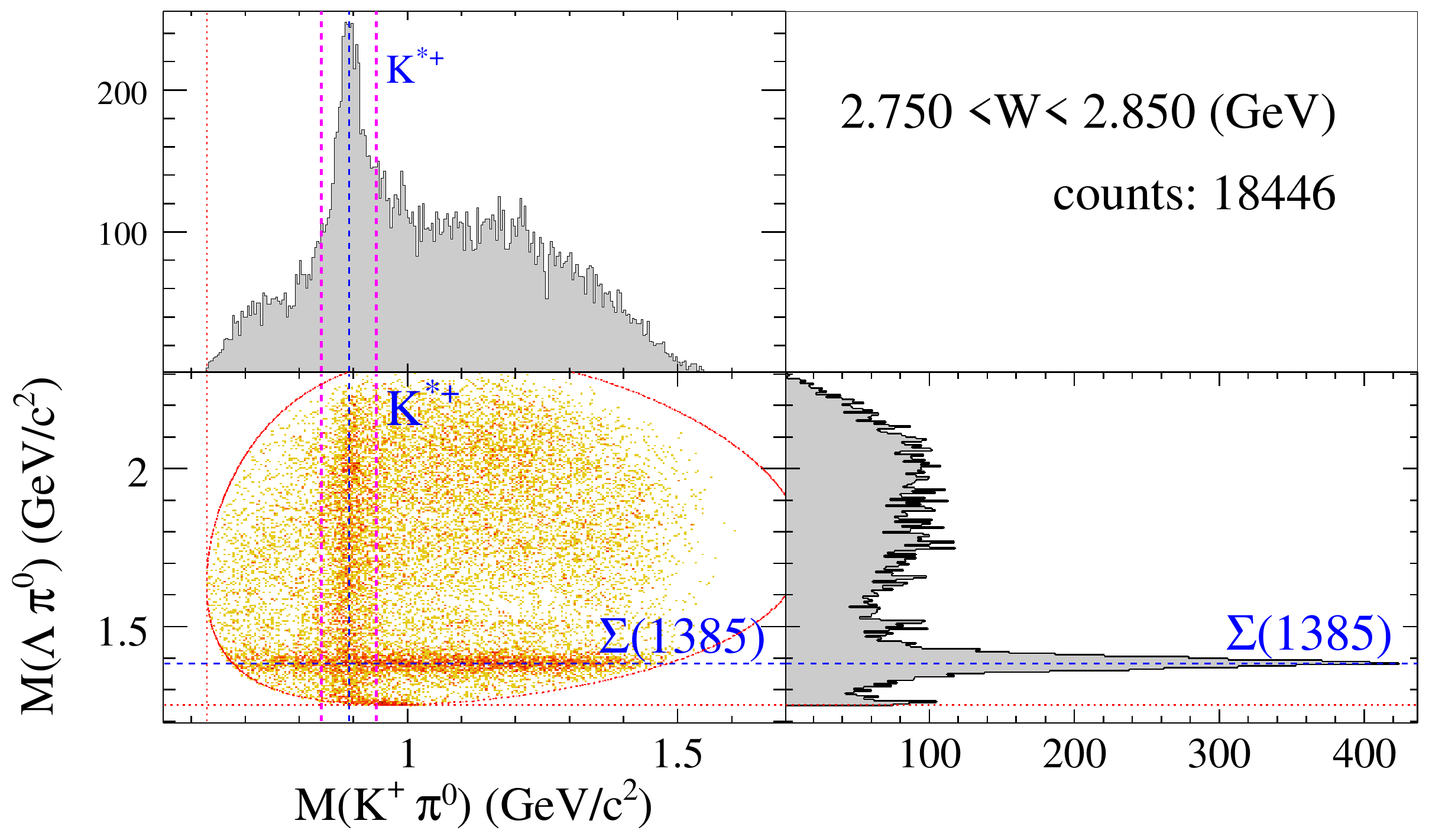}}
    \hfill
  \end{center}
    \caption[$\mathrm{M}(\Lambda \pizero)$ against $\mathrm{M}(\kp
      \pizero)$.]{(Color online) $\mathrm{M}(\Lambda \pizero)$ versus
      $\mathrm{M}(\kp \pizero)$ for four energy bins increasing from
      (a) to (d). A clear horizontal band corresponding to the
      $\Sigma(1385)$ is seen, as well as a vertical band corresponding
      to the $K^{\ast +}$. The contours, as well as the dashed lines,
      show the kinematic boundaries allowed in that energy bin.  The
      blue dashed lines show the masses of each resonance
      ($\Sigma(1385)$, \KstarPlus) as given by the PDG, while the
      vertical dashed lines show where the $\kp \pizero$ invariant
      mass is $M_{0} \pm \Gamma$, and $M_{0}$ and $\Gamma$ are the
      mass and width of the \KstarPlus{} as given by the PDG.  }
    \label{fig:Dalitz:case2Lambda}
\end{figure*}

\begin{figure*}[p!t!b!h!]
  \begin{center}
    \subfloat[]
    {\label{fig:Dalitz:case2SigmaPlus:W1}\includegraphics[width=0.45\textwidth]{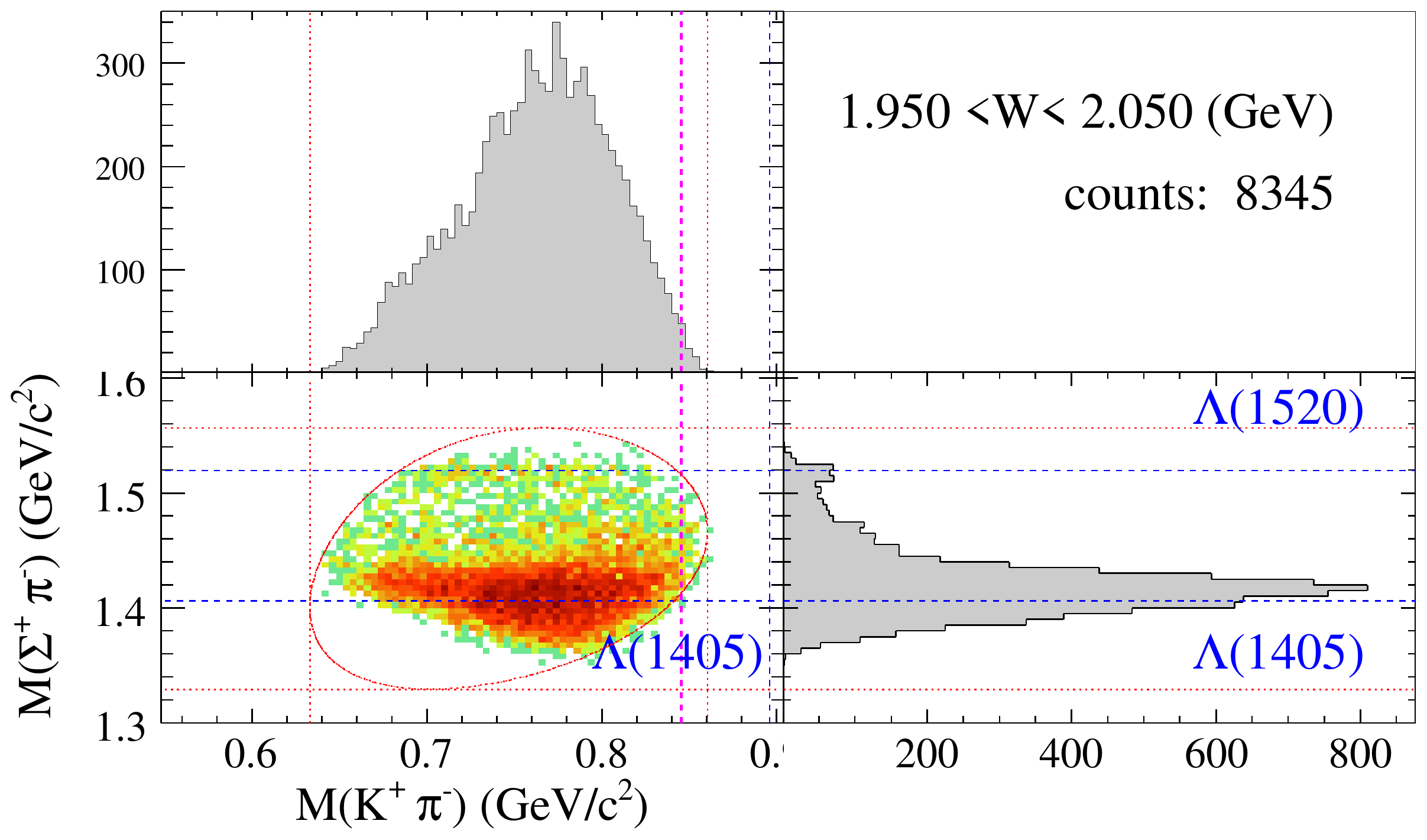}}
    \hfill
    %\subfloat[]
    %{\label{fig:Dalitz:case2SigmaPlus:W2}\includegraphics[width=0.45\textwidth]{4___W2_SigmaPlus-crop.pdf}}
    %\hfill
    %
    \subfloat[]
    {\label{fig:Dalitz:case2SigmaPlus:W3}\includegraphics[width=0.45\textwidth]{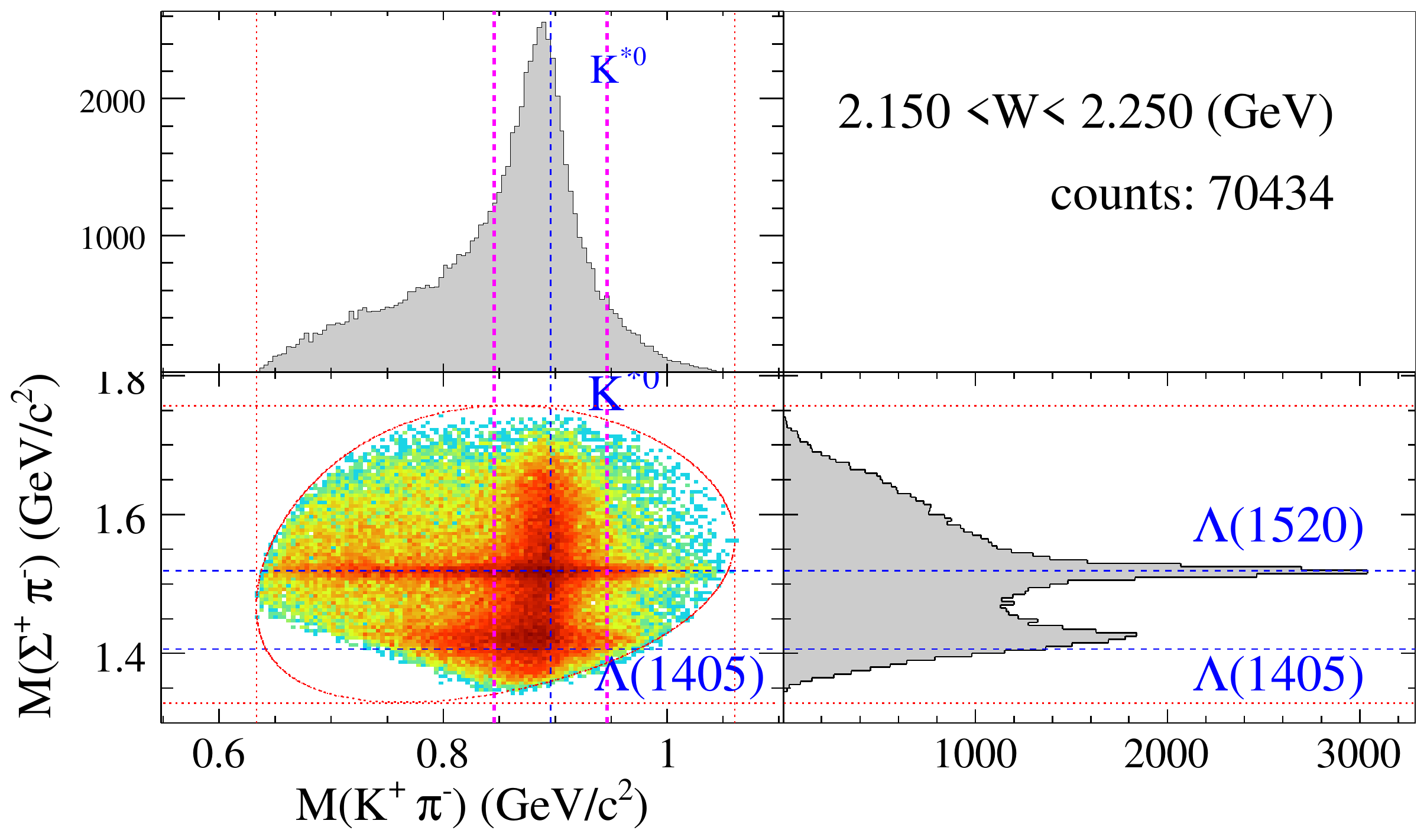}}
    \hfill
    %\subfloat[]
    %{\label{fig:Dalitz:case2SigmaPlus:W4}\includegraphics[width=0.45\textwidth]{4___W4_SigmaPlus-crop.pdf}}
    %\hfill
    %
    %\subfloat[]
    %{\label{fig:Dalitz:case2SigmaPlus:W5}\includegraphics[width=0.45\textwidth]{4___W5_SigmaPlus-crop.pdf}}
    %\hfill
    %\subfloat[]
    %{\label{fig:Dalitz:case2SigmaPlus:W6}\includegraphics[width=0.45\textwidth]{4___W6_SigmaPlus-crop.pdf}}
    %\hfill
    %
    \subfloat[]
    {\label{fig:Dalitz:case2SigmaPlus:W7}\includegraphics[width=0.45\textwidth]{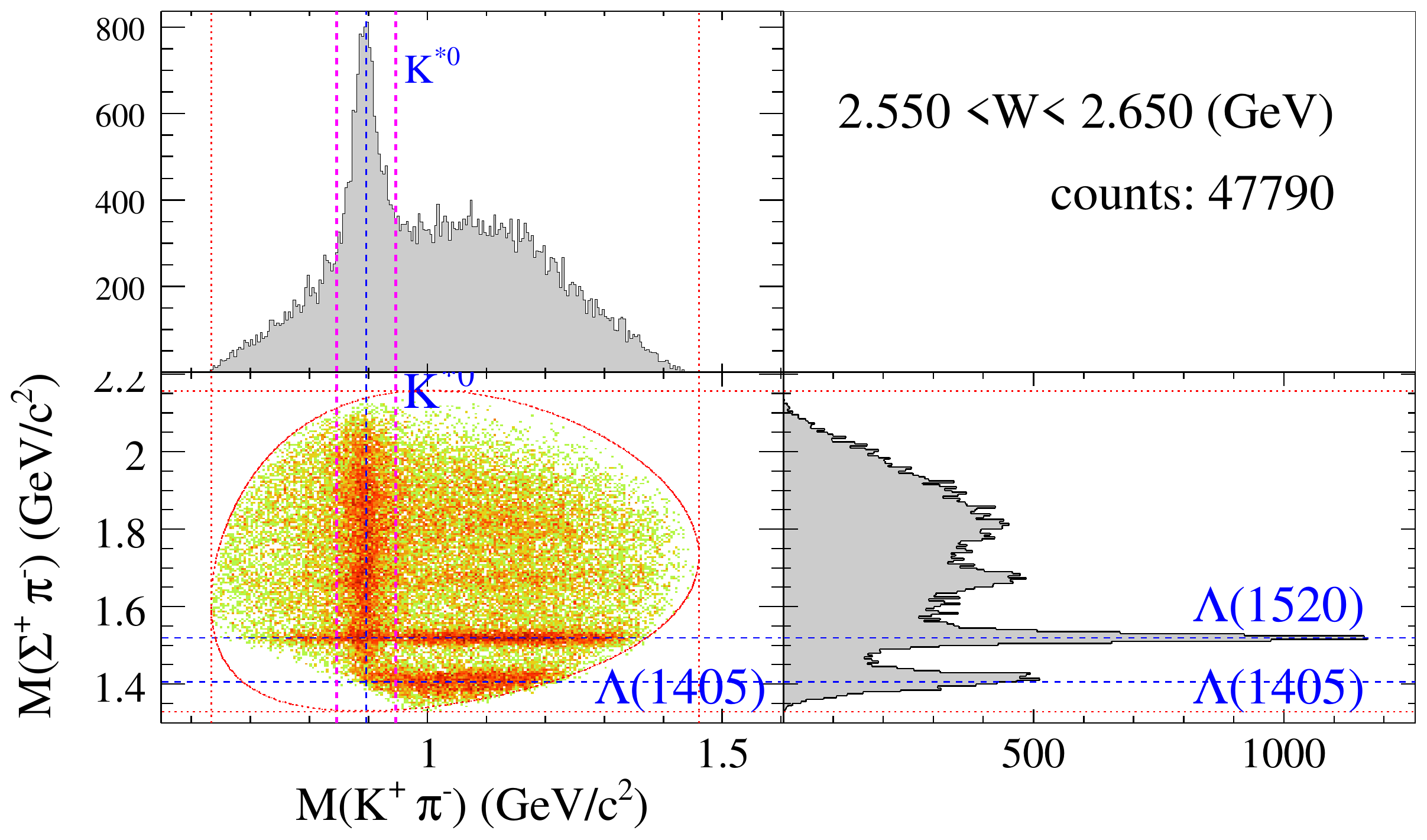}}
    \hfill
    %\subfloat[]
    %{\label{fig:Dalitz:case2SigmaPlus:W8}\includegraphics[width=0.45\textwidth]{4___W8_SigmaPlus-crop.pdf}}
    %\hfill
    %
    \subfloat[]
    {\label{fig:Dalitz:case2SigmaPlus:W9}\includegraphics[width=0.45\textwidth]{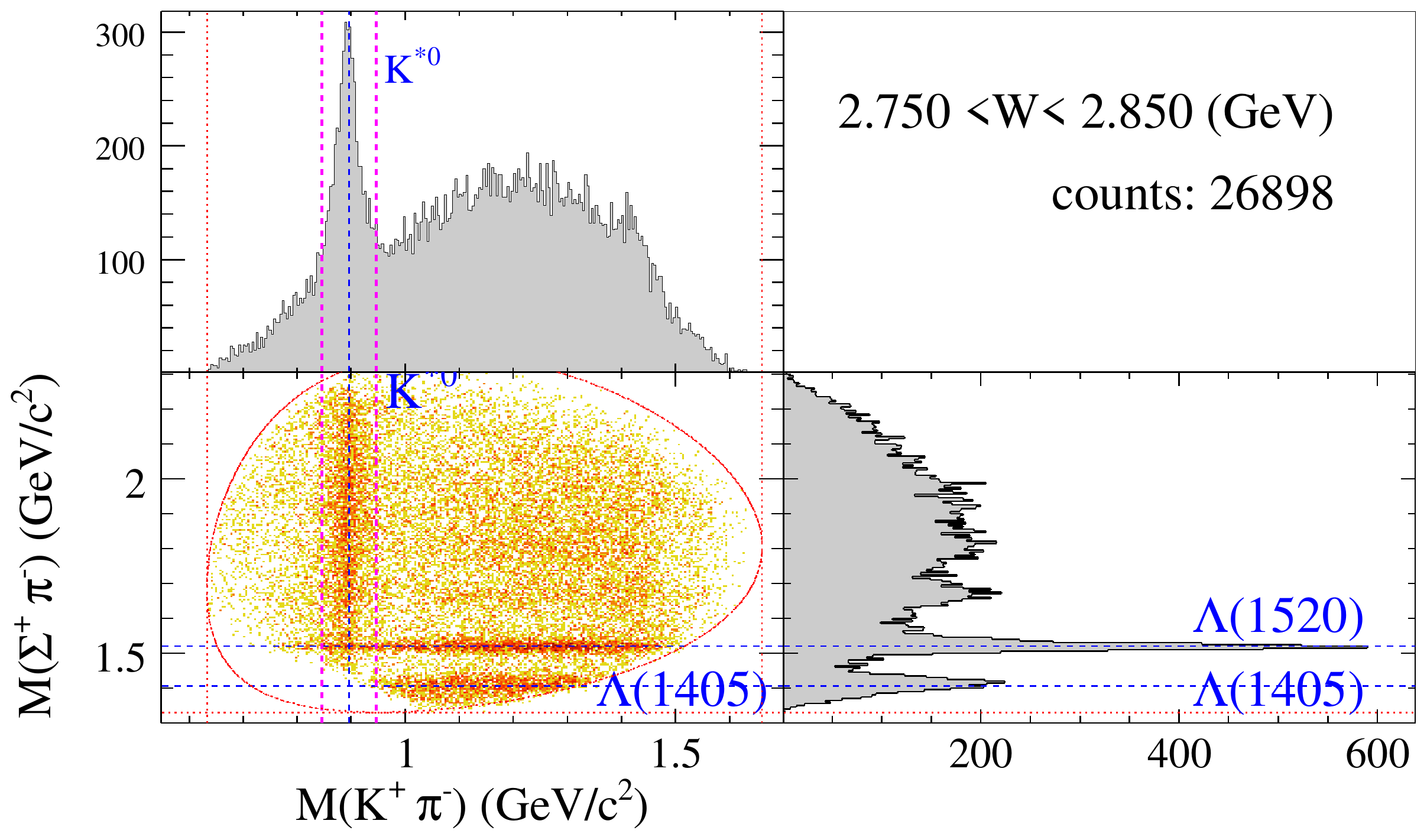}}
    \hfill
  \end{center}
  \caption[]{(Color online) $\mathrm{M}(\SigmaPlus \pim)$ versus
    $\mathrm{M}(\kp \pim)$ for four bins of increasing energy from (a)
    to (d). Clear horizontal bands corresponding to the \LambdaOne{},
    \LambdaTwo, and higher $Y^{\ast}$ resonances are seen, as well as
    a vertical band corresponding to the $K^{\ast 0}$. The contours as
    well as the dashed lines show the kinematic boundaries allowed in
    that energy bin. The blue dashed lines show the masses of each
    resonance (\LambdaOne, \LambdaTwo, \Kstar) as given by the PDG,
    while the vertical dashed lines show where the $\kp \pim$
    invariant mass is $M_{0} \pm \Gamma$, and $M_{0}$ and $\Gamma$ are
    the mass and width of the \KstarZero{} as given by the PDG.  }
  \label{fig:Dalitz:case2SigmaPlus}
\end{figure*}

After these steps, we arrive at the data set of the strong final
states of $\kp \Lambda \pizero$ and $\kp \SigmaPlus
\pim$. Figures~\ref{fig:Dalitz:case2Lambda} and
\ref{fig:Dalitz:case2SigmaPlus} show the invariant masses of $Y\pi$
against $\kp \pi$, where $Y$ and $\pi$ are the ground state hyperon
and pion in each strong final state, respectively. In each of these
Dalitz-like plots there are visible bands due to resonances in the $Y
\pi$ system and $\kp \pi$ systems.  In the four ranges of $W$ shown,
one sees the shifting overlap of the hyperons $\Sigma(1385)$,
$\Lambda(1405)$, and $\Lambda(1520)$ versus the $K^{*0}$ and $K^{*+}$.
In Section~\ref{section:YieldExtraction} the fits to extract the
yields of each excited hyperon will be discussed.

%%%%%%%%%%%%%%%%%%%%%%%%%%%%%%%% FIGURE 7 %%%%%%%%%%%%%%%%%%%%%%%%%%%%%%%%%%%%%%%%%%%%%%
%%%%%%%%%%%%%%%%%%%%%%%%%%%%%%%% END FIGURE %%%%%%%%%%%%%%%%%%%%%%%%%%%%%%%%%%%%%%%%%%%%%%

%%%%%%%%%%%%%%%%%%%%%%%%%%%%%%%% FIGURE 8 %%%%%%%%%%%%%%%%%%%%%%%%%%%%%%%%%%%%%%%%%%%%%%
%%%%%%%%%%%%%%%%%%%%%%%%%%%%%%%%  END FIGURE %%%%%%%%%%%%%%%%%%%%%%%%%%%%%%%%%%%%%%%%%%%%%%

\subsubsection{Event selection for \chSigmaPlusN{} and \chSigmaMinus}
\label{subsubsection:EventSelection:EventsForAnalysis:case3}

\begin{figure*}[t!b!hp!]
  \begin{center}
    \includegraphics[width=0.90\textwidth]{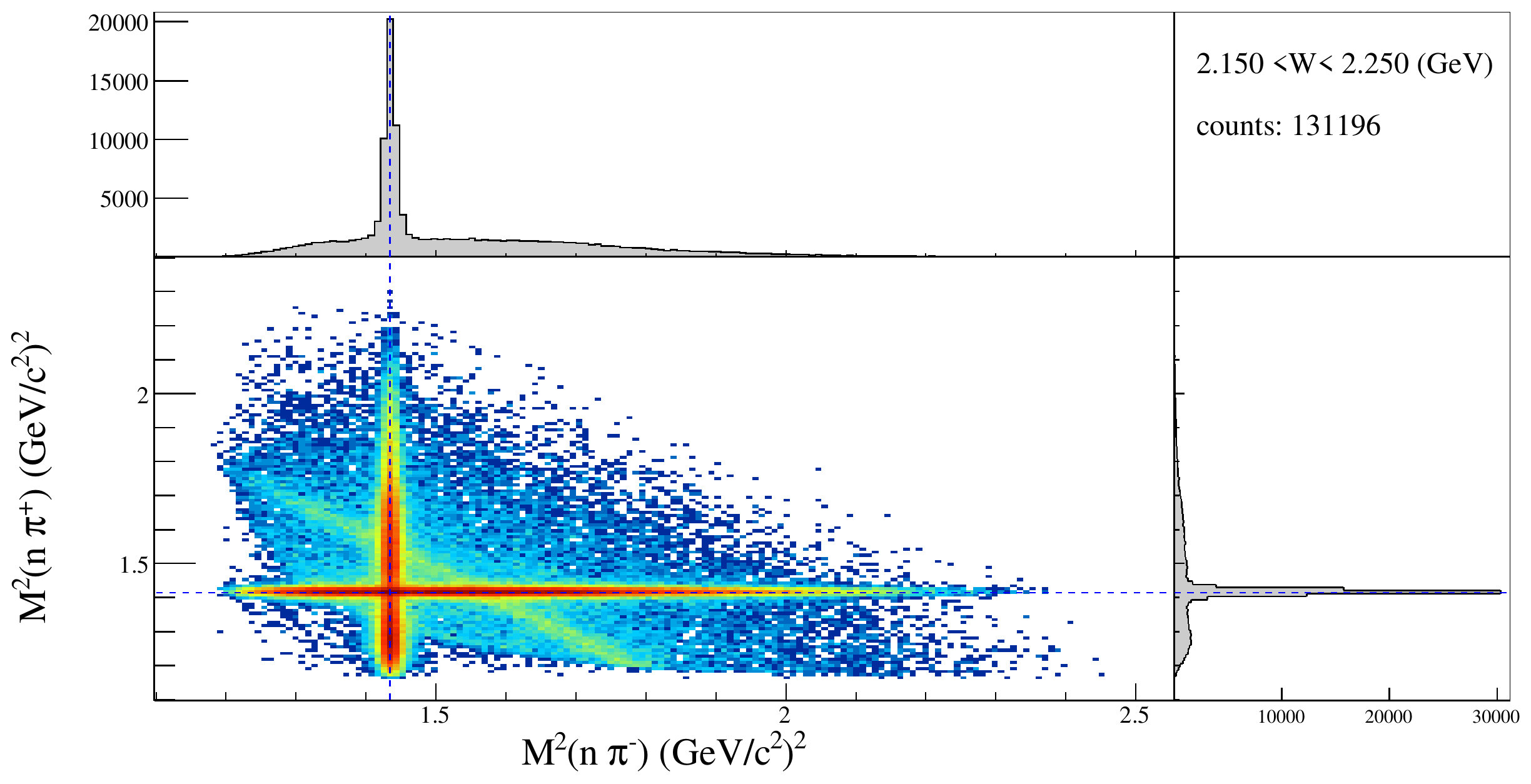}
  \end{center}
  \caption[$\mathrm{M}^{2}(\neutron \pip)$ against
    $\mathrm{M}^{2}(\neutron \pim)$.]{(Color online)
    $\mathrm{M}^{2}(\neutron \pip)$ against
    $\mathrm{M}^{2}(\neutron \pim)$ for a single energy bin, with
    projections. The bands corresponding to the $\SigmaMinus$ (vertical)
    and \SigmaPlus{} (horizontal) are clearly seen. The faint
    diagonal band corresponds to the events of $\neutron \kshort \to
    \neutron \pip \pim$. 
  }
  \label{fig:case3cross}
\end{figure*}

For these channels a final state of $\kp \pip \pim$ with a missing
neutron was required. A kinematic fit to the missing neutron mass was
applied to the selected events, retaining those with a confidence
level greater than $1\%$. Again, there are two possible hyperon
combinations, $\Sigma^{\pm} \to \neutron \pi^{\pm}$, and these
correspond to the bands shown in Fig.~\ref{fig:case3cross}, where the
plot of $\mathrm{M}^{2}(\neutron \pip)$ versus
$\mathrm{M}^{2}(\neutron \pim)$ is shown for a particular energy bin.

%%%%%%%%%%%%%%%%%%%%%%%%%%%%%%%% FIGURE 9 %%%%%%%%%%%%%%%%%%%%%%%%%%%%%%%%%%%%%%%%%%%%%%
%%%%%%%%%%%%%%%%%%%%%%%%%%%%%%%% END FIGURE %%%%%%%%%%%%%%%%%%%%%%%%%%%%%%%%%%%%%%%%%%%%%%

\begin{figure}[t!b!h!p!]
  \begin{center}
    \subfloat[Fit to $\SigmaPlus$.]
    {\label{fig:case3crossfits:SigmaPlus}\includegraphics[width=0.45\textwidth]{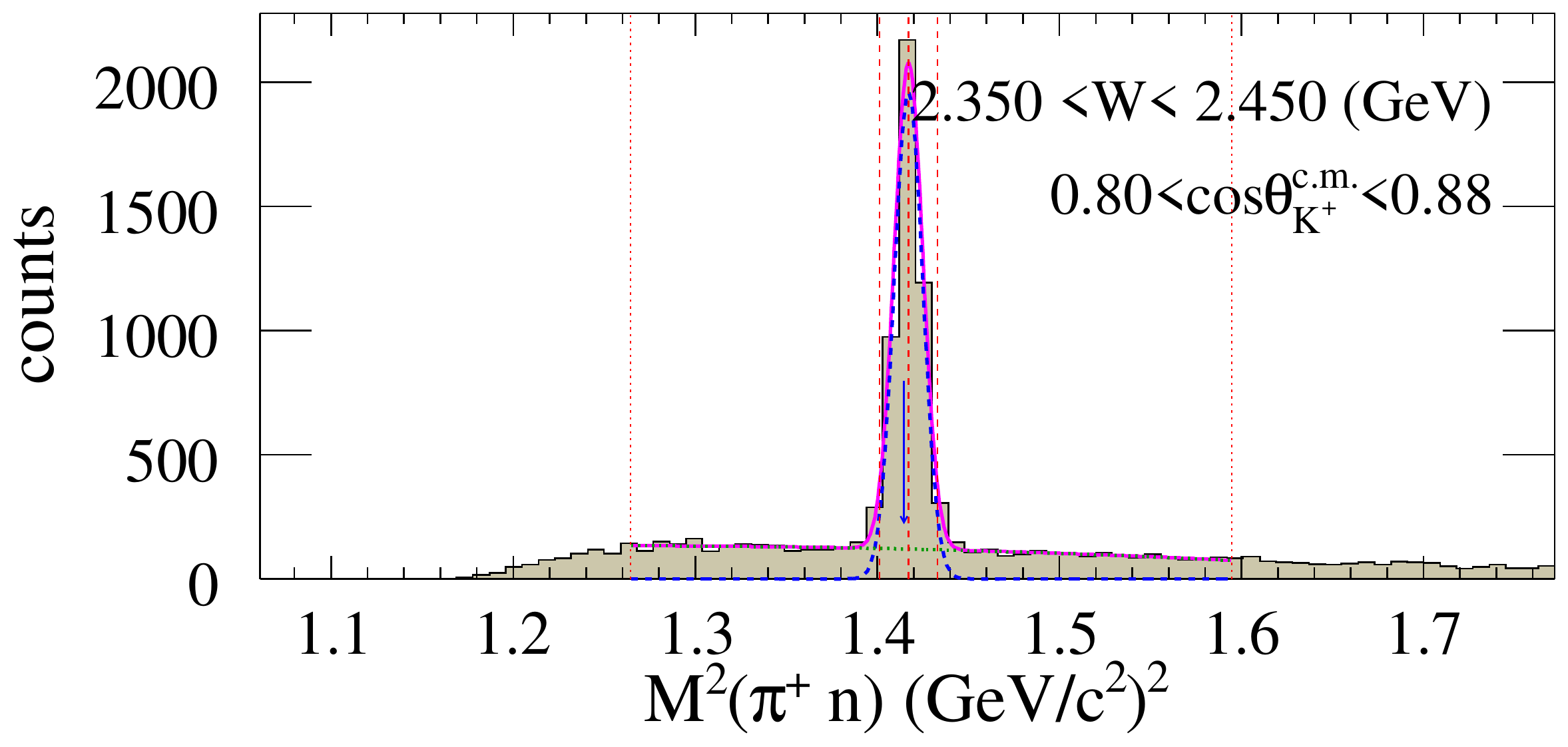}}
    \hfill
    \subfloat[Fit to \SigmaMinus.]
    {\label{fig:case3crossfits:SigmaMinus}\includegraphics[width=0.45\textwidth]{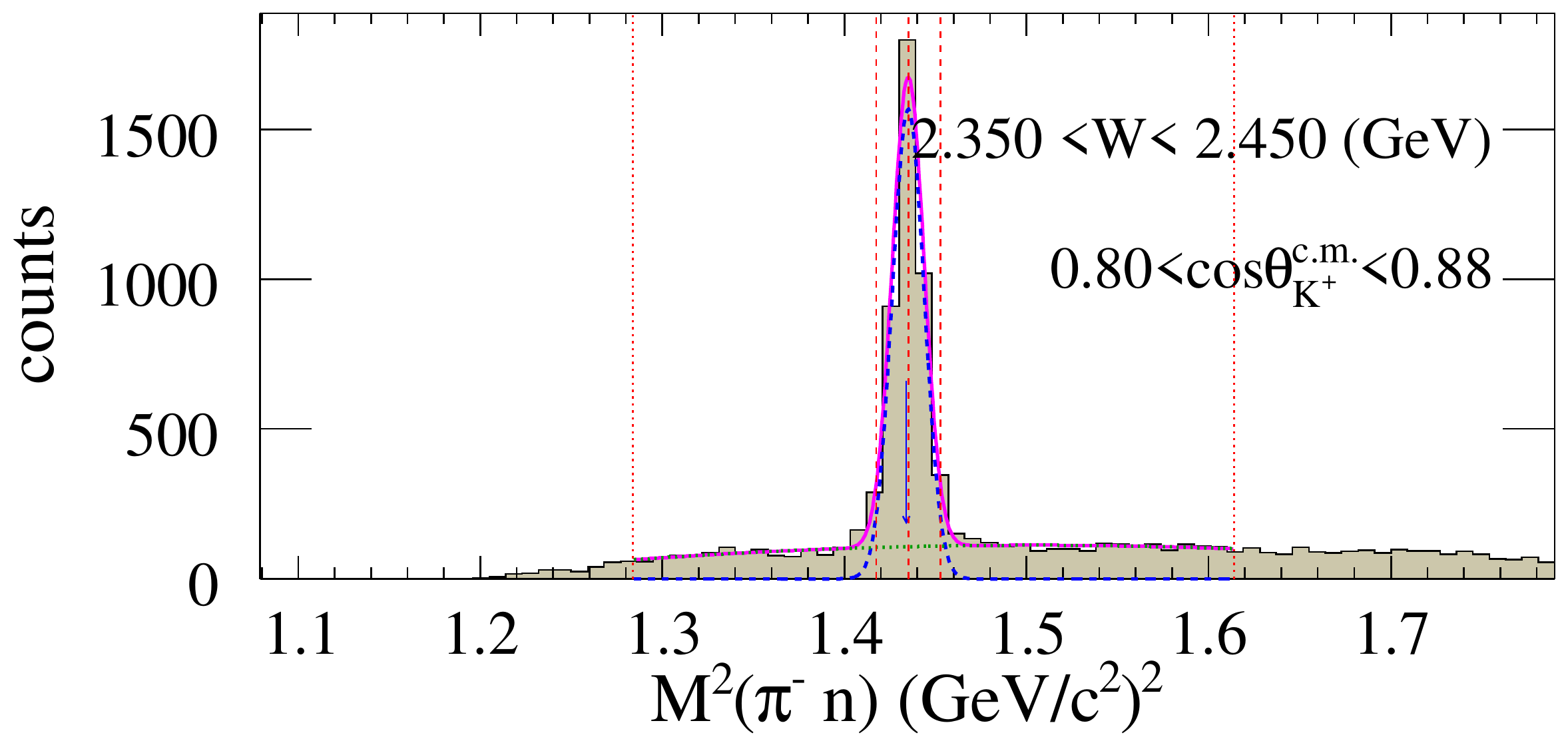}}
  \end{center}
  \caption[]{(Color online) Fits to the ground state hyperons
    \subref{fig:case3crossfits:SigmaPlus} $\SigmaPlus$ and
    \subref{fig:case3crossfits:SigmaMinus} \SigmaMinus~ for a single
    bin in energy and angle.  The data of each invariant mass squared
    are shown as the histograms, and the fits are shown as the solid
    curve (total), dashed curve (Gaussian), and dot-dashed curve
    (background). The outer dotted vertical lines show the range of
    the fit, and the inner lines show $\pm 2 \sigma$ around the peaks
    used to select events. The opposing signal region was excluded, as
    well as the $\pm 2 \sigma$ peak around the \kshort.  }
  \label{fig:case3crossfits}
\end{figure}

\begin{figure*}[p!t!b!h!]
  \begin{center}
    \subfloat[]
    {\label{fig:Dalitz:case3SigmaMinus:W1}\includegraphics[width=0.45\textwidth]{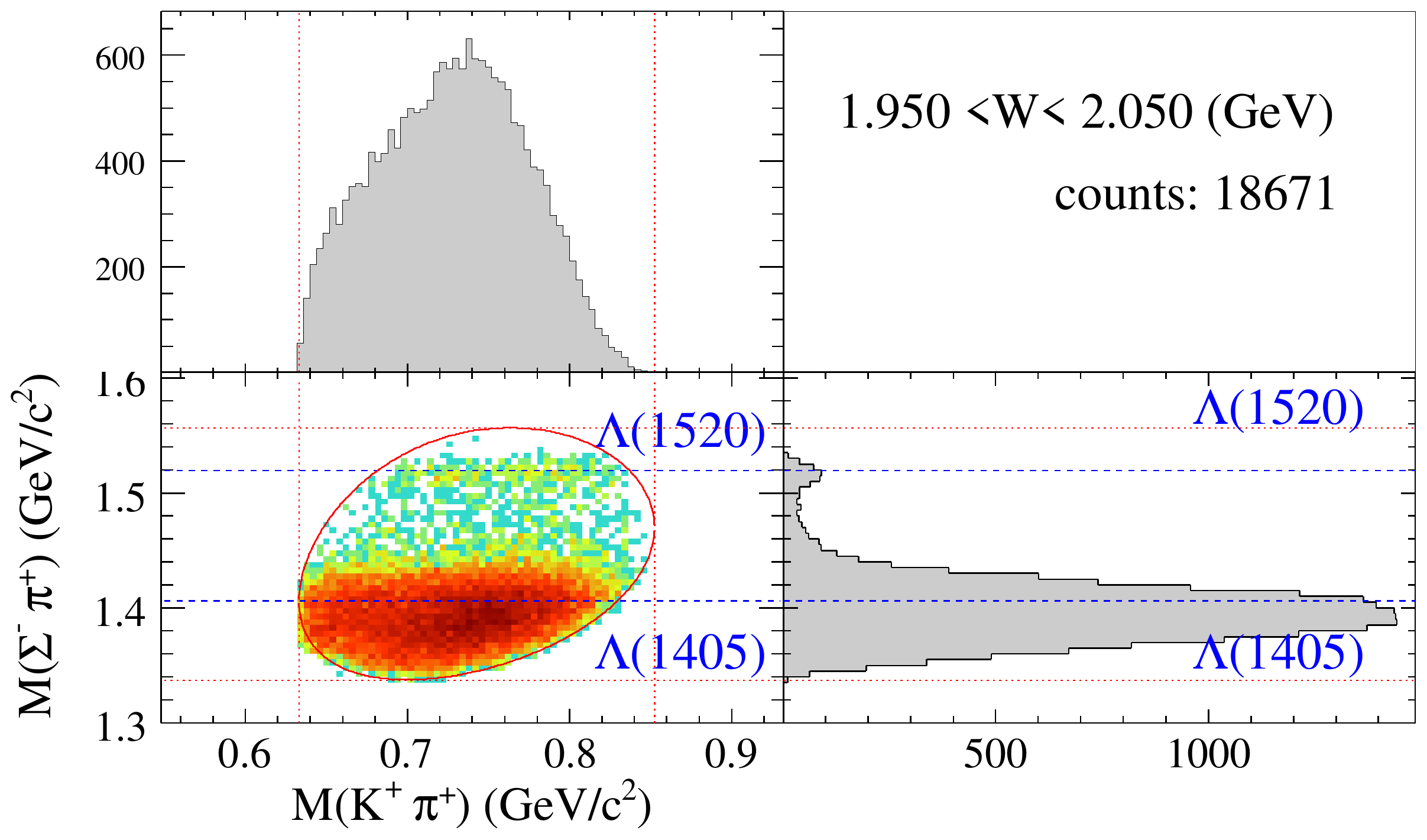}}
    \hfill
    %\subfloat[]
    %{\label{fig:Dalitz:case3SigmaMinus:W2}\includegraphics[width=0.45\textwidth]{4___Wproj2-crop.pdf}}
    %\hfill
    %
    \subfloat[]
    {\label{fig:Dalitz:case3SigmaMinus:W3}\includegraphics[width=0.45\textwidth]{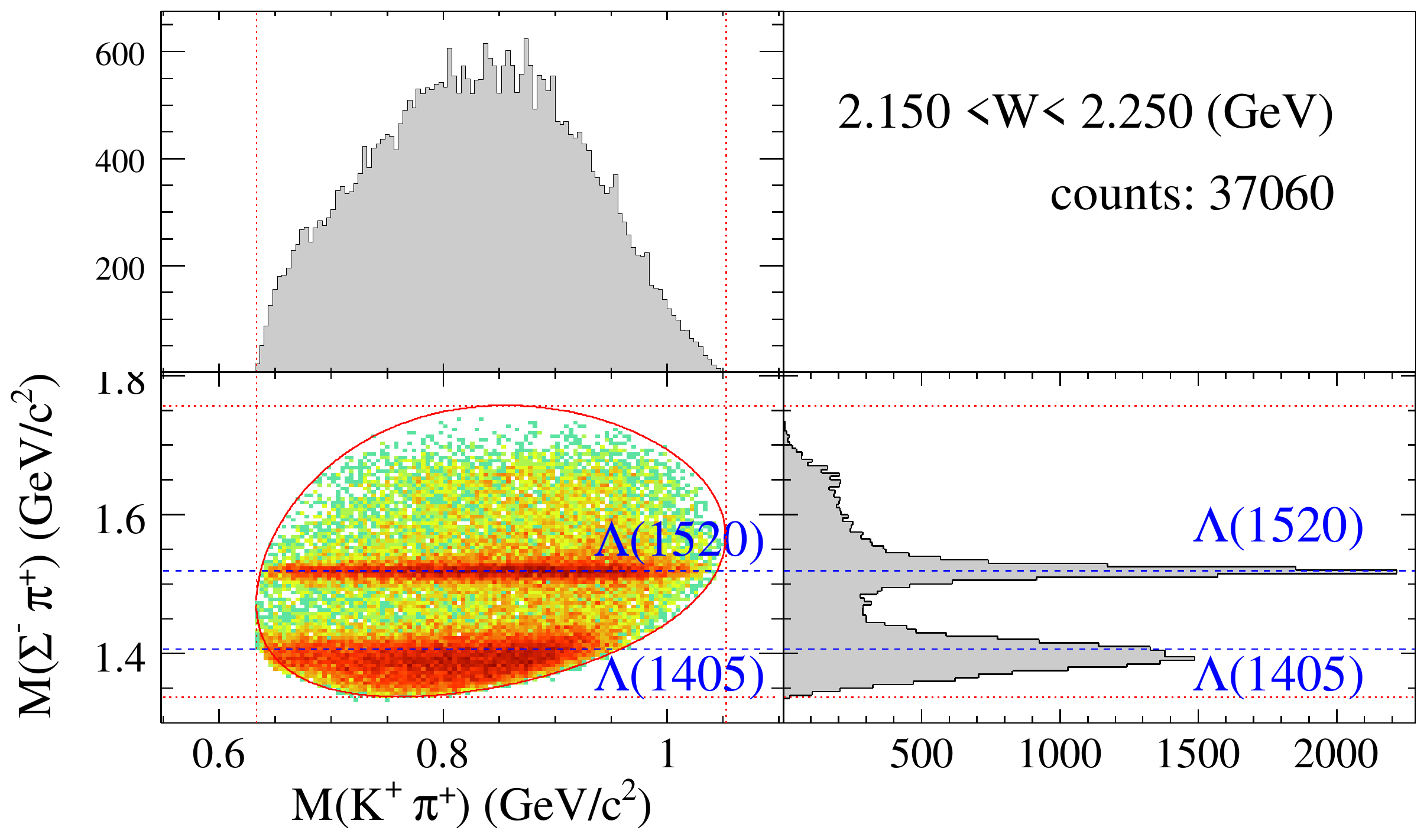}}
    \hfill
    %\subfloat[]
    %{\label{fig:Dalitz:case3SigmaMinus:W4}\includegraphics[width=0.45\textwidth]{4___Wproj4-crop.pdf}}
    %\hfill
    %
    %\subfloat[]
    %{\label{fig:Dalitz:case3SigmaMinus:W5}\includegraphics[width=0.45\textwidth]{4___Wproj5-crop.pdf}}
    %\hfill
    %\subfloat[]
    %{\label{fig:Dalitz:case3SigmaMinus:W6}\includegraphics[width=0.45\textwidth]{4___Wproj6-crop.pdf}}
    %\hfill
    %
    \subfloat[]
    {\label{fig:Dalitz:case3SigmaMinus:W7}\includegraphics[width=0.45\textwidth]{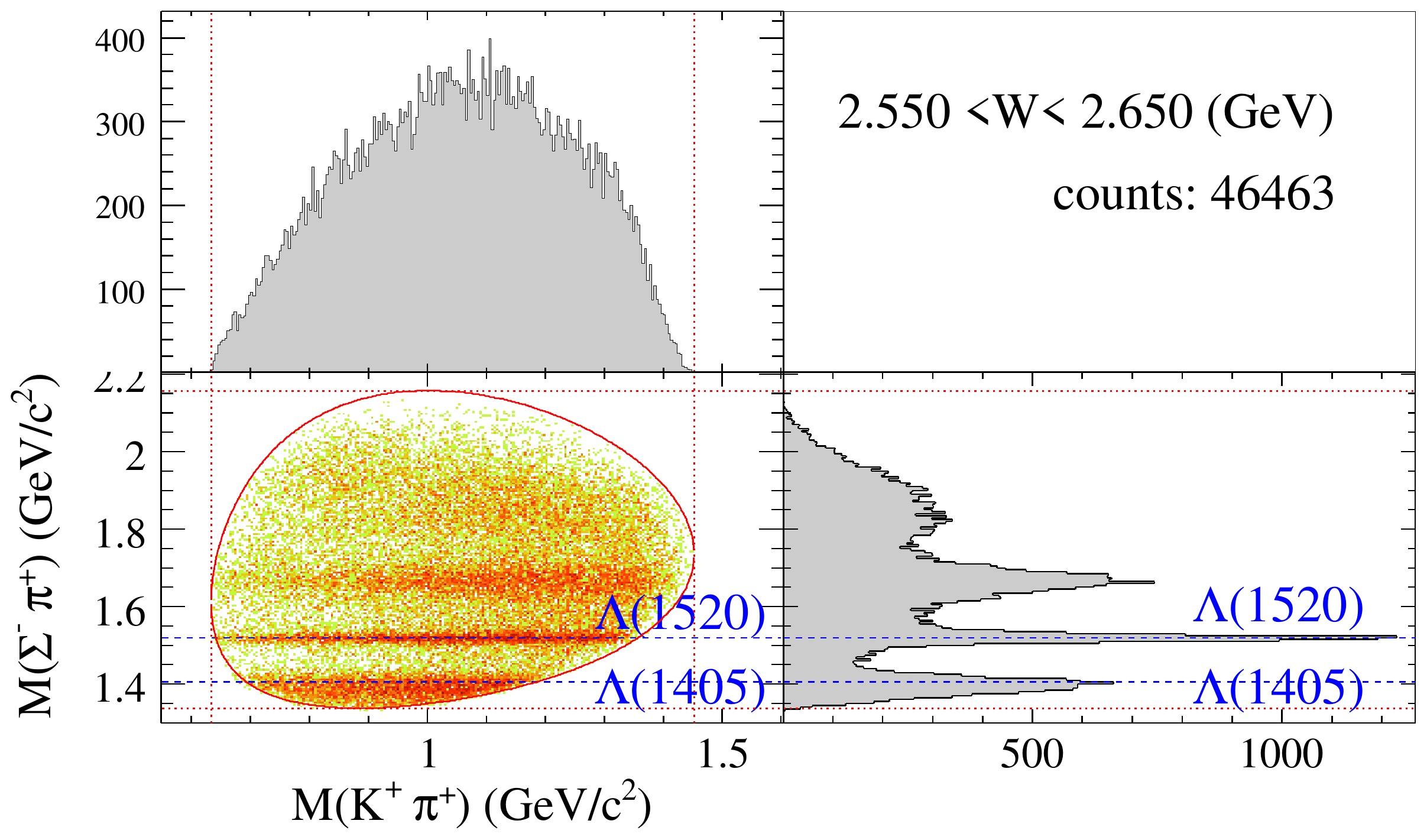}}
    \hfill
    %\subfloat[]
    %{\label{fig:Dalitz:case3SigmaMinus:W8}\includegraphics[width=0.45\textwidth]{4___Wproj8-crop.pdf}}
    %\hfill
    %
    \subfloat[]
    {\label{fig:Dalitz:case3SigmaMinus:W9}\includegraphics[width=0.45\textwidth]{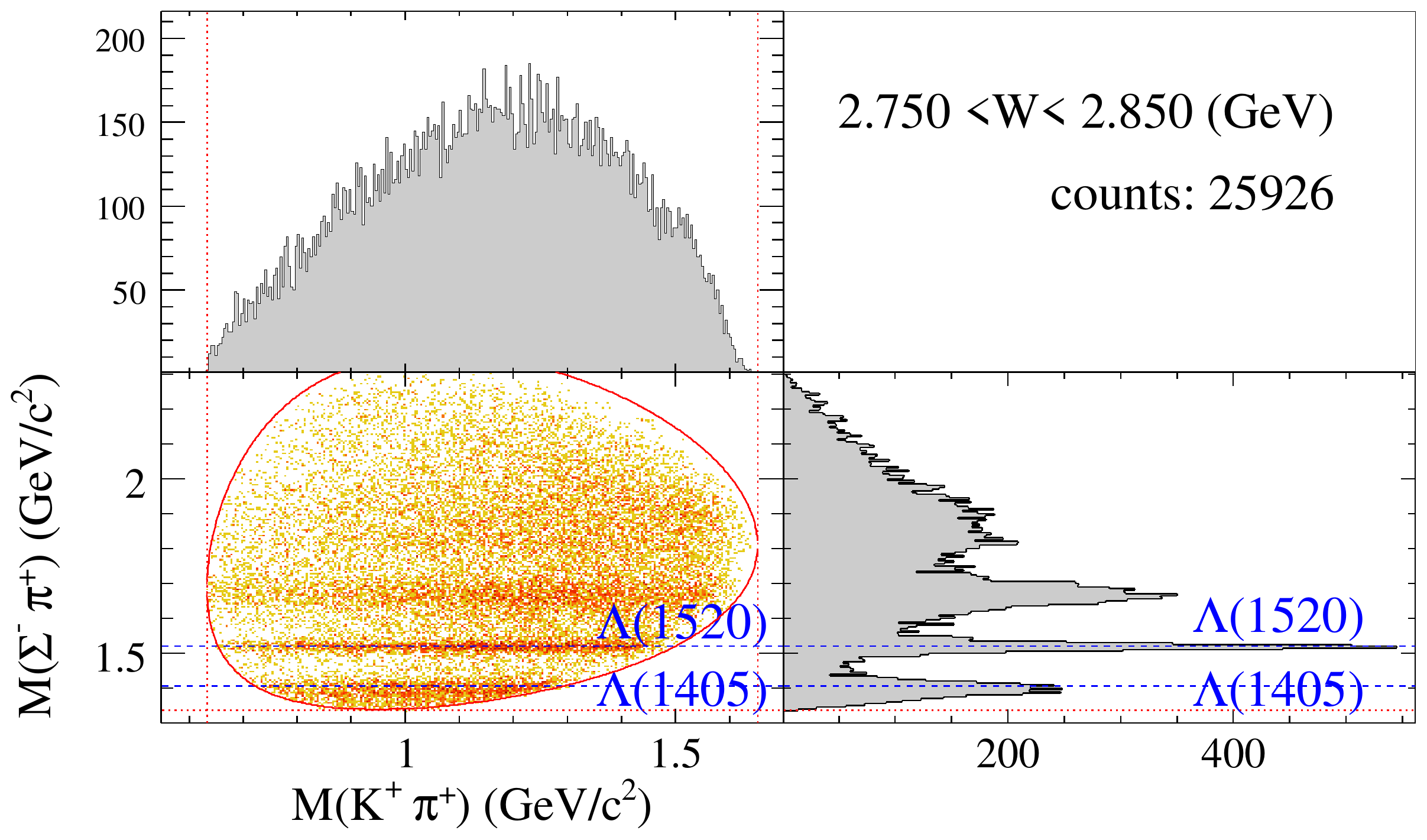}}
    %\hfill
    %
  \end{center}
  \caption[]{(Color online) $\mathrm{M}(\SigmaMinus \pip)$ versus
    $\mathrm{M}(\kp \pip)$ for four energy bins increasing from (a) to
    (d). Clear horizontal bands corresponding to the \LambdaOne{},
    \LambdaTwo, and higher $Y^{\ast}$ resonances are seen. The blue
    dashed lines show the nominal masses of the \LambdaOne{} and
    \LambdaTwo{} from the PDG. Note that in this channel the
    combination of $\kp \pip$ shows no resonant structure. The
    contours as well as the dashed lines show the kinematic boundaries
    allowed in that energy bin.  }
  \label{fig:Dalitz:case3SigmaMinus}
\end{figure*}

To isolate the events for \SigmaPlus{} (\SigmaMinus), we projected the
distributions onto $\mathrm{M}^{2}(\neutron \pip)$
($\mathrm{M}^{2}(\neutron \pim)$) and fit the hyperon peaks with a
Gaussian and second order background polynomial. Examples are shown in
Fig.~\ref{fig:case3crossfits}. A region of $\pm 2 \sigma$ around each
peak was chosen as the signal, and the overlap region of the two peaks
was excluded from each signal. Also, since there is a band
corresponding to $\kshort \to \pip \pim$ events seen in
Fig.~\ref{fig:case3cross}, we followed a similar procedure for
$\mathrm{M}^{2}(\pip \pim)$ and excluded events within $\pm 2 \sigma$
of the $\kshort$ peak also.  Figure~\ref{fig:Dalitz:case3SigmaMinus}
shows the invariant mass combinations of \chSigmaMinus versus
\kp\pip. Note that in the \kp \pip \SigmaMinus{} final state, the
combination of $\kp \pip$ has no resonant structure.  After this
selection of $\Sigma^{\pm}$ events, the strong final states of $\kp
\Sigma^{\pm} \pi^{\mp}$ were in hand.

%%%%%%%%%%%%%%%%%%%%%%%%%%%%%%%% FIGURE 10 %%%%%%%%%%%%%%%%%%%%%%%%%%%%%%%%%%%%%%%%%%%%%%
%%%%%%%%%%%%%%%%%%%%%%%%%%%%%%%% END FIGURE %%%%%%%%%%%%%%%%%%%%%%%%%%%%%%%%%%%%%%%%%%%%%%

\subsubsection{Event selection for \chSigmaZero}
\label{subsubsection:EventSelection:EventsForAnalysis:SigmaZero}

\begin{figure}[t!b!h!p!]
  \centering
  \subfloat{\label{fig:case2MM2:2_17}
    \includegraphics[width=0.48\textwidth]{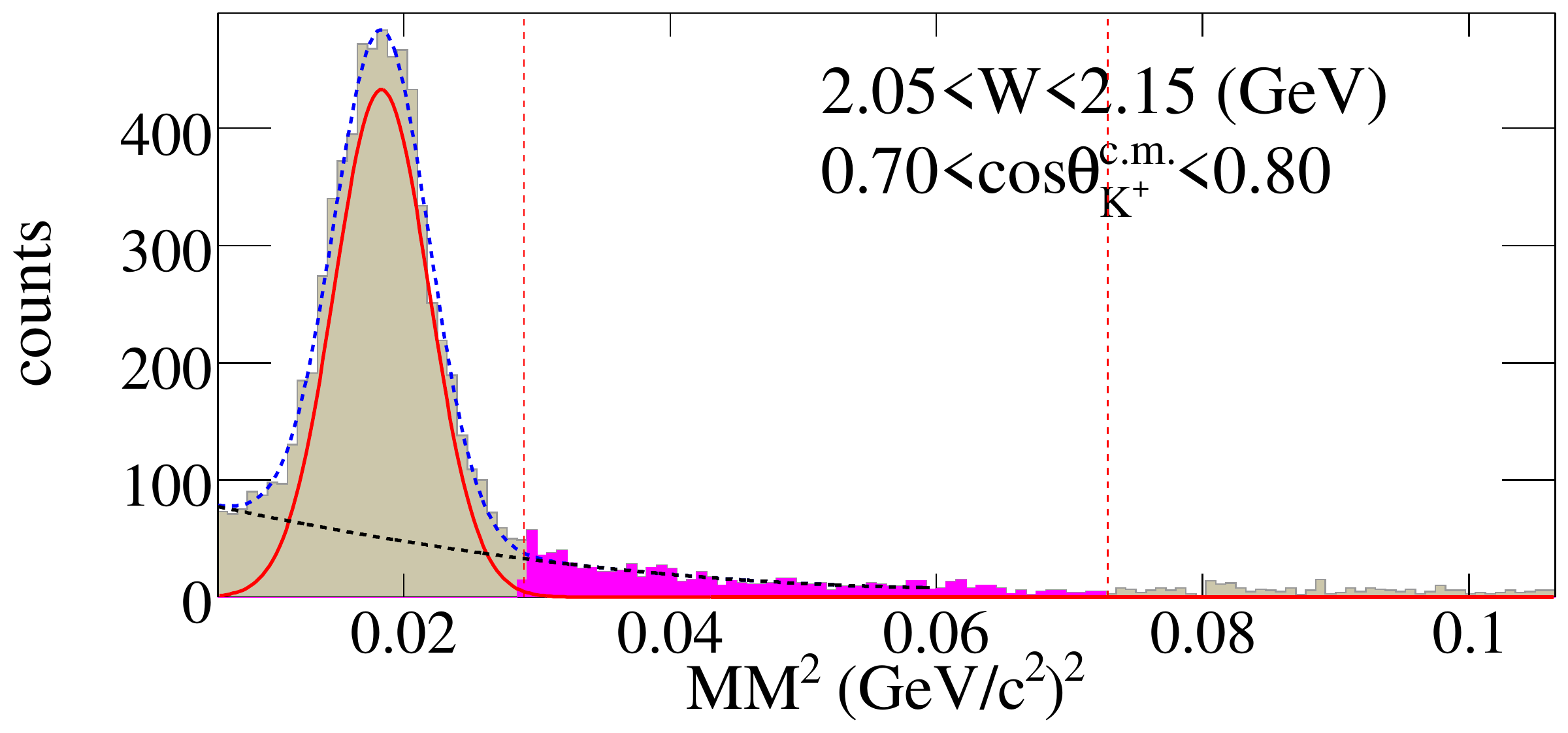}
  } \hfill
  \subfloat{\label{fig:case2MM2:7_18}
    \includegraphics[width=0.48\textwidth]{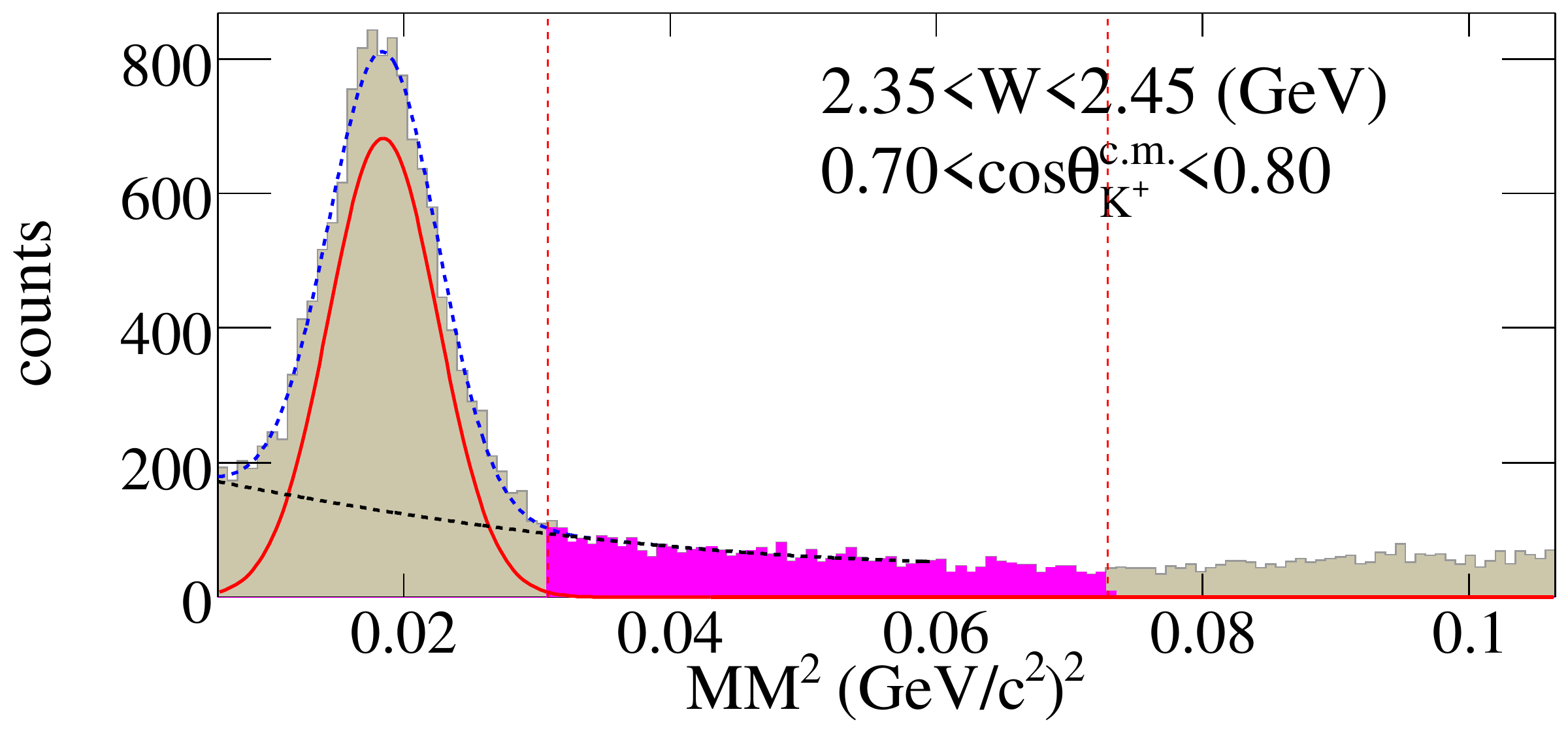}
  } \hfill
  \caption{(Color online) Examples of the $\gamma p \to K^+ \proton \pim
    (X)$ missing mass squared ($\mathrm{MM}^{2}$) spectrum for
    selected kinematic bins.  The vertical dashed lines show the
    selection range for the \chSigmaZero{} channel between the $\pi^0$
    and $2\pi^0$ limits.
    \label{fig:case2MM2}}
\end{figure}

For the \chSigmaZero{} channel, the reaction is $\gamma \proton \to
\kp \SigmaZero \pizero$, with $\SigmaZero \to \gamma \Lambda$, and
$\Lambda \to \proton \pim$. In this case, we were unable to detect the
$\pizero$ as well as the $\gamma$ from the \SigmaZero{} decay,
therefore making a kinematic fit impossible. Instead, we fitted the
missing mass squared $(\mathrm{MM}^{2})$ with a Gaussian peak for the
\pizero{} and a second order polynomial for the signal region, and
required that $\mathrm{MM}^{2}$ be more than $3\sigma$ above the
\pizero{} peak. Two examples of the selection of $\mathrm{MM}^{2}$ are
shown in Fig.~\ref{fig:case2MM2}, where the selection ranges are shown
by the dashed vertical lines.

To select the \SigmaZero{} events from this channel, the invariant
mass squared of the \proton{} and \pim{} was plotted for each bin, and
a fit with a Gaussian and a second order polynomial background was
performed. The $\pm 3 \sigma$ region around the $\Lambda$ peak was
retained.  The missing mass off the \kp{} then gave the
$\Sigma^0\pi^0$ line shape. For the strong final state $\kp \SigmaZero
\pizero$, there are possible hyperon as well as $K^{\ast +}$
resonances, and in Section~\ref{section:YieldExtraction} the
extraction of hyperon events is discussed.

Table~\ref{tab:num_events} shows the number of events for each channel
after the selections shown in this section. Further selections to
isolate the states of interest will be shown later.

%%%%%%%%%%%%%%%%%% end of file %%%%%%%%%%%%%%%%%%%%%%%%%%%%%%%%%%%%%%%%%%%%

%*************************************************************************************
%                                                                                    *
%                           Acceptance and Normalization                             *
%                                                                                    *
%*************************************************************************************
% Revisions:
% 8-28-12 RS just a few wording changes
% 9-19-12 RS wording changes...
% 1-11-13 RS After CLAS comments

\section{Acceptance and Normalization}
\label{section:AcceptanceAndNormalization}

To understand and correct for the CLAS detector acceptance, a large
number of Monte Carlo (MC) events were processed using the GEANT-based
standard CLAS simulation package GSIM. After generating the events of
interest, the events were passed through the detector simulation, and
the momenta were smeared to match the data. An earlier detailed
analysis of the g11a data showed~\cite{MW-thesis} that the trigger
condition for this run was not ideally simulated, so an ad hoc trigger
efficiency correction of $\sim 5\%$ was applied depending on the event
kinematics. After all corrections were made, the simulated events were
passed through the same analysis procedures as the data.

One final correction was applied for the events of interest that had a
$\Lambda$ in the strong final state. As mentioned in
Section~\ref{section:ExperimentalSetup}, the hardware trigger for this
run required that two particles register hits in separate sectors of
the Start Counter. In the case of an event involving a $\Lambda$ ($c
\tau = 7.89 ~\mathrm{cm}$~\cite{Beringer:1900zz}), there was a small
probability of the $\Lambda$ decaying outside of the Start Counter,
and this detail of the trigger was not simulated in software. To
remedy this, events in the simulation were removed based on whether
the secondary vertex was geometrically outside of the Start
Counter. The effect of this correction was stronger at higher
energies, and for events with the kaon going backward in the
center-of-mass frame, so that for most bins the correction was less
than $\sim 3\%$, while for some bins it was as high as $10\%$. For the
other ground-state hyperons \SigmaPlus{} and \SigmaMinus{} ($c \tau =
2.404 ~\mathrm{cm}$ and $4.434 ~\mathrm{cm}$, respectively), the
effect of the $\Sigma^{\pm}$ decaying beyond the Start Counter was
found to be negligible.

The photon flux in each energy bin was determined so that
differential cross sections could be computed. This was done using the
CLAS-standard method based on counting out-of-time electrons in the
photon tagger within well-defined time windows, and correcting for the
measured $\simeq 70\%$ transmission of photons from the tagger to the
physics target.  Other corrections were made to handle tagger counters
not in the primary trigger and to account for the measured $\simeq
85\%$ data acquisition livetime for this data set.

%%%%%%%%%%%%%%%%% end of file %%%%%%%%%%%%%%%%%%%%%%%%%%%%%%%%%%%%%%%%

%*************************************************************************************
%                                                                                    *
%                        Yield Extraction of Excited States                          *
%                                                                                    *
%*************************************************************************************
% Revisions:
% 8-28-12 RS lots of wording and grammar
% 8-29-12 RS improve discussion of 1385 fits
% 8-17-12 RS spelling
% 1-11-12 RS Final CLAS comments

\section{Yield Extraction of Excited Hyperons}
\label{section:YieldExtraction}

Our method of extracting the strong final state yields used
simulations of the signal reaction of interest and of the background
reactions in each channel. A fit in the excited hyperon spectrum was
performed independently in each bin of center-of-mass energy and kaon
angle to match the data.

As mentioned at the beginning of Section~\ref{section:EventSelection},
we extracted the $\Sigma^0(1385)$ yield in the dominant $\Lambda
\pizero$ decay channel, and with the appropriate acceptance and
branching fraction ratios scaled this down to determine the background
yields in the $\Sigma^{\pm} \pi^{\mp}$ channels. (Note that the
$\SigmaZero \pizero$ channel does not result from $\Sigma^0(1385)$
decay because the isospin coupling coefficient vanishes.)  Thus, the
$\Sigma^0(1385)$ yield to $\Sigma\pi$ was always known from indirect
measurement within any single bin of center-of-mass energy and
angle. For this reason, we first discuss extracting the
$\Sigma^0(1385) \to \Lambda \pizero$ events, and then move on to the
\LambdaOne{} yields in each $\Sigma \pi$ decay channel.

\subsection{\chLambda}
\label{subsection:YieldExtraction:chLambda}

\begin{figure*}[p!t!b!h!]
  \begin{center}
    \subfloat[$W=2.1$ GeV, $\costhetakp=.85$]
    {\label{fig:Fit1385:2_18}\includegraphics[width=0.45\textwidth]{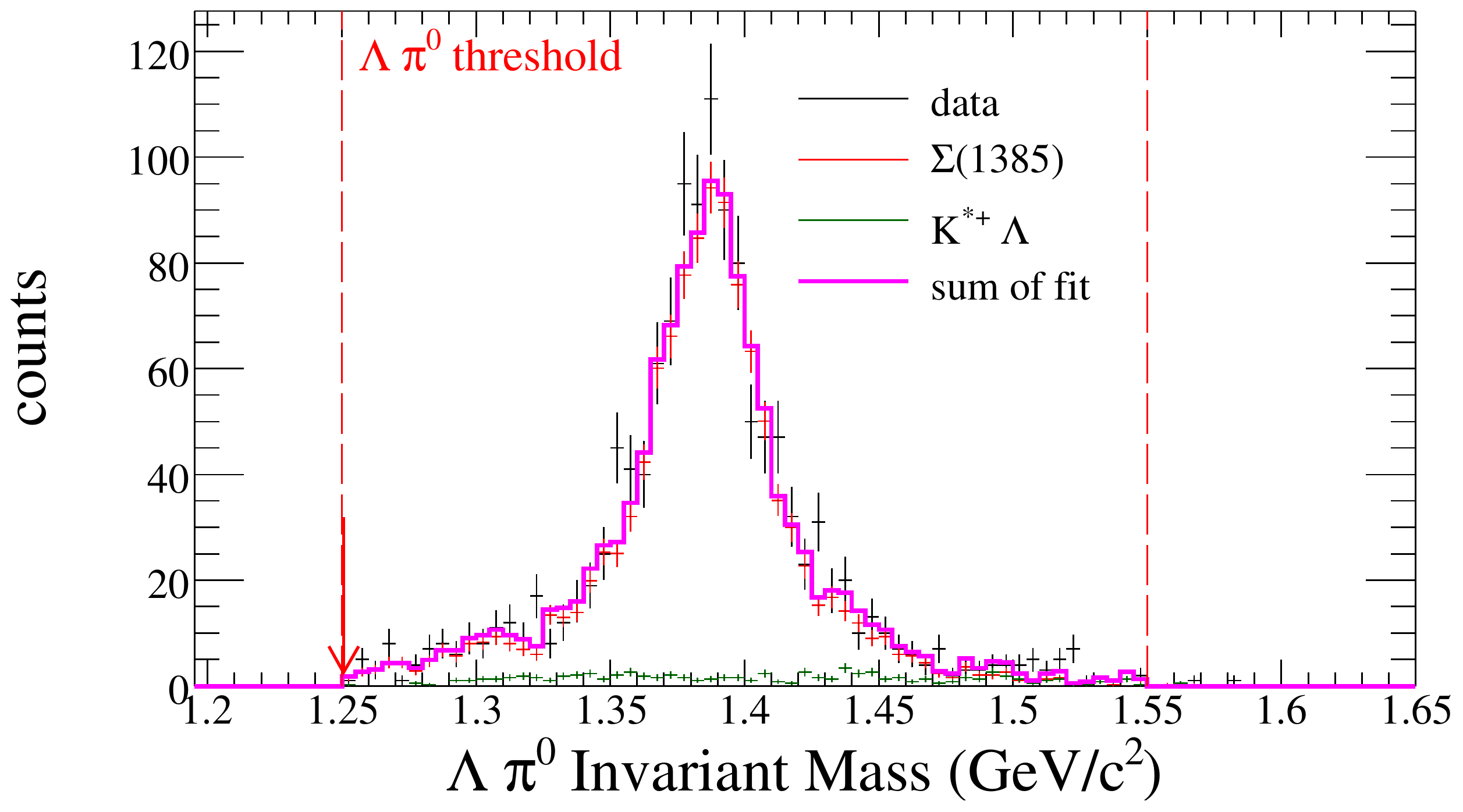}}
    \hfill
    %\subfloat[]
    %{\label{fig:Fit1385:3_17}\includegraphics[width=0.45\textwidth]{3_17-crop.pdf}}
    %\hfill
    %
    %\subfloat[]
    %{\label{fig:Fit1385:5_17}\includegraphics[width=0.45\textwidth]{5_17-crop.pdf}}
    %\hfill
    \subfloat[$W=2.6$ GeV, $\costhetakp=.65$]
    {\label{fig:Fit1385:7_16}\includegraphics[width=0.45\textwidth]{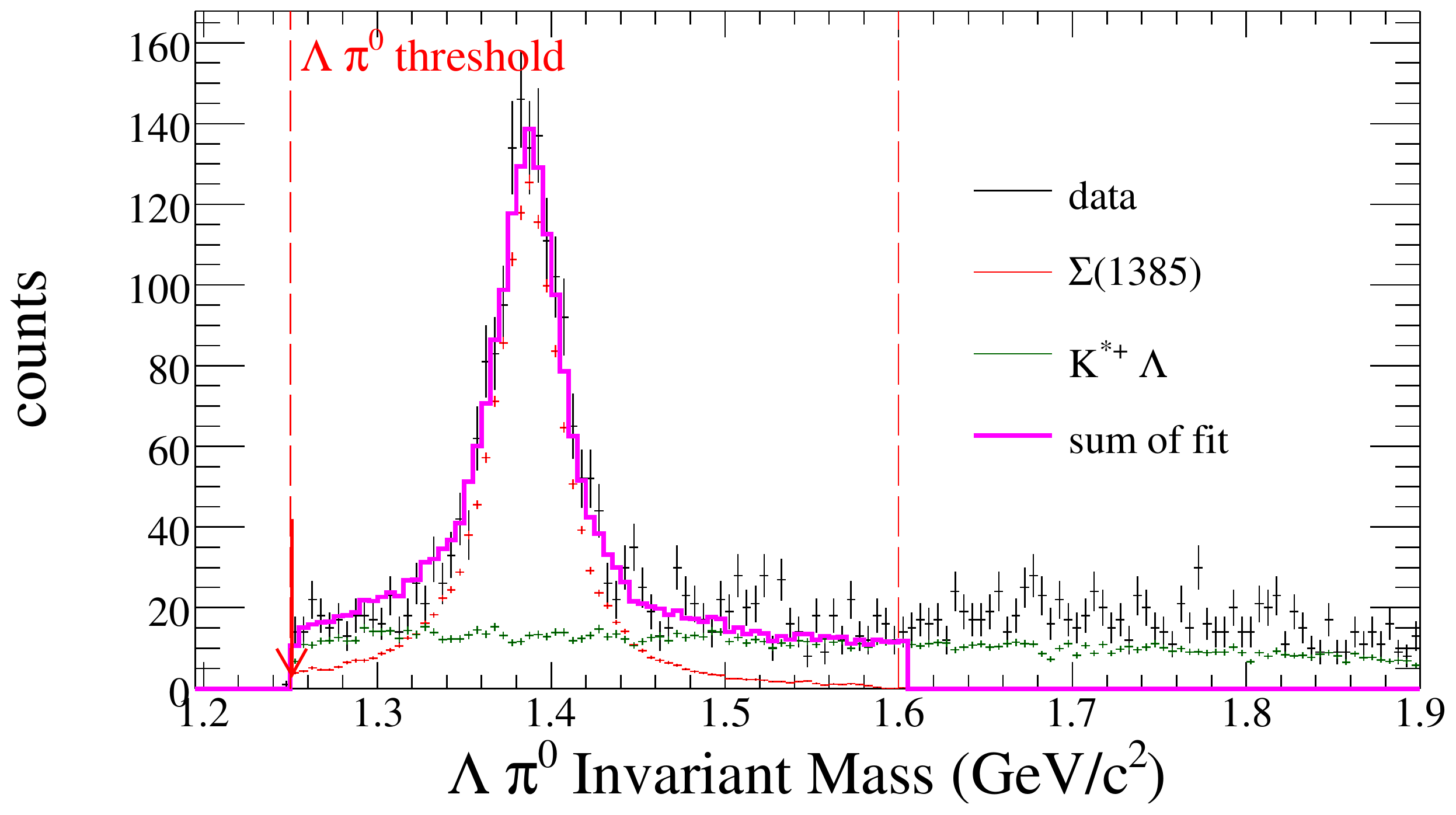}}
    \hfill
    %
    %\subfloat[]
    %{\label{fig:Fit1385:8_18}\includegraphics[width=0.45\textwidth]{8_18-crop.pdf}}
    %\hfill
    %
    %\subfloat[]
    %{\label{fig:Fit1385:9_18}\includegraphics[width=0.45\textwidth]{9_18-crop.pdf}}
    %\hfill
    %
  \end{center}
  \caption[]{(Color online) Sample fit results of the strong final
    state of $\kp \Lambda \pizero$. The events are plotted versus the
    missing mass from the \kp, which is equivalent to the invariant
    mass of the $\Lambda\pizero$ system. The data are shown by the
    black crosses, while the $\Sigma^0(1385)$ signal Monte Carlo and
    the $K^{\ast +} \Lambda$ background are shown by the red crosses
    and green circles, respectively. The sum of the Monte Carlo
    templates are shown by the solid magenta line.  }
  \label{fig:Fit1385}
\end{figure*}

For the strong final state of $\kp \Lambda \pizero$, large samples of
Monte Carlo events for the reactions $\gamma + \proton \to \kp +
\Sigma^0(1385)$ and $\gamma + \proton \to K^{\ast +} + \Lambda$ were
generated and processed. For each bin in center-of-mass energy and
kaon angle, the data and Monte Carlo events were kinematically fit and
plotted as the missing mass from the \kp, which is equivalent to the
invariant mass of the $\Lambda$ and \pizero. A fit to the data with
these Monte Carlo templates was performed by scaling each Monte Carlo
template by an overall factor. Figure~\ref{fig:Fit1385} shows the fit
results for some of these bins. In all bins the $K^{\ast +} \Lambda$
channel contributes as a smooth background.

%%%%%%%%%%%%%%%%%%%%%%%%%%%%%%%% FIGURE 13 %%%%%%%%%%%%%%%%%%%%%%%%%%%%%%%%%%%%%%%%%%%%%%
%%%%%%%%%%%%%%%%%%%%%%%%%%%%%%%% END FIGURE %%%%%%%%%%%%%%%%%%%%%%%%%%%%%%%%%%%%%%%%%%%%%%

\begin{figure*}[t!b!h!p!]
  \begin{center}
    \subfloat[Fit to $\Sigma^0(1385)$ with relativistic Breit-Wigner form.]
    {\label{fig:Fit1385cmp:rBW}\includegraphics[width=0.45\textwidth]{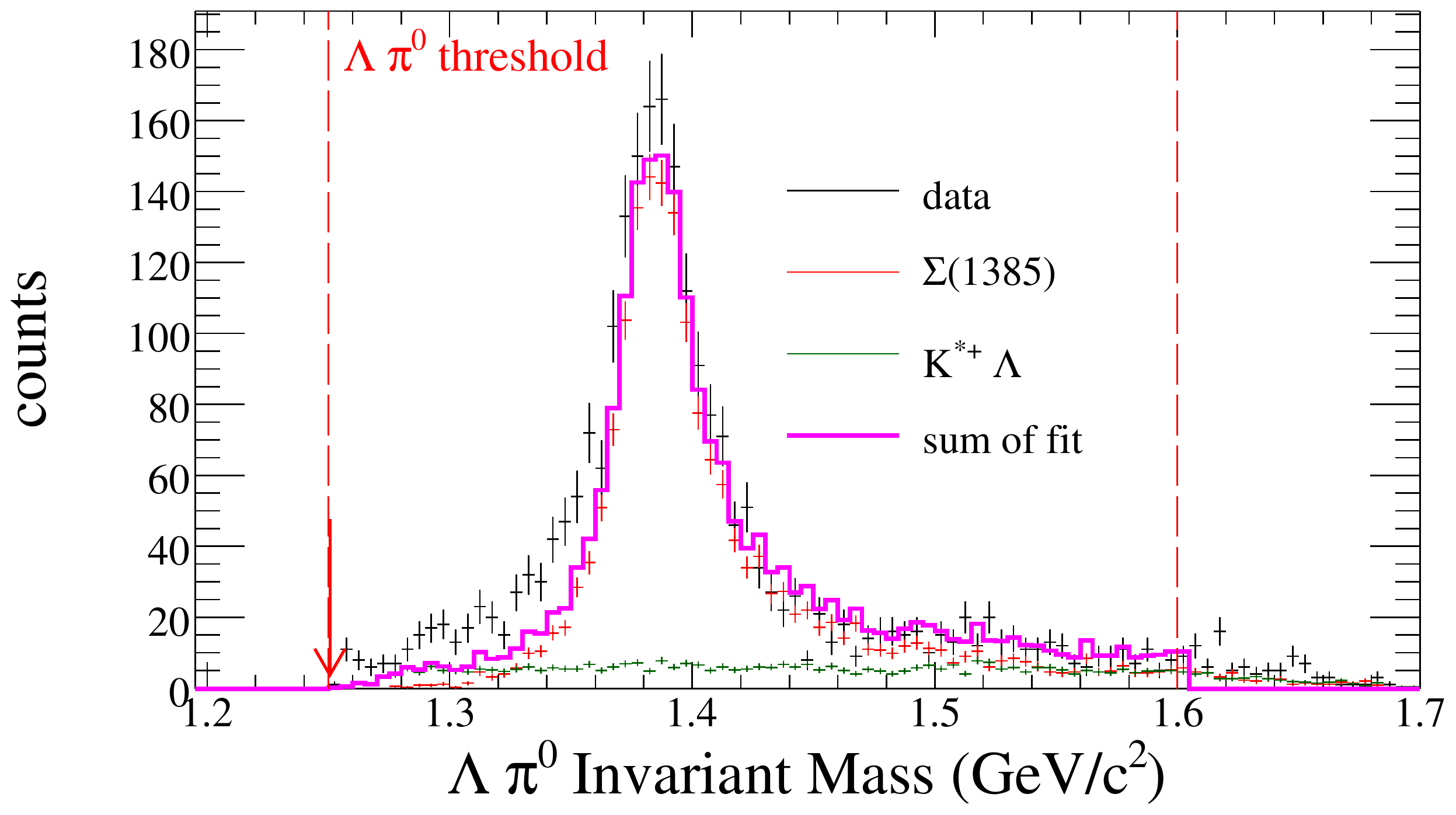}
    }
    \vspace{0cm}
    \subfloat[Fit to $\Sigma^0(1385)$ with non-relativistic Breit-Wigner form.]
    {\label{fig:Fit1385cmp:nonrBW}\includegraphics[width=0.45\textwidth]{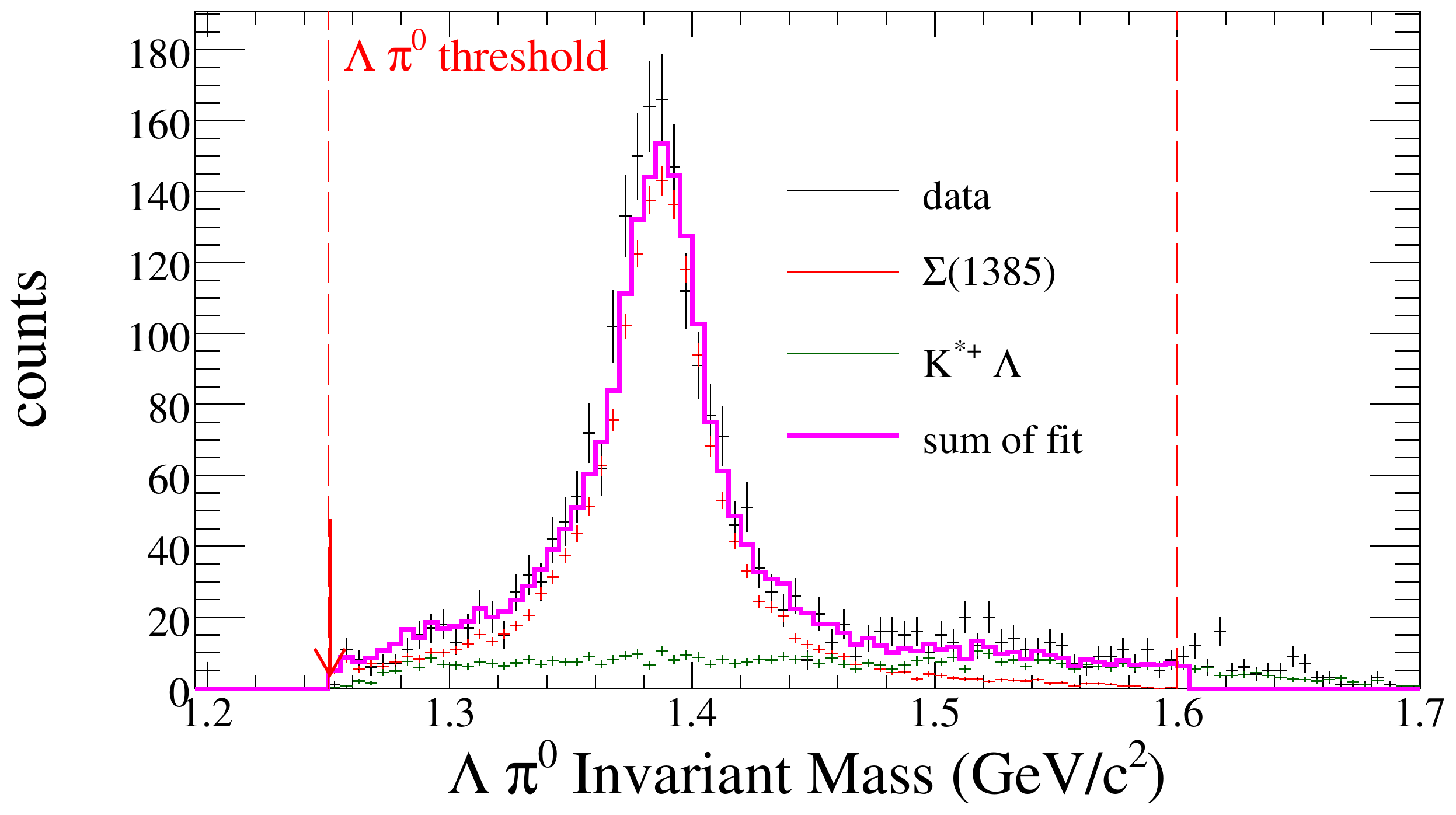}
    }
  \end{center}
  \caption[]{(Color online) Sample invariant mass spectra for W = 2.6
    GeV and $\costhetakp=.65$ showing the $\Sigma^0(1385)$ peak using
    a Monte Carlo template based on \subref{fig:Fit1385cmp:rBW}
    relativistic Breit-Wigner (mass-dependent width)
    \subref{fig:Fit1385cmp:nonrBW} non-relativistic Breit-Wigner
    (mass-independent width). The fit with the relativistic
    Breit-Wigner form clearly does not fit the data well.  }
  \label{fig:Fit1385cmp}
  % choices for bins are 2_18, 3_17, 5_17, 7_16, 8_18, 9_18
\end{figure*}

A peculiarity was noticed that the $\Sigma(1385)$ line shape could not
be fit well using a relativistic Breit-Wigner function with a
mass-dependent width. Rather, a non-relativistic Breit-Wigner function
with width independent of mass was seen to match the data much
better. The line shapes used as input to the CLAS Monte Carlo for
counts $dC_x(m)/dm$ as a function of $Y\pi$ mass $m$ were
\begin{equation}
\frac{d C_{\textrm{non-rel}}(m)}{d m} \sim \frac{\Gamma_0 / 2
  \pi}{(m_0 - m)^2 + (\Gamma_0 / 2)^2}
\label{eq:c_non_rel}
\end{equation}
and
\begin{equation}
\frac{d C_{\textrm{rel}}(m)}{d m}     \sim \frac{(2/\pi) m m_0
  \Gamma(q)}{(m_0^2 - m^2)^2 + (m_0\Gamma(q))^2}
\label{eq:c_rel}
\end{equation}
for the non-relativistic and relativistic cases, respectively.  The
mass-dependent width was $\Gamma(q) = \Gamma_{0} (q/q_{0})^{2L+1}$, in
which $q$ ($q_{0}$) is the breakup momentum of the $\Lambda \pizero$
or $\Sigma \pi$ system in the resonance rest frame at mass $m$
($m_{0}$).  The orbital angular momentum in this case is $L=1$.
Figure~\ref{fig:Fit1385cmp} shows a comparison of the fit results of
the $\Sigma^0(1385)$ peak using Monte Carlo templates generated with
the forms of Eqs.~\ref{eq:c_non_rel} and \ref{eq:c_rel}. Clearly the
relativistic Breit-Wigner template is not able to fit the data well,
essentially because the $q^3$ factor in the numerator suppresses the
yield near threshold too much.  Therefore for our present purpose, we
used the very simple non-relativistic Breit-Wigner form for fitting
the $\Sigma^0(1385)$ data in each bin.

%%%%%%%%%%%%%%%%%%%%%%%%%%%%%%%% FIGURE 14 %%%%%%%%%%%%%%%%%%%%%%%%%%%%%%%%%%%%%%%%%%%%%%
%%%%%%%%%%%%%%%%%%%%%%%%%%%%%%%% END FIGURE %%%%%%%%%%%%%%%%%%%%%%%%%%%%%%%%%%%%%%%%%%%%%%

The reason why the non-relativistic Breit-Wigner form fits better to the
data is not clear, but we note that previous experiments that
determined the $\Sigma(1385)$ mass and width based on hadronic
reactions~\cite{Beringer:1900zz}, also had difficulties in fitting to
a relativistic $P$-wave Breit-Wigner line shape, and tested
non-relativistic forms with mass-independent
widths~\cite{Borenstein,Cameron_1385,Holmgren}. Because these papers
measured the charged $\Sigma(1385)$ line shapes, where leakage due to
the \LambdaOne{} or other $\Lambda^{\ast}$ states cannot occur, this
seems to be an inherent feature of the $\Sigma(1385)$, and not due to
some unaccounted leakage in our data. Furthermore, the effect is seen
across all of our energy bins, even when below the nominal $K^{\ast
+}$ threshold or when kinematically separated from the $K^{\ast +}$.
Therefore, we conclude that this effect is not due to interference
with the $K^{\ast+}$
~\footnote{Refs.~\cite{Borenstein,Cameron_1385,Holmgren} used the
reaction $\km + \proton \to \Lambda + \pip + \pim$, where there is no
$K^{\ast +}$ background.}.

After the yields of the $\Sigma^0(1385)$ were extracted in each bin of
center-of-mass energy and angle, the differential cross sections were
calculated using the acceptance based on simulations and the photon
flux normalization. The $\Sigma^0(1385)$ differential cross section
results will be discussed in a separate
paper~\cite{crosssectionpaper}, along with those for the \LambdaOne{}
and \LambdaTwo. In this paper we focus on extracting the yields for
the \LambdaOne, for which the yields of the $\Sigma^0(1385)$ decaying
to $\Sigma^{\pm} \pi^{\mp}$ are necessary.

For each of the charged $\Sigma \pi$ channels, the
acceptance-corrected yield of the $\Lambda \pizero$ channel ($BR =
87.0 \%$) was scaled down by using the branching ratio ($BR = 11.7
\%$) and acceptance for each bin. Because the $\Sigma^0(1385)$ yield
was based on a measurement of the $\Lambda \pizero$ channel, it was
not allowed to vary when extracting the yields of the \LambdaOne.

\subsection{\SigmaPlus \pim}
\label{subsection:YieldExtraction:chSigmaPlus}

We next focus on the \chSigmaPlusP{} channel, although the other
$\Sigma \pi$ channels are quite similar in procedure.  As can be seen
from the plots of $\mathrm{M}(\SigmaPlus\pim)$ versus
$\mathrm{M}(\kp\pim)$ in Fig.~\ref{fig:Dalitz:case2SigmaPlus}, there
are contributions from the \LambdaOne, \LambdaTwo, and other excited
hyperon states, as well as from the $\KstarZero$. We model each of
these contributions separately with Monte Carlo event templates. Each
template is generated according to a relativistic Breit-Wigner form
with its resonance mass $M_{0}$ and width $\Gamma_{0}$ taken from the
PDG~\cite{Beringer:1900zz}.  We assumed a mass-dependent width of
$\Gamma(M) = \Gamma_{0} (q/q_{0})^{2L+1}$, where $q$ ($q_{0}$) is the
breakup momentum of the daughter particles in the resonance rest frame
at mass $M$ ($M_{0}$) with $L$ the orbital angular momentum.  In the
fitting procedure, only the normalization of each template was allowed
to change to get the best agreement with the data. For the
$\Sigma^0(1385)$ contribution, the yield was fixed by the \chLambda{}
channel discussed above, and therefore the yield was not allowed to
vary.

\begin{figure*}[t!b!p!h!]
    \subfloat[]{
      \label{fig:fitcase2SigmaPlus:6_13:nominal}\includegraphics[width=0.48\textwidth]{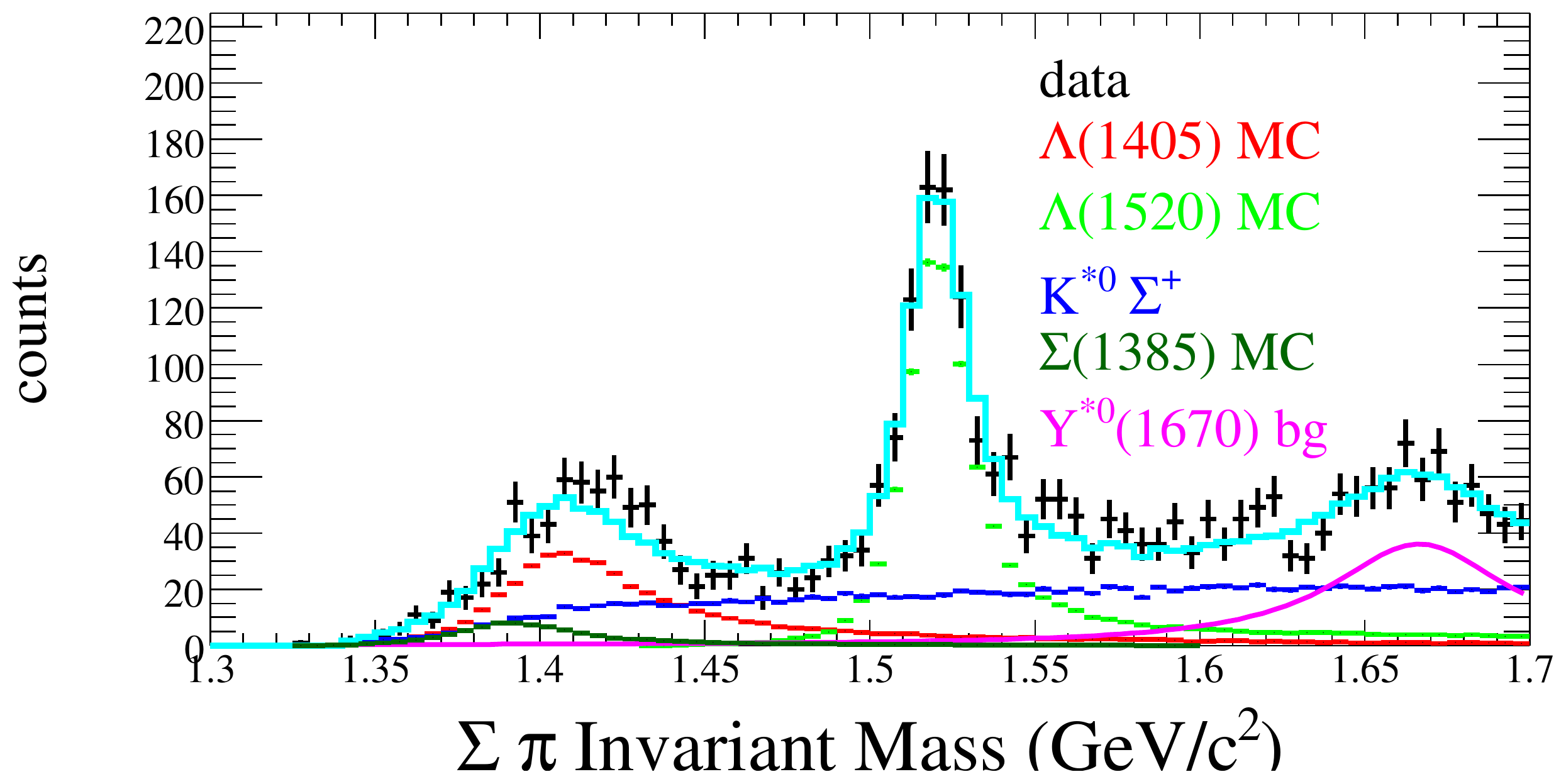}
    }
    \hfill
    \subfloat[]{
      \label{fig:fitcase2SigmaPlus:6_13:dataWithTail}\includegraphics[width=0.48\textwidth]{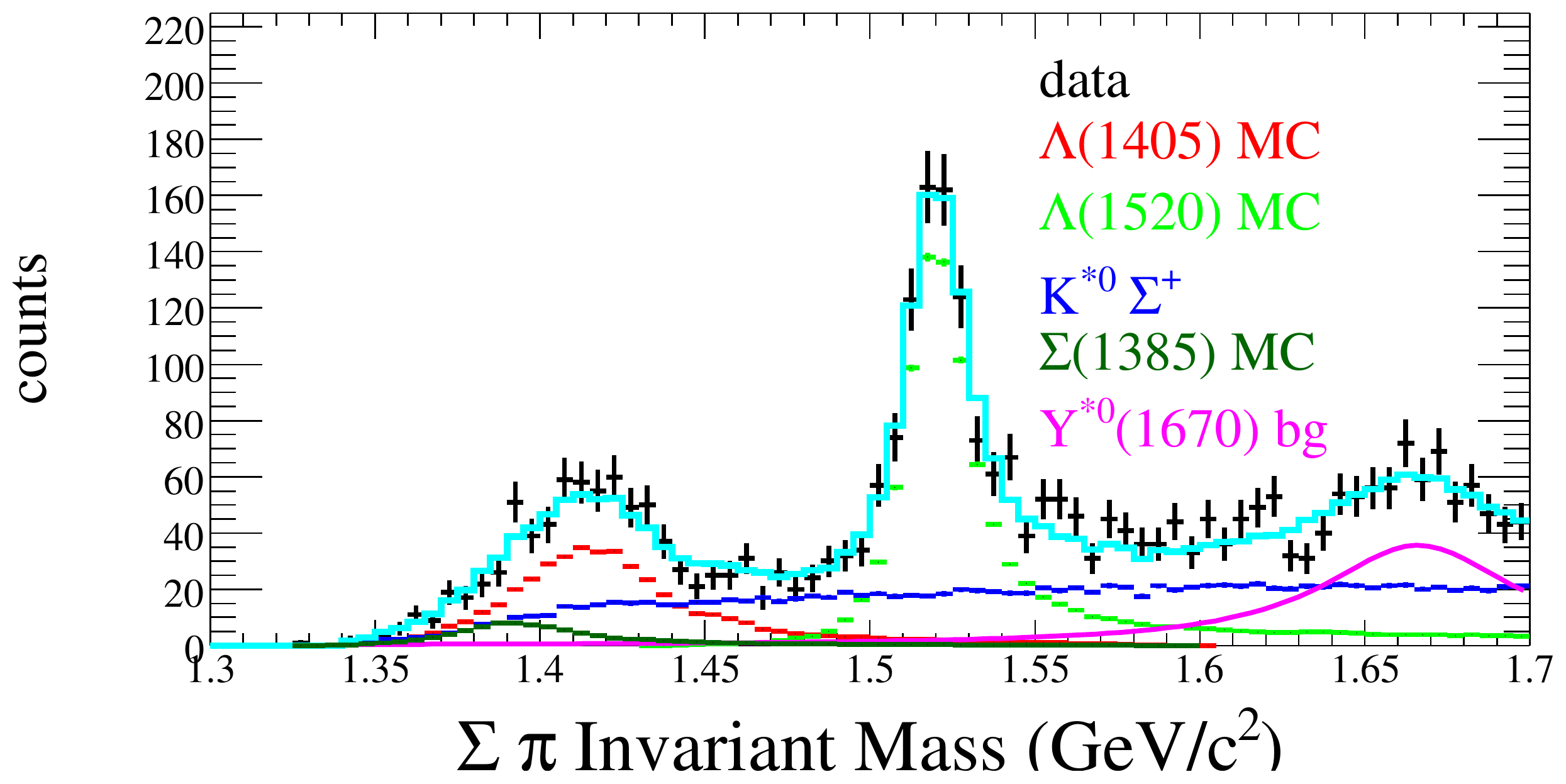}
      %\label{fig:residualcase2SigmaPlus:6_13:nominal}\includegraphics[width=0.48\textwidth]{nominal_residual_6_13-crop.pdf}
    }
    \caption{(Color online) Fit result to the strong final state of $
    \SigmaPlus\pim$ before and after Monte Carlo iteration, for
    $W=2.5$~GeV, $\costhetakp=.35$, as a function of the $\SigmaPlus
    \pim$ invariant mass. The data are shown with black crosses.
    \protect\subref{fig:fitcase2SigmaPlus:6_13:nominal} Before
    iteration.  Each Monte Carlo template and the Breit-Wigner
    function for the $Y^{\ast}(1670)$ is shown by a separate color.
    The total simulation is shown in cyan.
    \protect\subref{fig:fitcase2SigmaPlus:6_13:dataWithTail} After
    iterations of the \LambdaOne{} template.  }
    \label{fig:fitcase2SigmaPlus}
\end{figure*}

Figure~\ref{fig:fitcase2SigmaPlus} shows a fit result for the
$\SigmaPlus \pim$ invariant mass spectrum using the above templates
for a single bin in center-of-mass energy and angle, along with a
background Breit-Wigner function that fits the $Y^{\ast}$ resonance
around $1670$ \mevcc. Since our goal is to extract the \LambdaOne{}
line shape in the most model-independent way, we start with a
relativistic Breit-Wigner form based on the PDG values of mass and
width for the \LambdaOne, and this is shown in
Fig.~\ref{fig:fitcase2SigmaPlus}\subref{fig:fitcase2SigmaPlus:6_13:nominal}
as the red points.  The fit is inadequate around the \LambdaOne{}
region, showing that a simple Breit-Wigner function is not able to
describe the data well. For this reason, the template form of the
\LambdaOne{} was modified in an iterative way, as explained below.

%%%%%%%%%%%%%%%%%%%%%%%%%%%%%%%% FIGURE 15 %%%%%%%%%%%%%%%%%%%%%%%%%%%%%%%%%%%%%%%%%%%%%%
%%%%%%%%%%%%%%%%%%%%%%%%%%%%%%%% END FIGURE %%%%%%%%%%%%%%%%%%%%%%%%%%%%%%%%%%%%%%%%%%%%%%

Once an initial fit was obtained, we subtracted incoherently the
contributions due to the $\Sigma^0(1385)$, \LambdaTwo, \KstarZero, and
$Y^{\ast}(1670)$ so that the only remaining contribution was from what
should be the \LambdaOne. We call this the residual distribution for
the \LambdaOne. Because this residual distribution is the best measure
for the raw \LambdaOne{} yield, we applied an acceptance correction
based on the Monte Carlo simulation of CLAS. A large number of events
was generated flat in the $\kp \SigmaPlus \pim$ three-body phase
space, and the residual distribution was corrected as a function of
the $\SigmaPlus \pim$ invariant mass based on the acceptance of these
events. After acceptance correction, the true line shape of the
\LambdaOne{} was obtained for each energy and angle bin.

As noted above, the \LambdaOne{} was not adequately described by the
initial template, so we used the acceptance-corrected line shape
obtained with the above procedure to iterate the Monte Carlo template
for the \LambdaOne{} region.  The iteration process made use of data
summed over all kaon angles within each energy bin.
%As the residual that we
%obtain tends to have some extra strength left over in the \LambdaTwo{}
%region, we dampen the obtained line shape by an exponential above
%$1.46$ GeV.
Figure~\ref{fig:fitcase2SigmaPlus}\subref{fig:fitcase2SigmaPlus:6_13:dataWithTail}
shows the third and final iteration.  Note that the total fit is now
closer to the data, and we see how the iteration converged to stable
line shapes based on the data. Since the residual is determined by
subtracting off components such as the $\Sigma(1385)$ and the
\LambdaTwo, the residual shapes do not depend strongly on the exact
template shape we used for the \LambdaOne.
%For each iteration, it was confirmed that
%the residuals did not change appreciably for different templates.
%As the extracted \LambdaOne{} region yield converged, we used the
%third iteration of the line shape as the final template. 
The residual that was obtained from the fit using this final template
was acceptance-corrected and normalized to the photon flux, yielding
our intermediate result for $d^2\sigma / dm d\costhetakp$ in bins of
energy, angle, and $\SigmaPlus \pim$ mass $m$.

The procedure for the \chSigmaPlusN{} channel was exactly the same as
for the \chSigmaPlusP{} channel, because the physics is identical
except for the final decay of $\SigmaPlus \to \neutron \pip$. Line
shapes were obtained in each energy and kaon angle bin.  By comparing
the two $\SigmaPlus \pim$ channels we were able to check our results,
as will be shown in Section~\ref{section:LineshapeResults}.

\subsection{\chSigmaMinus}
\label{subsection:YieldExtraction:chSigmaMinus}

For the \chSigmaMinus{} channel, we followed the same procedures as
above, but in this case, with the strong final state of $\kp
\SigmaMinus \pip$, the $K \pi$ combination is exotic, and therefore we
expect no resonance. However, to accommodate the broadly distributed
events seen in Fig.~\ref{fig:Dalitz:case3SigmaMinus}, a phase-space
distribution of $\kp \SigmaMinus \pip$ was generated, and this was
used as a fit component. The line shapes were iterated as before and
then acceptance-corrected.

An interesting feature of this channel is the presence of the
$Y^{\ast}(1670)$, which shows up much more strongly compared to the
other $\Sigma \pi$ channels, as seen in
Fig.~\ref{fig:Dalitz:case3SigmaMinus}.  The PDG lists several
candidate resonances in this region, but we have not made an effort to
further identify this state.

\subsection{\chSigmaZero}
\label{subsection:YieldExtraction:chSigmaZero}

For the remaining \chSigmaZero{} channel, we did fits to the \kp{}
missing mass distribution similar to the previous cases, but since the
$\Sigma^0(1385)$ cannot decay to $\SigmaZero \pizero$ due to the
vanishing isospin factor, there is no $\Sigma^0(1385)$
contribution. The fits were performed with templates for \LambdaOne,
\LambdaTwo, and $\KstarPlus \SigmaZero$. As the $Y^{\ast}(1670)$
region does not show any prominent peaks, the Breit-Wigner function
for $Y^{\ast}(1670)$ was not used.

In summary, all $\Sigma \pi$ channels were isolated in order to
extract the line shape of the \LambdaOne{} region based on fits to the
data. The line shape templates for the region of interest were
generated in an iterative way using the data for each channel
independently, and in all cases the results showed convergence after
several iterations.

%%%%%%%%%%%%%%%%%%%%%%%%%%%%%% end of file %%%%%%%%%%%%%%%%%%%%%%%%%%%%

%*************************************************************************************
%                                                                                    *
%                                     Lineshapes                                     *
%                                                                                    *
%*************************************************************************************
% Revisions
% 08-28-12 RS Lots of text changes
% 09-06-12 RS after KM:  wordsmithing some of the discussion
% 09-20-12 RS small edits on wording
% 11-16-12 RS revisions after comments from Ad Hoc
% 01-11-13 RS final CLAS comments

\section{Line shape Results}
\label{section:LineshapeResults}

The $\Sigma\pi$ mass distributions or line shapes, $d \sigma /d
\costhetakp dm$, were obtained in each bin of center-of-mass energy
and kaon production angle, but due to limited statistics we have
summed over all angles within each energy bin to obtain a single line
shape, $\d \sigma / \d m$, for each energy bin.  Alternatively, we can
sum over mass to obtain $d\sigma / d \costhetakp$, the differential
cross section. These results will be shown in a separate paper.  Here
we compare the results of the two $\SigmaPlus \pim$ channels for
consistency, then proceed to a comparison of all three $\Sigma \pi$
channels.

\subsection{Line shape results for \SigmaPlus \pim{} channels}
\label{subsection:LineShapeResults:chSigmaPlus}

\begin{figure*}[t!b!p!h!]
  \begin{center}
    \subfloat[]
    {\label{fig:lineshape_SigmaPlus:W1}\includegraphics[width=0.49\textwidth]{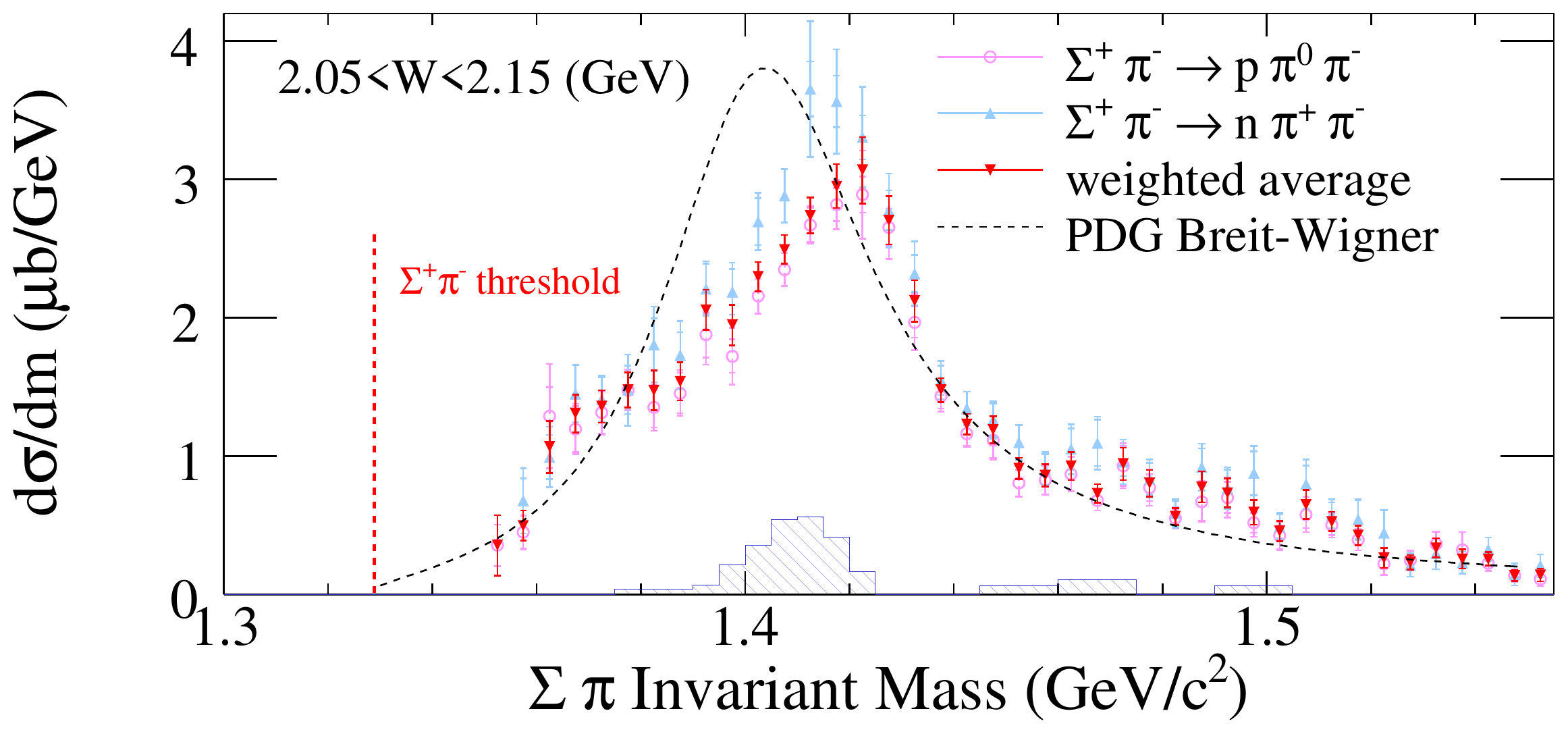}}
    \hfill
    \subfloat[]
    {\label{fig:lineshape_SigmaPlus:W4}\includegraphics[width=0.49\textwidth]{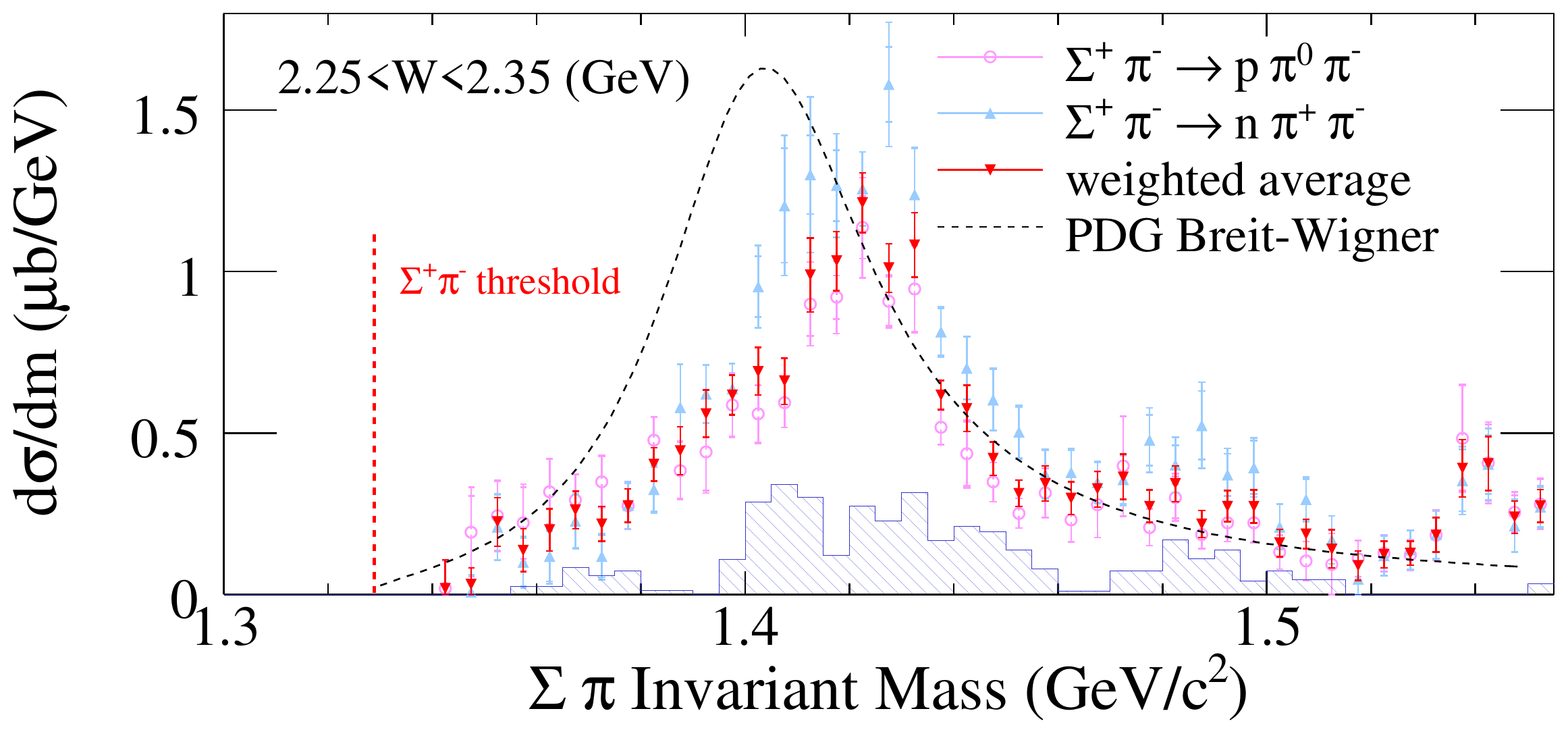}}
    \hfill
    \subfloat[]
    {\label{fig:lineshape_SigmaPlus:W6}\includegraphics[width=0.49\textwidth]{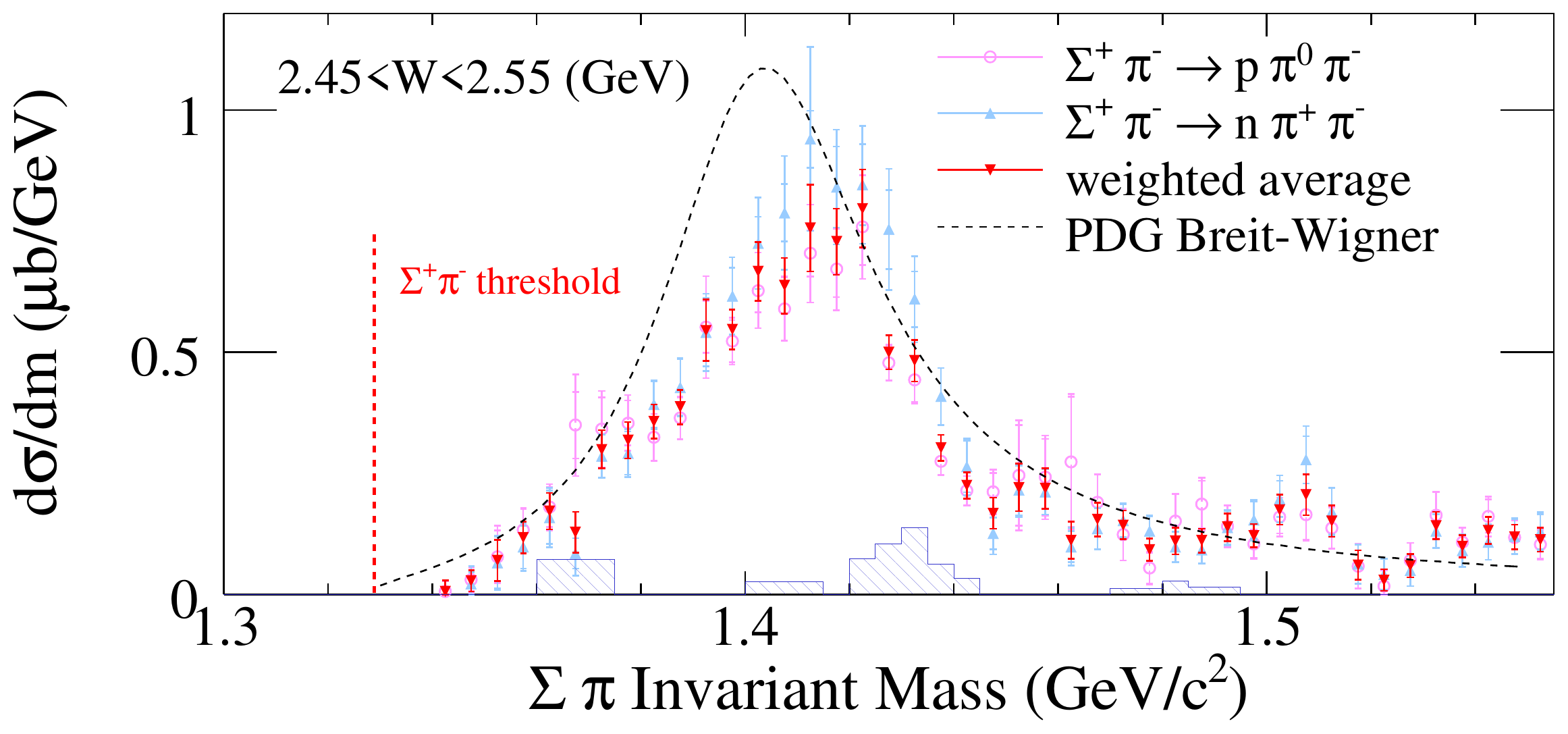}}
    \hfill
    \subfloat[]
    {\label{fig:lineshape_SigmaPlus:W8}\includegraphics[width=0.49\textwidth]{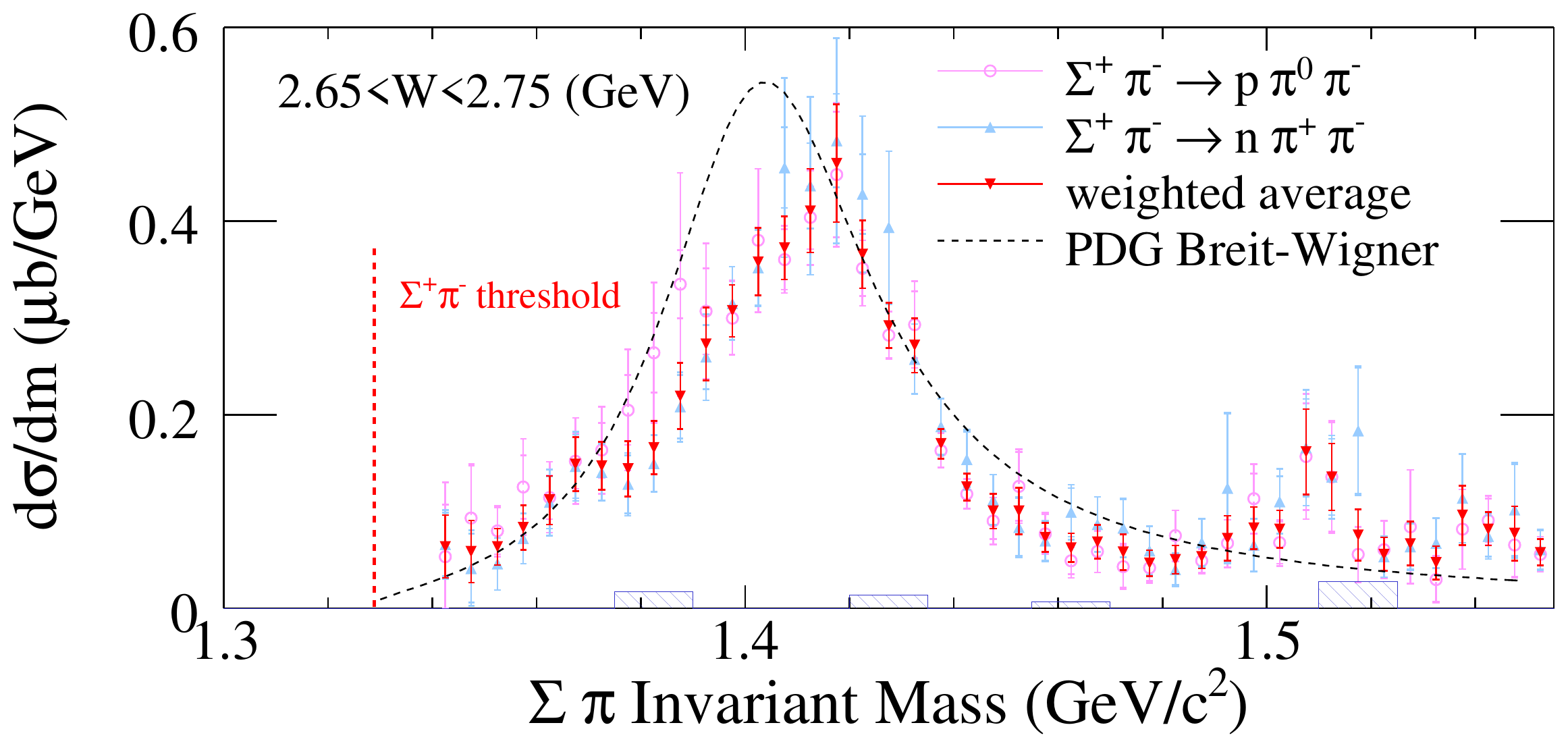}}
    \hfill
  \end{center}
  \caption[]{(Color online) Line shape results for the two $\SigmaPlus
    \pim$ channels. The \chSigmaPlusP{} channel is shown with light
    magenta open circles, while the \chSigmaPlusN{} channel is shown
    with light blue triangles. The weighted average of the two line
    shapes is taken as the final $\SigmaPlus \pim$ line shape, and is
    shown as red downward triangles. The dashed line represents a
    relativistic Breit-Wigner function with a mass-dependent width,
    with the mass and width taken from the PDG.  The blue-hatched
    histogram at the bottom shows the averaged estimated systematic
    discrepancy between the two reconstructed decay modes.  }
  \label{fig:lineshape_SigmaPlus}
\end{figure*}

We begin by studying the two channels \chSigmaPlusP{} and
\chSigmaPlusN, which share the same strong final state, and differ
only in the decay of the \SigmaPlus. Comparing these two channels
gives a measure of the reconstruction accuracy of the analysis.
Figure~\ref{fig:lineshape_SigmaPlus} shows a comparison of the line
shapes obtained for each $\SigmaPlus \pim$.  channel.  The inner error
bars are the combined statistical uncertainty of the data and of the
Monte Carlo samples that were used in the background subtraction.  Our
fits to the raw invariant mass spectra using Monte Carlo templates did
not always perfectly reproduce the data, even after iterating.  To
account for this possible systematic error in our analysis, we summed
the data within each energy bin over all kaon angles and compared to
the summed fit result.  Any discrepancy in each mass bin was taken as
an additional uncertainty, and a portion was added in quadrature with
the statistical errors above.  These are shown as the outer error bars
in Fig.~\ref{fig:lineshape_SigmaPlus}.  Thus, the outer errors bars
represent the combined point-to-point statistical and systematic
uncertainty.

Beyond these estimated uncertainties on single decay modes, any large,
possibly-nonstatistical difference between the two measured
$\Sigma^+\pi^-$ modes could also signal a systematic discrepancy in
the analysis. Therefore, for each mass bin, we took the difference of
the two measured values and subtracted the summed errors in
quadrature, obtaining a mass-dependent error that estimates this
systematic discrepancy.  The shaded histogram at the bottom of the
plots shows these uncertainties when the difference of the two
measured points is larger than the sum of the two errors.  The
agreement between the two decay mode reconstruction channels is
generally good. The average of these two measurements will be used in
the subsequent comparisons with the other charge decay modes.

%%%%%%%%%%%%%%%%%%%%%%%%%%%%%%%% FIGURE 16 %%%%%%%%%%%%%%%%%%%%%%%%%%%%%%%%%%%%%%%%%%%%%%
%%%%%%%%%%%%%%%%%%%%%%%%%%%%%%%% END FIGURE %%%%%%%%%%%%%%%%%%%%%%%%%%%%%%%%%%%%%%%%%%%%%%

In all cases the $\Sigma^+\pi^-$ mass distribution clearly peaks at a
mass of around $1420$ \mevcc{} which is higher than the nominal mass
of the $\LambdaOne$ at $1405.1$ \mevcc{} listed by the
PDG~\cite{Beringer:1900zz}. We also note the sharp drop or break of
the mass distributions at the $N \kbar$ threshold near 1.435
GeV/c$^2$, which is a signature of the opening of a new threshold for
$S$-wave resonances.  This will be discussed in
Section~\ref{section:isospin}.

\subsection{Line shape results for all $\Sigma \pi$ channels}
\label{subsection:LineShapeResults:all}

\begin{turnpage}
\begin{figure*}[t!b!p!h!]
    \subfloat[]
    {\label{fig:lineshape:W1}\includegraphics[width=0.32\textheight]{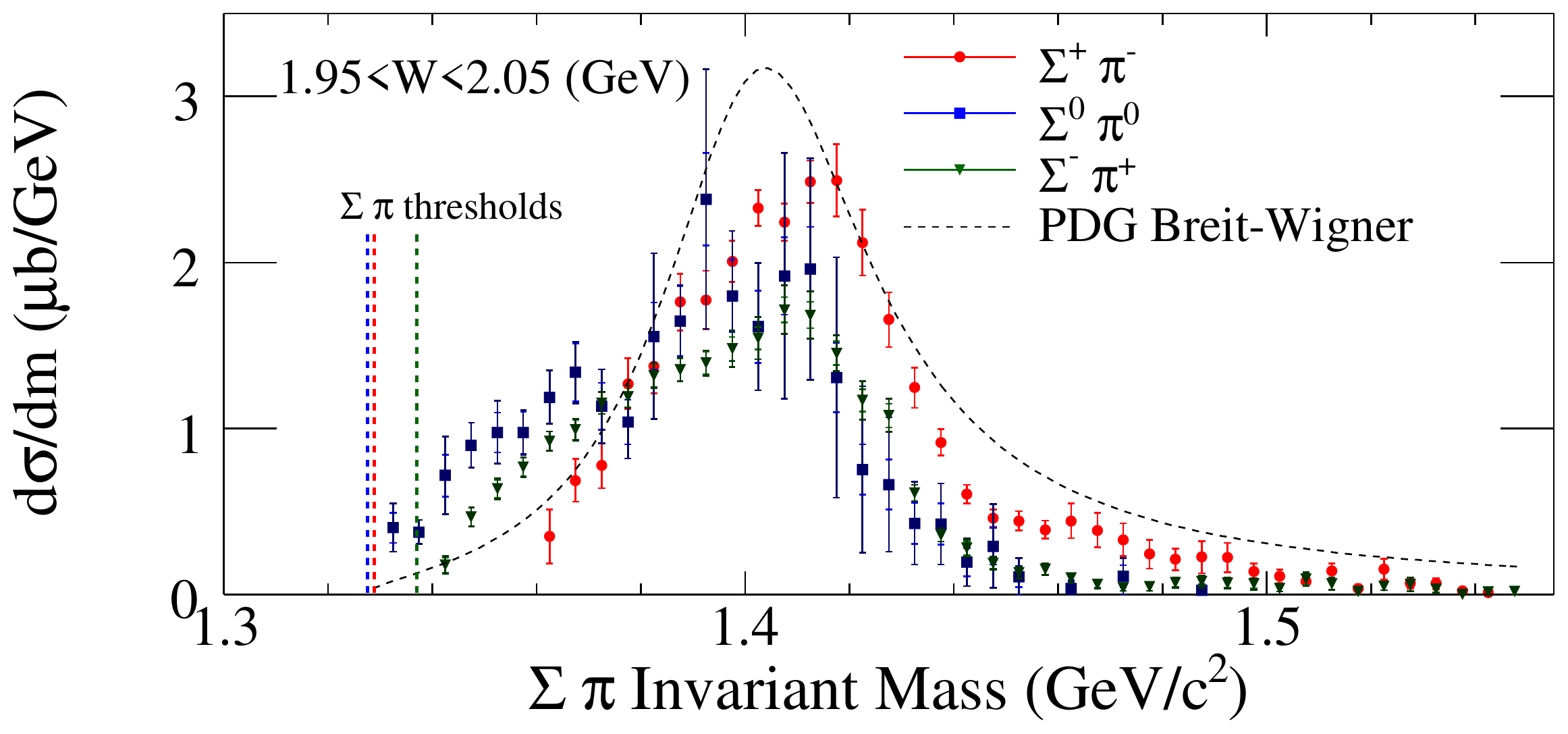}}
    \hfill
    \subfloat[]
    {\label{fig:lineshape:W2}\includegraphics[width=0.32\textheight]{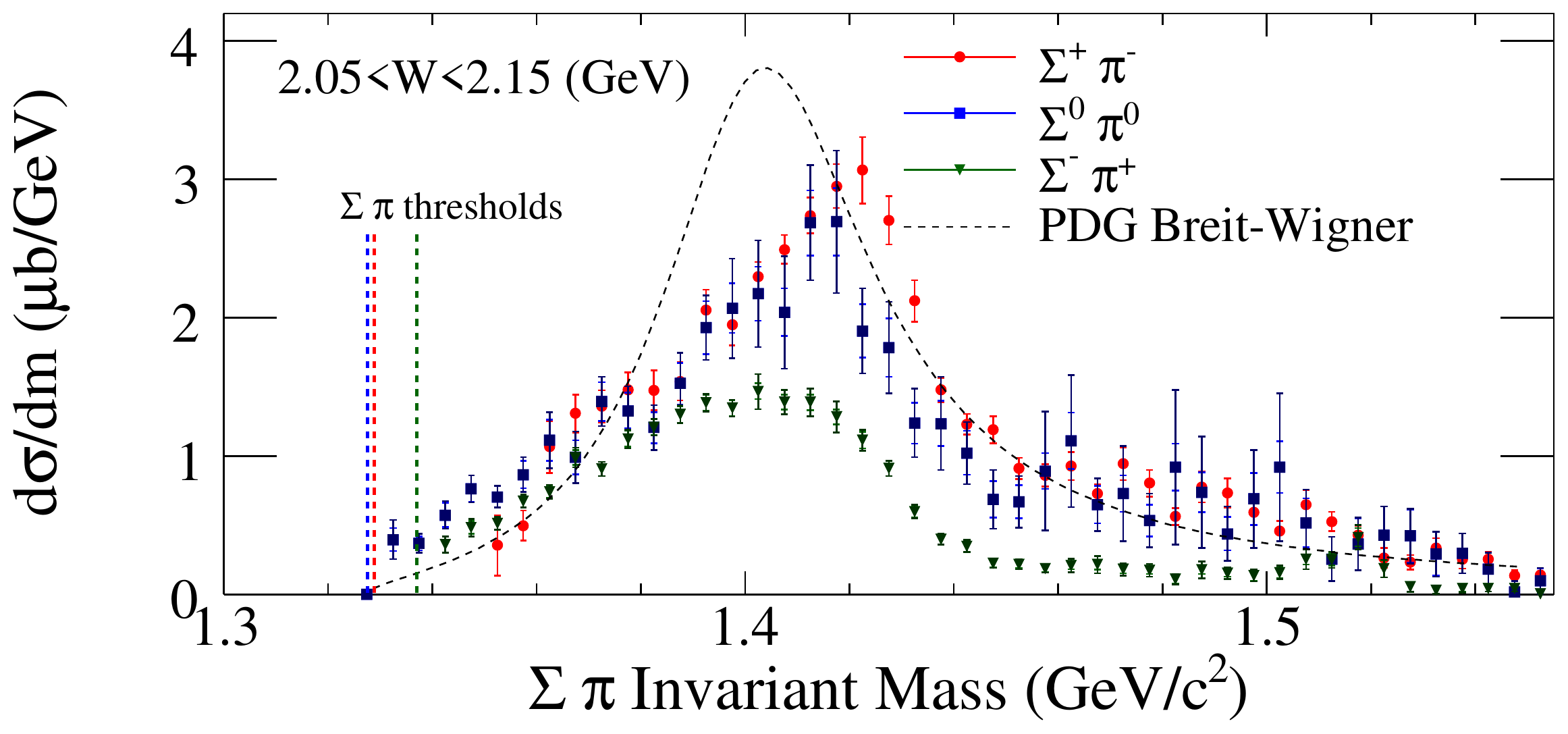}}
    \hfill
    \subfloat[]
    {\label{fig:lineshape:W3}\includegraphics[width=0.32\textheight]{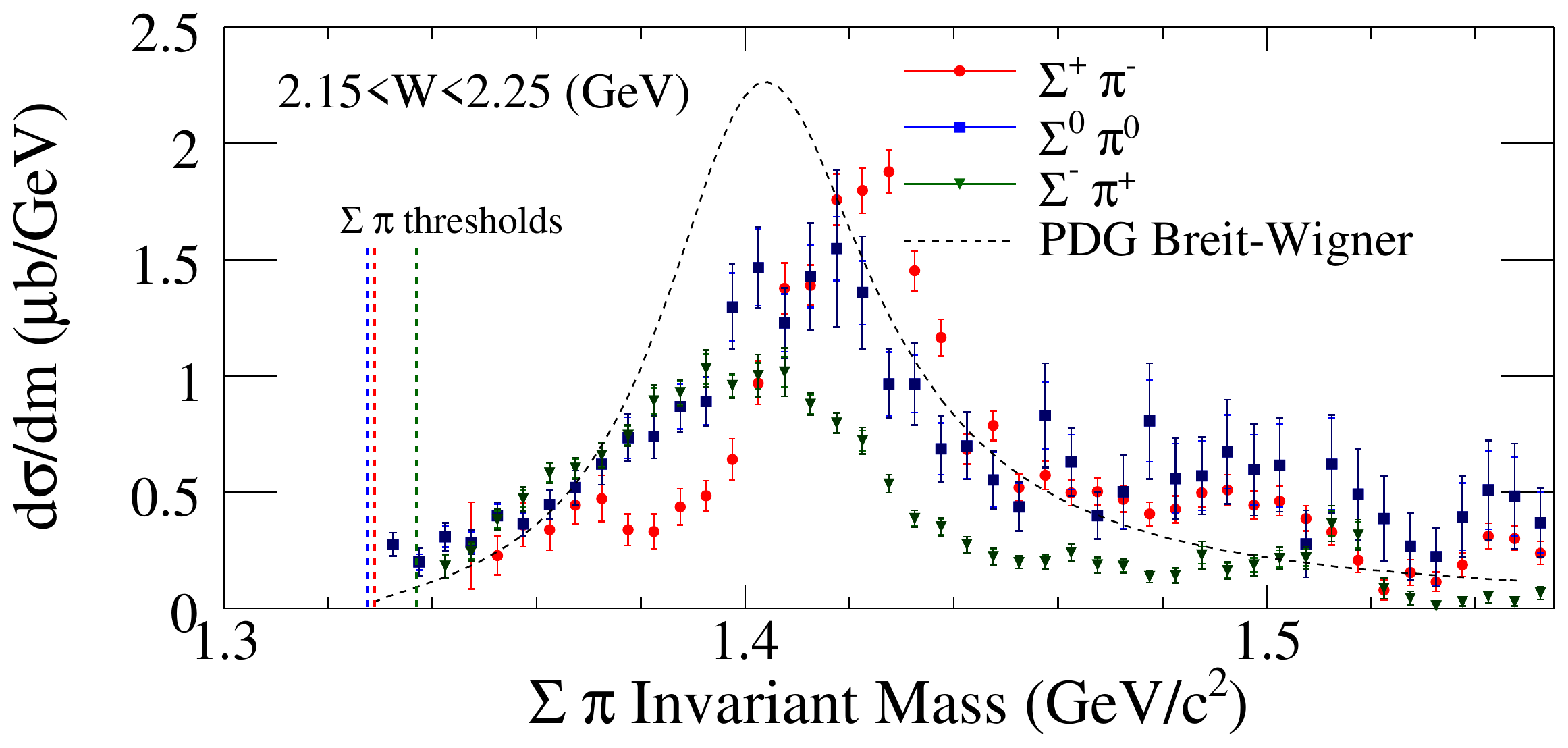}}
    \hfill
    \subfloat[]
    {\label{fig:lineshape:W4}\includegraphics[width=0.32\textheight]{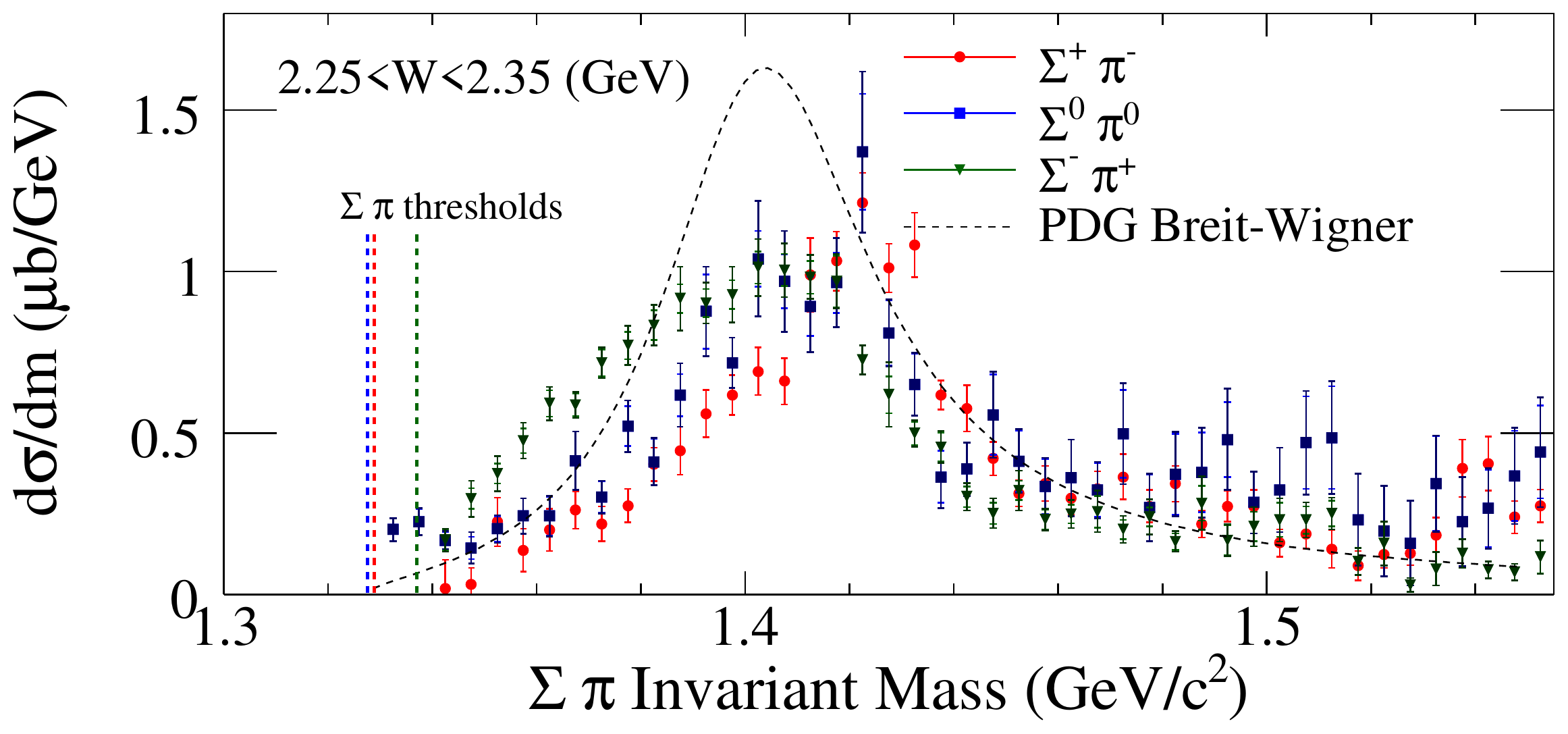}}
    \hfill
    \subfloat[]
    {\label{fig:lineshape:W5}\includegraphics[width=0.32\textheight]{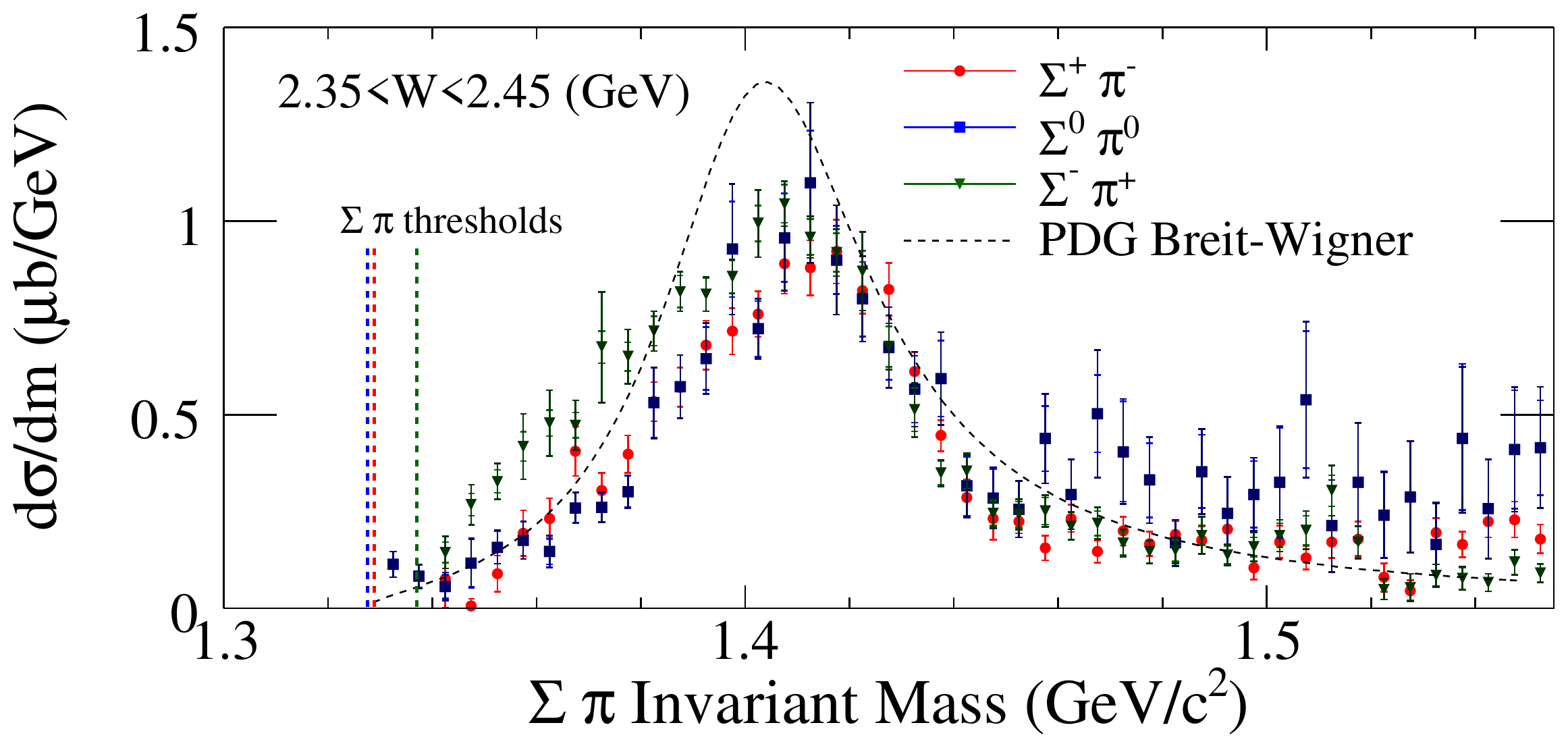}}
    \hfill
    \subfloat[]
    {\label{fig:lineshape:W6}\includegraphics[width=0.32\textheight]{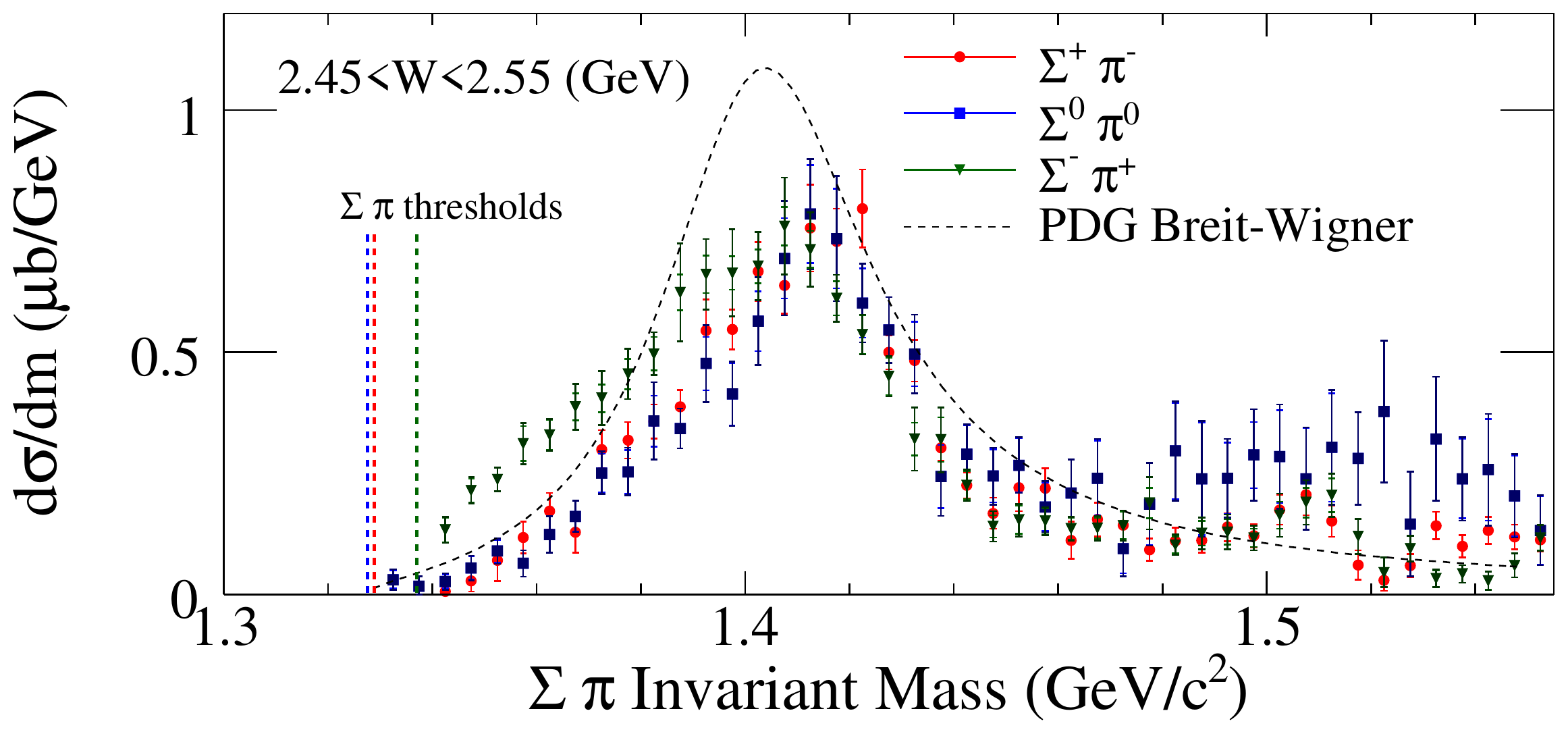}}
    \hfill
    \subfloat[]
    {\label{fig:lineshape:W7}\includegraphics[width=0.32\textheight]{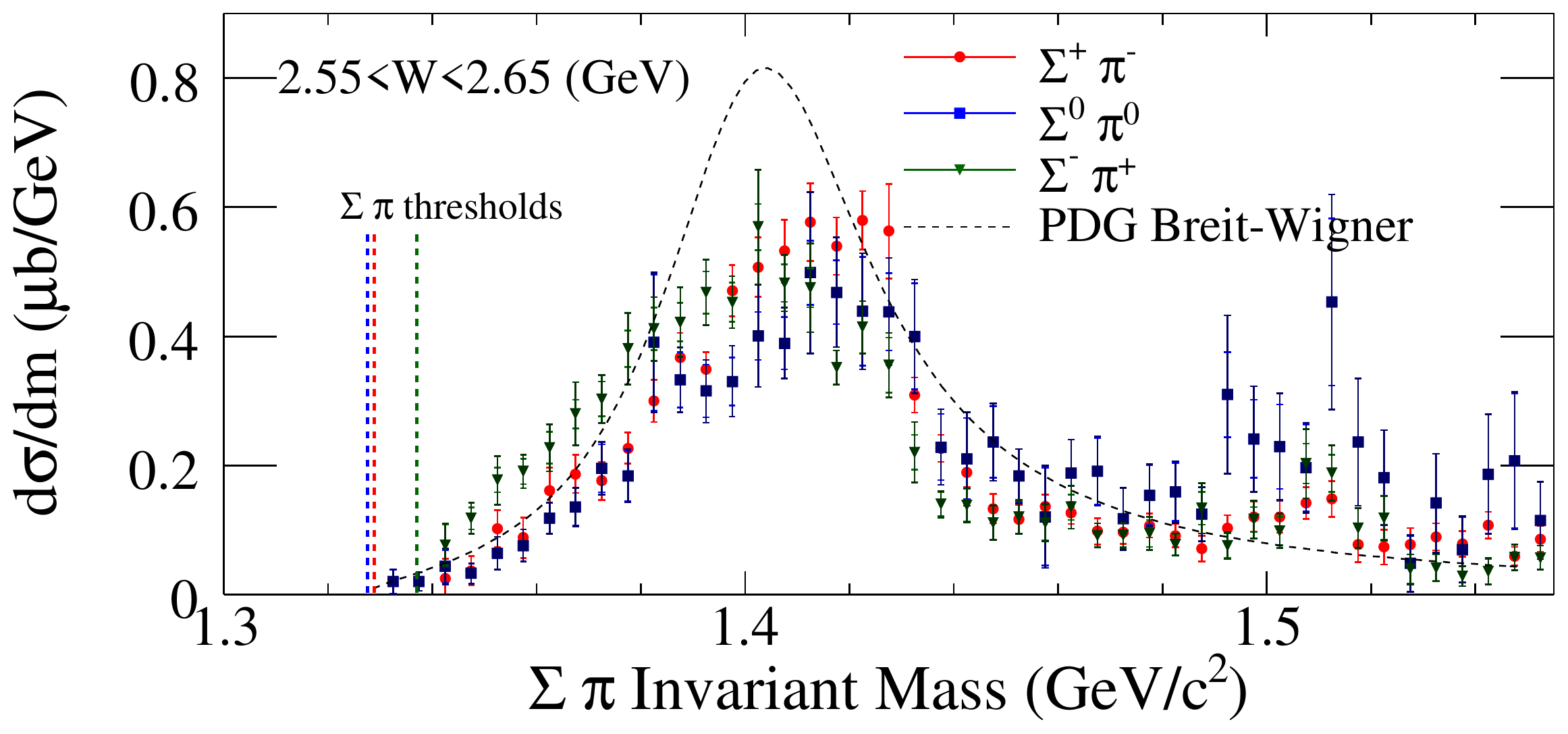}}
    \hfill
    \subfloat[]
    {\label{fig:lineshape:W8}\includegraphics[width=0.32\textheight]{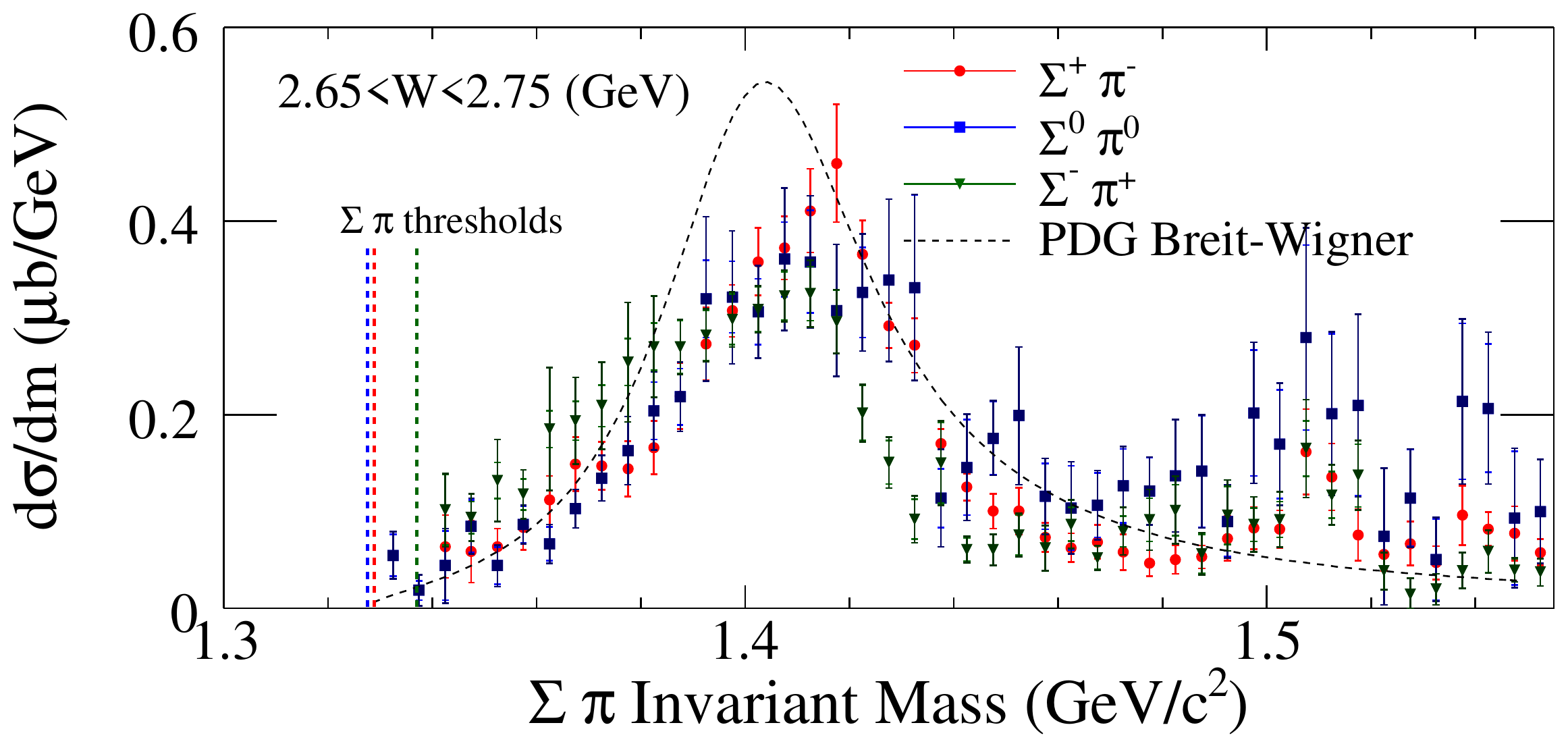}}
    \hfill
    \subfloat[]
    {\label{fig:lineshape:W9}\includegraphics[width=0.32\textheight]{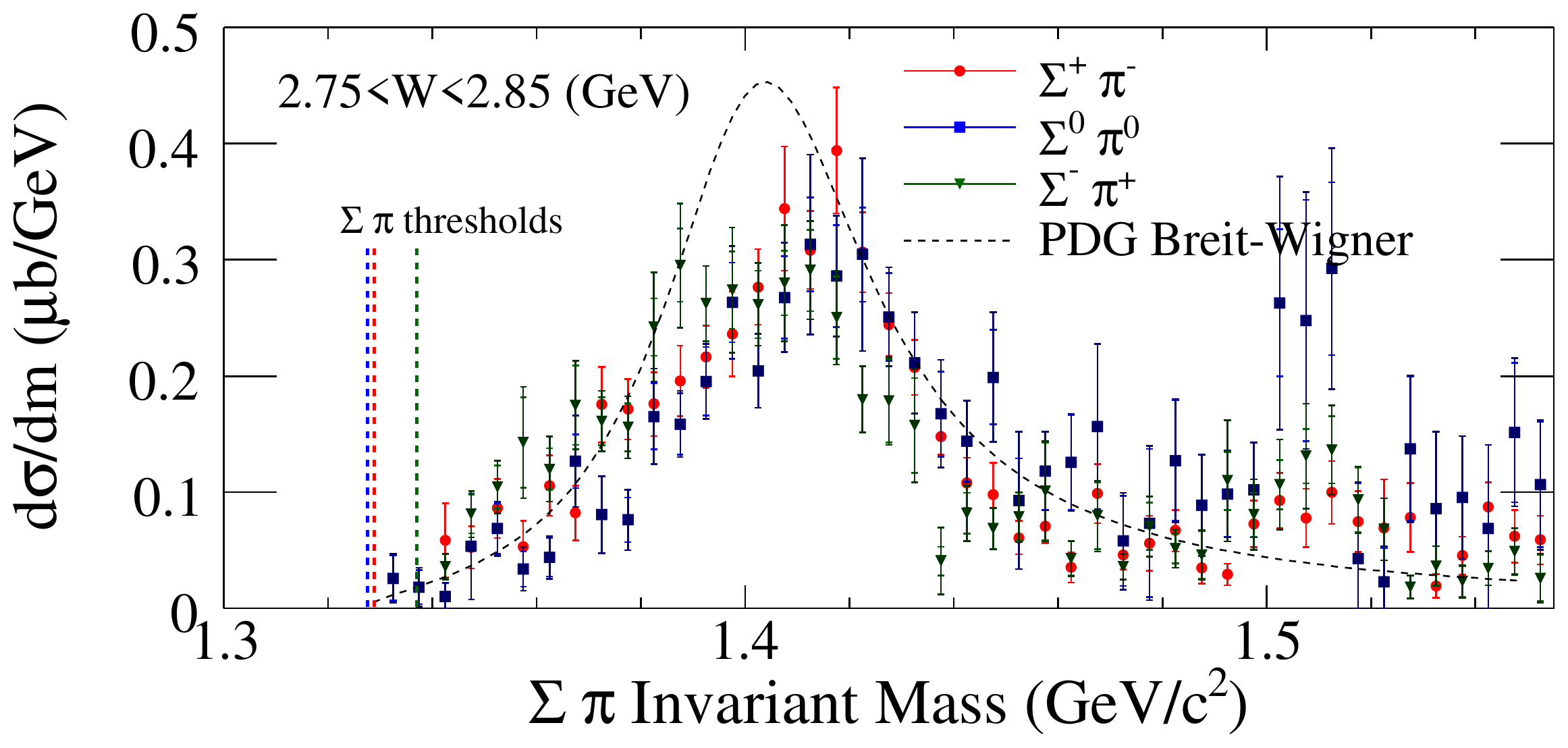}}
    \hfill
  \caption[]{ (Color online) Mass distribution results for all three
    $\Sigma \pi$ channels. The weighted average of the two $\SigmaPlus
    \pim$ channels is shown in the red circles, the \chSigmaZero{}
    channel is shown as the blue squares, and the \chSigmaMinus{}
    channel is shown as the green triangles. The \chSigmaZero{} and
    \chSigmaMinus{} channels have inner error bars representing the
    statistical errors, and outer error bars that have the estimated
    residual discrepancy of the data and fit results added in
    quadrature.  The dashed line represents a relativistic
    Breit-Wigner function with a mass-dependent width, with the mass
    and width taken from the PDG and with arbitrary normalization. The
    vertical dashed lines show the opening of each $\Sigma \pi$
    threshold. 
    }
  \label{fig:lineshape}
\end{figure*}
\end{turnpage}

Our main results~\cite{CLASdatabase}, the line shape comparison for
all three $\Sigma \pi$ channels, is shown in
Fig.~\ref{fig:lineshape}. As noted, the $\SigmaPlus \pim$ channel is
the weighted average of the two measured final states. The $\SigmaZero
\pizero$ channel and $\SigmaMinus \pip$ channels are again shown with
inner and outer error bars, where the inner bars are statistical, and
the outer bars include the estimated residual discrepancy in the fits
added in quadrature to the inner bars. For each of nine bins in
invariant energy $W$, we show the $\Sigma\pi$ mass distribution in
each of three charge states.  The data have been summed over the full
range of measured kaon production angles.  The large-angle cut-offs
were not quite identical for all charge states because of differing
acceptances, but since the cross sections get very small at large
angles ($\costhetakp<-0.5$) we can neglect these differences.

%%%%%%%%%%%%%%%%%%%%%%%%%%%%%%%% FIGURE 17 %%%%%%%%%%%%%%%%%%%%%%%%%%%%%%%%%%%%%%%%%%%%%%
%%%%%%%%%%%%%%%%%%%%%%%%%%%%%%%% FIGURE  %%%%%%%%%%%%%%%%%%%%%%%%%%%%%%%%%%%%%%%%%%%%%%

For all energies, it is evident that the line shapes differ markedly
between charge states; in some regions they differ by well over
$5\sigma$.  This occurs far away from the indicated reaction
thresholds, making it unlikely that the effects are due to mere mass
differences.  None of the mass distributions are reproduced by the
simple relativistic Breit-Wigner line shape with PDG-given centroid
and width.  The $\SigmaPlus \pim$ channel peaks at a higher mass than
the \chSigmaMinus{} channel, while having a width that is
significantly smaller.  The charge-dependence of the mass
distributions is largest for $W$ between 2.0 and 2.4 GeV.  For $W$
approaching 2.8 GeV the mass distributions tend to merge together.
This hints that whatever $I \neq 0$ coherent admixture of isospin
states is at work here, it fades away at higher total energy.  Our own
fit to the line shapes to extract our best estimates for the mass and
width of the \LambdaOne{} and other structures causing this
charge-dependence of the mass distributions will be shown in
Section~\ref{section:isospin}.

\begin{figure}[t!b!p!h!]
  \resizebox{0.45\textwidth}{!}{\includegraphics[angle=0.0]{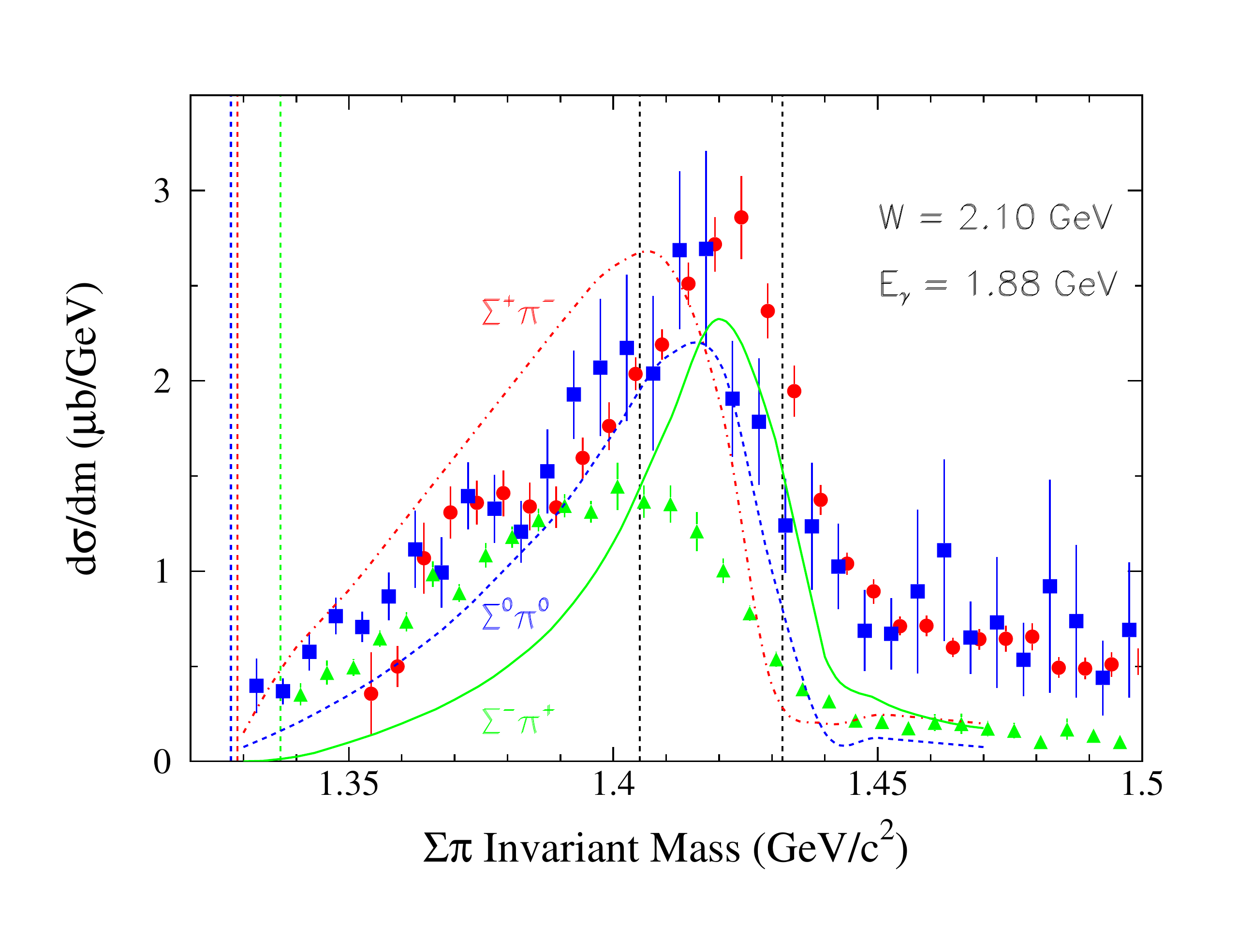}}
  \caption[]{ (Color online) Mass distributions at $W=2.10$~GeV and
    $E_\gamma=1.88$~GeV in comparison to the model of Nacher
    \etal~\cite{Nacher:1998mi} scaled down by a factor of 2.0.  The
    $\SigmaPlus \pim$ channel is shown as red circles and the red
    dot-dashed line; the \chSigmaZero{} channel is shown as the blue
    squares and the blue dashed line; the \chSigmaMinus{} channel is
    shown as the green triangles and the green solid line.  The dashed
    vertical colored lines at the left side show the reaction
    thresholds, and the vertical dashed lines at 1.405 GeV and 1.437
    GeV mark the nominal centroid and the $N\bar{K}$ thresholds,
    respectively.  The error bars on the data points are combined
    statistical and point-to-point systematic uncertainties.  }
    \label{fig:lineshape_Nacher}
\end{figure}

Comparing our line shape results to the prediction of Nacher
\etal~\cite{Nacher:1998mi} computed in a chiral unitary model
approach, we see in Fig.~\ref{fig:lineshape_Nacher} that they are
indeed different for each $\Sigma \pi$ channel. In the chiral unitary
theory this was explained as an $I=1$ amplitude interfering with the
$I=0$ \LambdaOne{} amplitude in such a way that the $\SigmaPlus \pim$
and $\SigmaMinus \pip$ channels were shifted in opposite directions
due to the interference term.  The model curves were computed for
$E_\gamma = 1.7$~GeV, but we compare with our results at $E_\gamma =
1.88$~GeV since our statistics are better there.  The model
calculation uses a Weinberg-Tomozawa contact interaction that is
energy and angle independent, allowing us to compare the model to the
data in any energy bin. In our results it is the $\Sigma^+ \pi^-$
channel that is shifted to higher mass with a narrower width, and the
$\Sigma^- \pi^+$ channel is smaller and wider, in contrast to the
model calculation.  Also, the model curves have been scaled down by a
factor of 2.0 to match the data, suggesting that the model
overestimates the strength of the photocouplings by that amount.  In
Section~\ref{section:isospin} we will make our own phenomenological
isospin decomposition to find a plausible explanation of what is seen.

%%%%%%%%%%%%%%%%%%%%%%%%%%%%%%%% FIGURE 18 %%%%%%%%%%%%%%%%%%%%%%%%%%%%%%%%%%%%%%%%%%%%%%
%%%%%%%%%%%%%%%%%%%%%%%%%%%%%%%% END FIGURE %%%%%%%%%%%%%%%%%%%%%%%%%%%%%%%%%%%%%%%%%%%%%%

The other existing prediction for the mass distribution of the
$\Sigma\pi$ final states is that of Lutz and
Soyeur~\cite{Lutz:2004sg}.  In their so-called double kaon pole model,
the combined effects of the $\Sigma(1385)$ and the $\Lambda(1405)$
were considered, and this produced some variation among the three
charge combinations we have presented.  However, as has been
discussed, we subtracted off the effect of the $\Sigma(1385)$ and
still are left with a substantial variation in the three final states.
We do not compare our results directly to theirs because they are
qualitatively similar in shape to those of Ref.\cite{Nacher:1998mi},
and also because they are about a factor of four too large in cross
section, indicating a serious quantitative discrepancy when comparing
to our results.

%%%%%%%%%%%%%%%%%%%%%%%%%% end of file %%%%%%%%%%%%%%%%%%%%%%%%%%%%%%%%%%%%%%%%%%%%

%*************************************************************************************
%                                                                                    *
%                                 Systematic Errors                                  *
%                                                                                    *
%*************************************************************************************
% Revisions:
% 09-06-12 RS First revision of KM draft
% 09-10-12 RS Fix bugs and typos
% 11-20-12 RS after comments from Ad Hoc group
% 01-11-13 RS after final CLAS comments

\section{Systematic Uncertainties and Tests}
\label{section:SystematicUncertainties}

\subsection{Overall systematics of the run}
\label{subsection:systematics:overall}

For systematic uncertainties, there were global contributions from the
yield extraction, acceptance corrections, flux normalization, and the
line shape fitting procedure.  The main cuts that influenced the yield
extraction were the \deltaTOF{} cuts, the confidence level cuts in the
kinematic fit, and the selection of intermediate the ground-state
hyperon. All of these cuts were varied within each bin of
center-of-mass energy and angle, and the total yields were checked for
any differences due to the cuts. Variation in the \deltaTOF{} width by
$0.2$ ns changed the acceptance-corrected yield between $2$--$6\%$ in
each bin. Changing the confidence level from the nominal $1\%$ to
$10\%$, changed the acceptance-corrected yields by $3$--$12\%$ in each
bin. For the majority of the bins, the final yields changed by less
than $4\%$ for the \deltaTOF{} cuts, less than $7\%$ for the
confidence level cuts, and less than $2\%$ for the ground-state
hyperon selection.

Stability of the normalization was monitored throughout the run.  The
fluctuations in target density were determined to be a negligible
$0.11\%$, while the target length was measured to $0.125\%$. The
photon normalization was examined on an hour-by-hour basis by
measuring the $\omega$ production yields~\cite{Williams2}, and the
uncertainty for the normalization was determined to be $7.3\%$. The
live-time correction that was necessary to determine the photon flux
introduced an additional uncertainty of $3\%$, and the photon
transmission efficiency added $0.5\%$, so that the total uncertainty
for the photon normalization was $7.9$.

For the final systematic uncertainty, all of the above global
uncertainties were added in quadrature to yield a final value of
$11.6\%$. A summary of each uncertainty is shown in
Table~\ref{tab:sys_uncertainties}.

\begin{table}
  \caption{\label{tab:sys_uncertainties} The global systematic
    uncertainties in the experiment. They arise from yield extraction,
    acceptance calculation, target characteristics, photon flux
    normalization, and branching ratios~\cite{Beringer:1900zz}. The
    total was calculated by summing all in quadrature.  }
%\resizebox{10cm}{!}{
%  \begin{ruledtabular}
    \begin{tabular}{l r}
      \hline 
      \hline
      Source           & Value \\ \hline
      % yield and acceptance
      % \multicolumn{2}{l}{Yield extraction and acceptance} \\
      \deltaTOF{} cuts & $2$--$6\%$ \\
      Confidence level on kinematic fit & $3$--$12\%$ \\
      Selection of intermediate hyperons & $2$--$3\%$ \\
      % \multicolumn{2}{l}{Target} \\
      Target density & $0.11\%$ \\
      Target length  & $0.125\%$ \\
      % photon normalization
      % \multicolumn{2}{l}{Photon normalization} \\
      Photon normalization & $7.3\%$ \\
      Live-time correction & $3\%$ \\
      Photon transmission efficiency & $0.5\%$ \\
      % \multicolumn{2}{l}{Branching ratios for each channel} \\
      $\Sigma(1385) \to \Sigma \pi, \Lambda \pi$ & $1.5\%$ \\
      $\Lambda \to \proton \pim$ & $0.5\%$ \\
      $\SigmaPlus \to \proton \pizero, \neutron \pip$ & $0.30\%$ \\
      $\SigmaMinus \to \neutron \pim$ & $0.005\%$ \\ \hline
      Total & 11.6\% \\
      \hline 
      \hline
    \end{tabular}
%  \end{ruledtabular}
%}
\end{table}

The mass resolution of the line shape results was investigated by
generating Monte Carlo samples of zero width centered at $\Sigma \pi$
invariant masses of $1.406, 1.450$, and $1.500$ \gevcc. Because the
$\Sigma \pi$ invariant mass is equivalent to the missing mass off the
detected \kp, the $\Sigma \pi$ mass resolution was related to the
momentum resolution of the \kp. However, kinematic fitting of most of
the channels improved the overall mass resolution.  For all generated
events in all bins of center-of-mass energy and angle, the $\Sigma
\pi$ invariant mass for the accepted events was fit with a Gaussian to
determine the resolution. This showed that for the lower energy $W$
bins, the resolution $(\sigma)$ was better than $6$ \mevcc, while for
the higher energy bins it was up to $8$ \mevcc{}, with worse
resolution in the backward kaon angles, where the CLAS magnetic field
is weaker.  Without the kinematic fit, as in the $\Sigma^0\pi^0$
results, the mass resolution averaged about 2 MeV wider at high $W$
and large angles.  No shift of the center of the Gaussian larger than
$1$ \mevcc{} was seen.  Since our results are shown with 5 \mevcc{}
bins, the mass resolution of the line shapes is one to two bins.  We
also remark that the absolute mass accuracy of the experiment for
hyperons such as the $\Sigma(1385)$ and the \LambdaTwo, and of meson
states in this mass range is $\alt 1$ MeV/c$^2$.

\subsection{Removal of \Kstar}
\label{subsection:systematics:kstar}

A concern in the photoproduction line shape analysis of the
\LambdaOne{} region is the effect that the \Kstar{} may have. As seen
in Fig.~\ref{fig:Dalitz:case2SigmaPlus}, the \LambdaOne{} has a
kinematic overlap with the \Kstar{} in the strong final-state phase
space, so that the difference in line shapes seen in the various
$\Sigma \pi$ channels could be due to interference with the \Kstar.
Below we argue that this is not the case.

We measure the line shape of the \LambdaOne{} in bins of
center-of-mass energy, $W$, and the kinematic overlap of the \Kstar{}
depends strongly on this energy.  Fig.~\ref{fig:Dalitz:case2SigmaPlus}
shows no \Kstar{} overlap at low $W$ below the \Kstar{} threshold,
strong overlap at intermediate $W$ and again no overlap at high $W$.

\begin{figure*}[p!t!b!h!]
  \begin{center}
    \subfloat[]
    {\label{fig:lineshape_Kstar:W1}\includegraphics[width=0.49\textwidth]{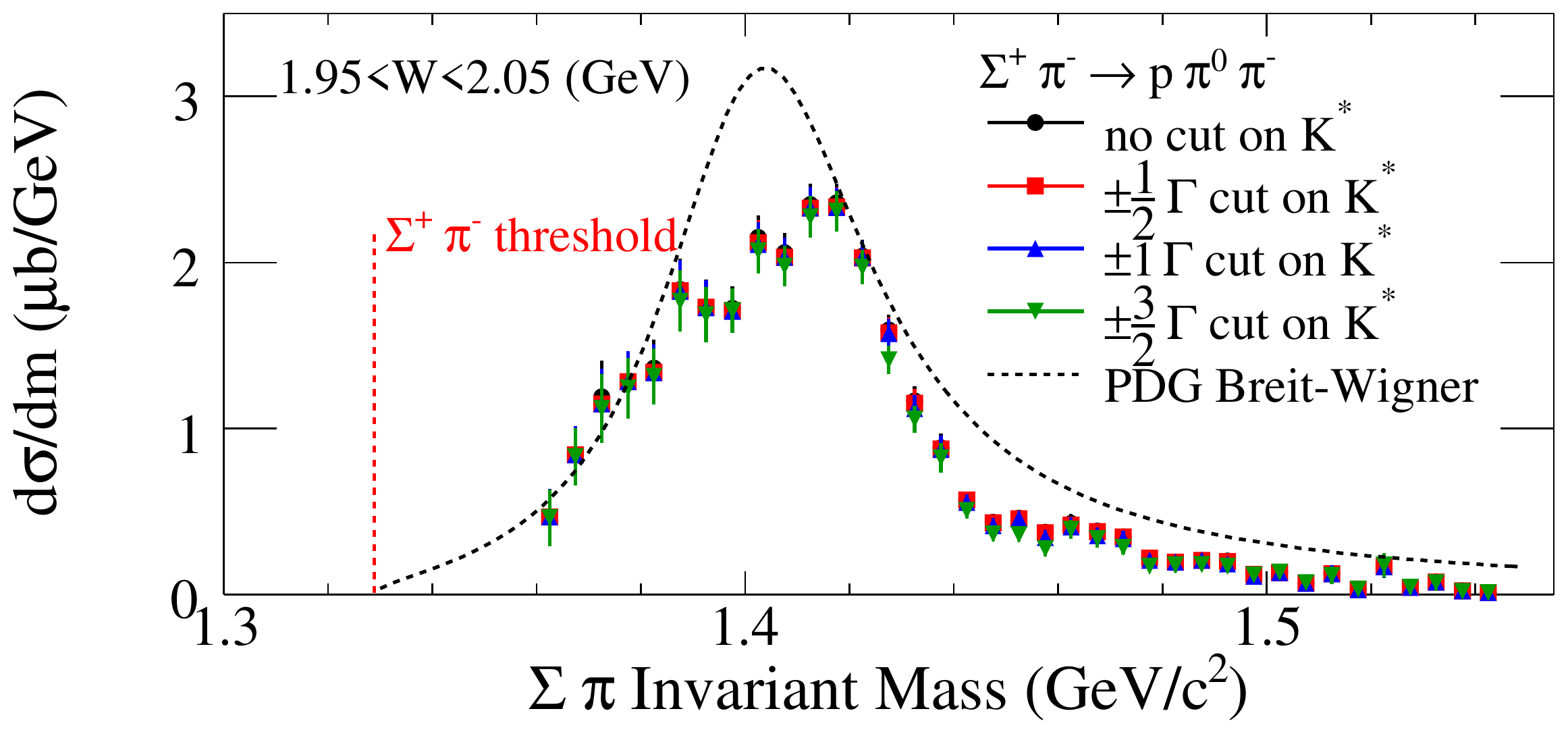}}
    \hfill
    %\subfloat[]
    %{\label{fig:lineshape_Kstar:W2}\includegraphics[width=0.49\textwidth]{dsigmadM_SigmaPlus_2-crop.pdf}}
    %\hfill
    %
    \subfloat[]
    {\label{fig:lineshape_Kstar:W3}\includegraphics[width=0.49\textwidth]{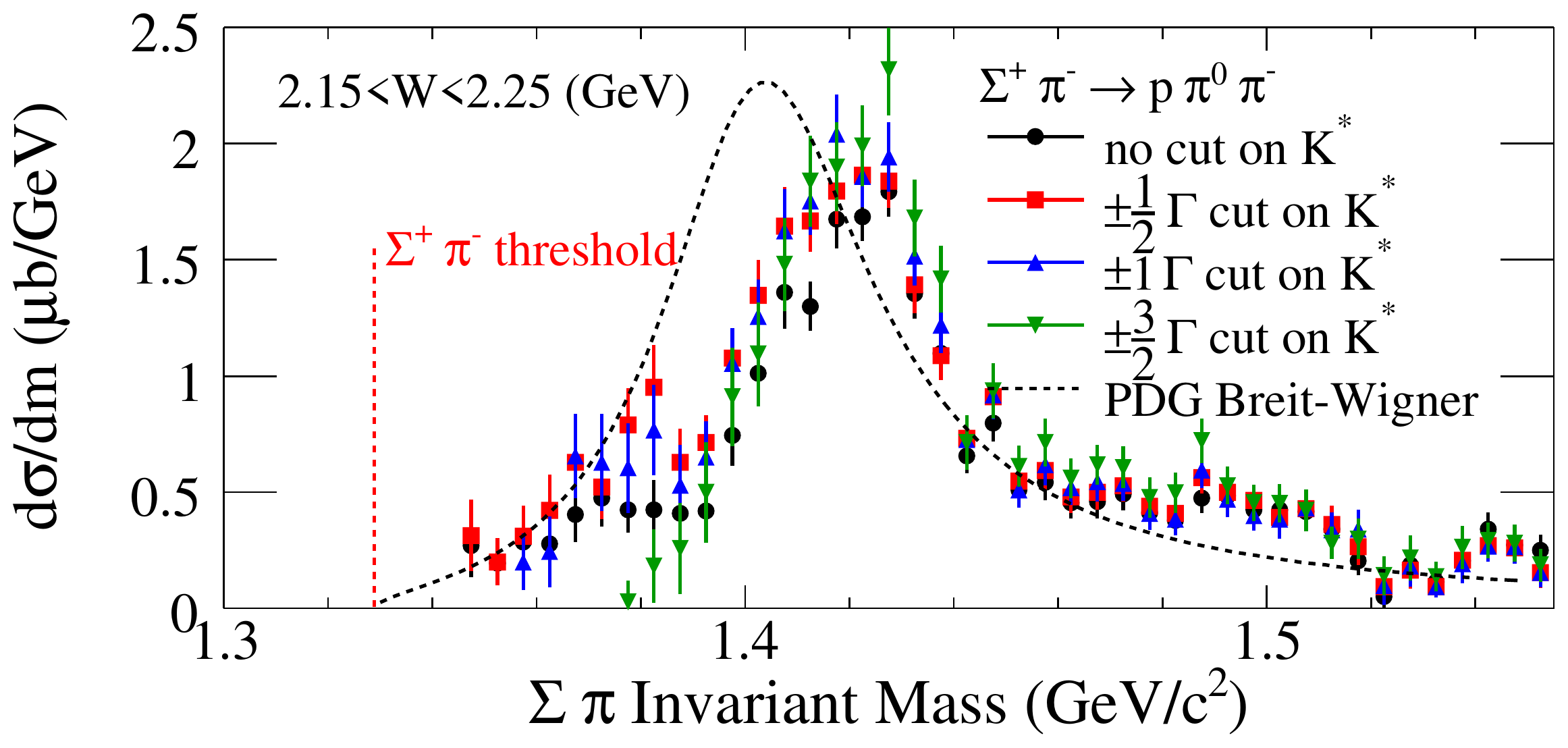}}
    \hfill
    %\subfloat[]
    %{\label{fig:lineshape_Kstar:W4}\includegraphics[width=0.49\textwidth]{dsigmadM_SigmaPlus_4-crop.pdf}}
    %\hfill
    %
    %\subfloat[]
    %{\label{fig:lineshape_Kstar:W5}\includegraphics[width=0.49\textwidth]{dsigmadM_SigmaPlus_5-crop.pdf}}
    %\hfill
    %\subfloat[]
    %{\label{fig:lineshape_Kstar:W6}\includegraphics[width=0.49\textwidth]{dsigmadM_SigmaPlus_6-crop.pdf}}
    %\hfill
    %
    \subfloat[]
    {\label{fig:lineshape_Kstar:W7}\includegraphics[width=0.49\textwidth]{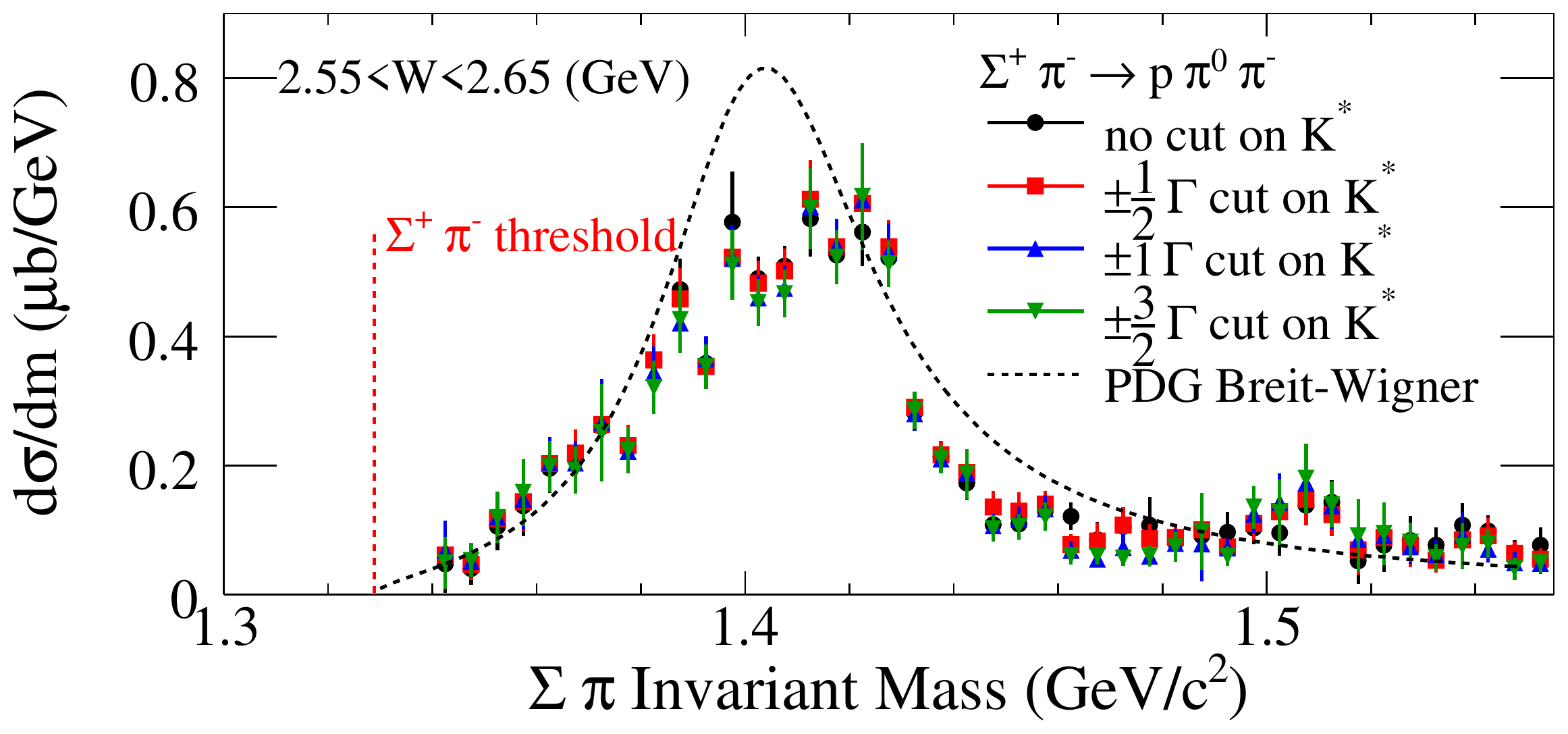}}
    \hfill
    %\subfloat[]
    %{\label{fig:lineshape_Kstar:W8}\includegraphics[width=0.49\textwidth]{dsigmadM_SigmaPlus_8-crop.pdf}}
    %\hfill
   %
    \subfloat[]
    {\label{fig:lineshape_Kstar:W9}\includegraphics[width=0.49\textwidth]{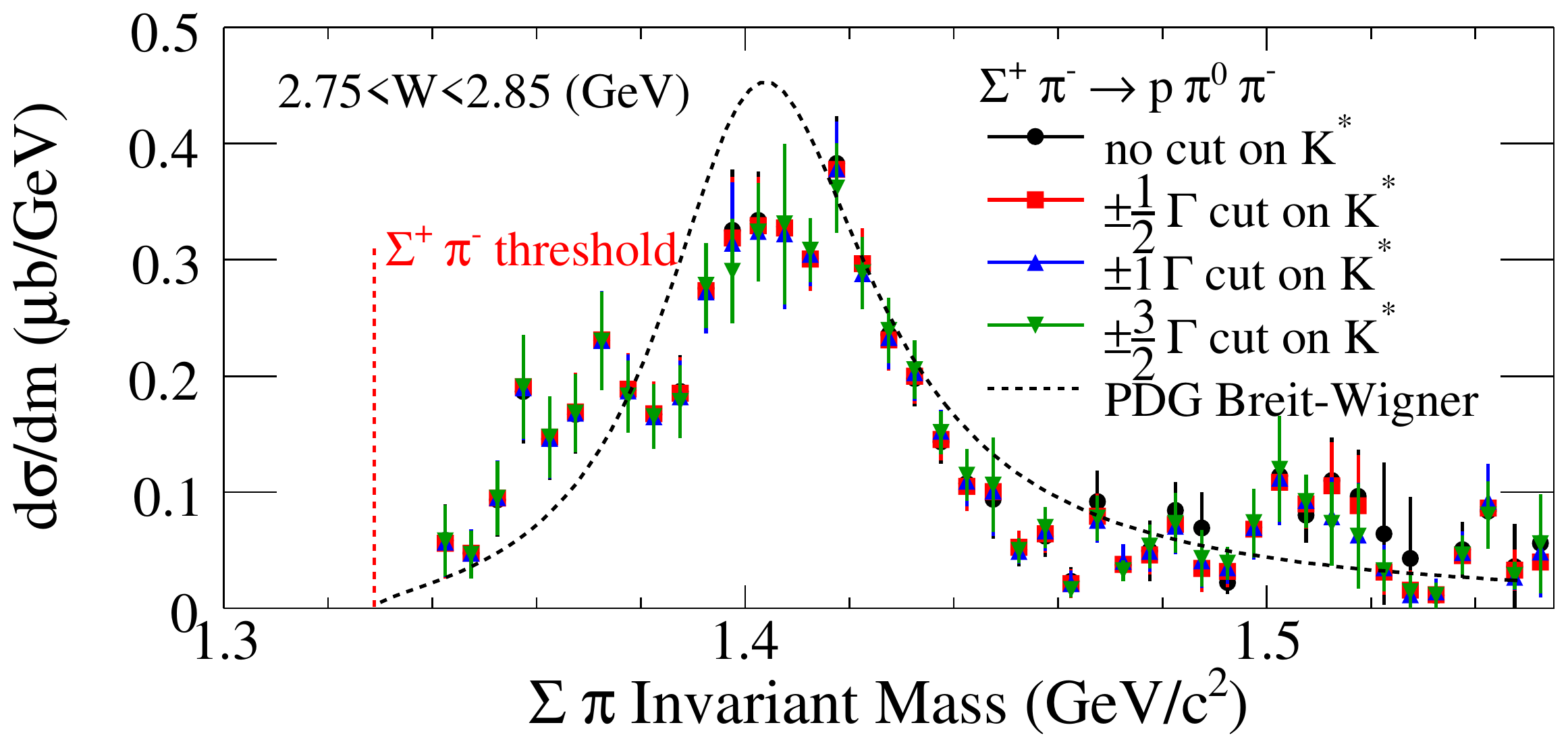}}
    \hfill
  \end{center}
  \caption[]{ (Color online) Final results for the line shape in the
    \chSigmaPlusP{} channel when the \Kstar{} is removed successively
    in steps of $\pm \half \Gamma$, where $\Gamma$ is the width of the
    \Kstar{} quoted in the PDG. The black circles represent our final
    results without a cut on the \Kstar, while the red squares, blue
    triangles, and green downward triangles represent cuts of $\pm
    \Gamma/2$, $\pm \Gamma$, $\pm \frac{3}{2} \Gamma$ centered around
    the \Kstar, respectively.  }
  \label{fig:lineshape_Kstar}
\end{figure*}

We tested for the presence of \Kstar{} interference by cutting out
regions of $\kp \pim$ invariant mass centered around the \Kstar{} mass
and in multiples of $\Gamma/2$, where $\Gamma$ is the width of the
\Kstar{} listed in the
PDG. Figure~\ref{fig:lineshape_Kstar}\subref{fig:lineshape_Kstar:W1}
shows the effect of each \Kstar{} rejection cut up to $\pm \frac{3}{2}
\Gamma$ at $W=2.0$~GeV.  For each cut, we reprocessed all of the Monte
Carlo samples of the other channels used in the template fit, redid
the fit, and applied acceptance corrections. If there were any
interference between the \Kstar{} and the \LambdaOne, we would expect
it to be strongest in the region where the \Kstar{} is strongest,
whereas our results show that even with the overlap region removed,
the final result is remarkably unchanged by this drastic removal. In
Fig.~\ref{fig:lineshape_Kstar}\subref{fig:lineshape_Kstar:W3}, the
line shape changes significantly only for the cut at $\pm \frac{3}{2}
\Gamma$ (green downward triangles), but this is simply due to the loss
of phase space and acceptance, since the cut removes about 300
\mevcc{} of $\kp \pim$ invariant mass centered around the \Kstar.
This is reflected in the Dalitz-like plot of the strong final state in
Fig.~\ref{fig:Dalitz:case2SigmaPlus}, where the boundaries of $\pm
\Gamma$ around the \Kstar{} mass are shown as vertical dashed lines.
Figures~\ref{fig:lineshape_Kstar}\subref{fig:lineshape_Kstar:W7} and
\ref{fig:lineshape_Kstar}\subref{fig:lineshape_Kstar:W9} are more
evidence of the insensitivity of the $\Sigma\pi$ mass distributions to
the \Kstar.  A similar study was done to test for possible coherent
interference between $K^{*+}\Lambda$ and $K^+\Sigma(1385)$, and again
no such effect was detected.

We conclude that, although we cannot completely rule out interference
due to the \Kstar, our results are not significantly altered even when
we apply a drastic cut on the \Kstar{} region, thereby removing most
of its strength. Because the photoproduction line shape of the
\LambdaOne{} is not known to any accuracy, we do not attempt any
further analysis of the interference with the \Kstar. We anticipate
our measurement will further stimulate theoretical interest in this
state, and with more theoretical input, a more elaborate analysis may
be possible in future experiments.

%%%%%%%%%%%%%%%%%%%%%%%%%%%%%%% FIGURE 6 %%%%%%%%%%%%%%%%%%%%%%%%%%%%%%%%%%%%%%%%%%%%%%
%%%%%%%%%%%%%%%%%%%%%%%%%%%%%%%%  END FIGURE %%%%%%%%%%%%%%%%%%%%%%%%%%%%%%%%%%%%%%%%%%%%%%

%%%%%%%%%%%%%%%%%%%%%%%%%%%%%%%% end of file %%%%%%%%%%%%%%%%%%%%%%%%%%%%%%%%%%%%%%%%%%%%%%

%*************************************************************************************
%                                                                                    *
%                                Model Fits                                          *
%                                                                                    *
%*************************************************************************************
% Revisions
% 9-13-12 RS move this section to later in the paper.
% 1-11-13 RS after CLAS comments

\section{Model for Isospin Decomposition}
\label{section:ModelFitting}

We have shown that the line shapes of the $\Sigma\pi$ final states are
far from those of a simple Breit-Wigner form.  Indeed, there are
two main modifications to the picture of a simple single resonance for
the $\Lambda(1405)$ mass region that we will consider in order to gain
a reasonable representation of the experimental results.  The first
arises from the channel-coupling between the detected $\Sigma\pi$
final state and the undetected $N\bar{K}$ final state.  This is done
by using a Flatt\'e-like formalism~\cite{Flatte} to enforce
two-channel unitarity and analyticity of the production amplitude.
The second arises because we find that the different charge states
have markedly different mass distributions, implying that amplitudes
other than $I=0$ must participate in the reaction mechanism.  This is
treated by including coherent $I=1$ amplitudes that interfere with the
$I=0$ amplitude.

Since the electromagnetic interaction does not conserve isospin, the
initial $\gamma p$ state in this reaction can have both $I=1/2$ or
$3/2$ character, and will lead to a final $K^+ (\Sigma\pi)$ state
wherein the $\Sigma\pi$ system is in a superposition of
$I_{\Sigma\pi}=0, 1,$ and $2$ states.  The three measured $\Sigma\pi$
final states all have their third component of isospin,
$I_{\Sigma\pi}^3$, equal to zero.  If we denote the isospin state of
the system as $|I_{\pi\Sigma},I_{\pi\Sigma}^3\rangle$, we can write
each of the three measured final charge combinations using
Clebsch-Gordon coefficients as
\begin{eqnarray}
|\pi^+\Sigma^-\rangle &=& 
\frac{1}{\sqrt{3}}|0,0\rangle +
\frac{1}{\sqrt{2}}|1,0\rangle +
\frac{1}{\sqrt{6}}|2,0\rangle \\
|\pi^0\Sigma^0\rangle &=& 
-\frac{1}{\sqrt{3}}|0,0\rangle +
0 |1,0\rangle + 
\sqrt{\frac{2}{3}}|2,0\rangle \\
|\pi^-\Sigma^+\rangle &=& 
\frac{1}{\sqrt{3}}|0,0\rangle -
\frac{1}{\sqrt{2}}|1,0\rangle + 
\frac{1}{\sqrt{6}}|2,0\rangle.
\end{eqnarray}

%Let $t_0$ be the complex matrix element that takes the initial $\gamma p$
%state via a transition operator $\hat{T}^{(0)}$ to the final state that
%contains the kaon and the $\Sigma\pi$ system in the $I_{\pi\Sigma}=0$
%state, so that
%\begin{equation}
%|t_0|^2 \equiv |\langle 0,0 |\hat T^{(0)}|\gamma p\rangle|^2.
%\end{equation}
%Similarly, for the $I=1$ and $I=2$ transitions we define
%\begin{equation}
%|t_1|^2 \equiv |\langle 1,0 |\hat T^{(1)}|\gamma p\rangle|^2 \text{ and }
%|t_2|^2 \equiv |\langle 2,0 |\hat T^{(2)}|\gamma p\rangle|^2. 
%\end{equation}

Let $t_I$ be the complex matrix element that takes the initial $\gamma p$
state via a transition operator $\hat{T}^{(I)}$ to the final state that
contains the kaon and the $\Sigma\pi$ system in the $I_{\Sigma\pi}=I$
state, so that
\begin{equation}
|t_I|^2 \equiv |\langle I,0 |\hat T^{(I)}|\gamma p\rangle|^2.
\end{equation}
The magnitude-squared matrix element for creating a particular charged
final-state pair, $T_{\pi^a\Sigma^b}$ $(a,b \in \{+-, 00, -+\})$, can
then be obtained by combining these expressions.  For example, the
probability of populating the $|\pi^- \Sigma^+ \rangle$ state is
proportional to
\begin{widetext}
\begin{eqnarray}
|T_{\pi^- \Sigma^+}|^2 &\equiv&
|\langle\pi^- \Sigma^+|\hat T^{(0)}+\hat T^{(1)} + \hat T^{(2)}|
\gamma p\rangle|^2 \\
                       &=& \frac{1}{3}|t_0|^2 + \frac{1}{2}|t_1|^2 + \frac{1}{6}|t_2|^2 - 
\frac{2}{\sqrt{6}}|t_0||t_1|\cos{\phi_{01}} -
\frac{1}{\sqrt{3}}|t_1||t_2|\cos{\phi_{12}} +
\frac{\sqrt{2}}{3}|t_0||t_2|\cos{\phi_{02}},  
\end{eqnarray}
\end{widetext}
in which the real relative phases between the three isospin amplitudes
are $\phi_{01}(m)$, $\phi_{12}(m)$, and $\phi_{02}(m)$.  The other two
charge combinations have similar forms.  Thus there are five real
parameters, assuming one phase is set to zero.  We expect the matrix
element to have the kinematic dependence $T_{\pi^- \Sigma^+} =T_{\pi^-
\Sigma^+}(W,m)$, where $W$ is the available overall center-of-mass
invariant energy and $m$ is the $\Sigma\pi$ invariant mass.  At a
given value of $W$ and $m$ we have three measured cross sections that
are proportional to the three quantities $T_{\pi^a\Sigma^b}$, so we
cannot determine all five numbers uniquely.

Before going on, we chose at this point to apply the assumption that
the $I=2$ amplitude is negligible, and that all of the interference in
this reaction is between $I=0$ and $I=1$ amplitudes only.  This
assumption is consistent with all previous work on this subject, for
example Refs.~\cite{Nacher:1998mi,Jido:2003cb}, in which the dynamics
of the $\Lambda(1405)$ is presumed to be all within $I=0$ and/or both
$I=0$ and $1$.  With this assumption, we can write the expressions for
the production strength of the three $\Sigma\pi$ channels as
\begin{eqnarray}
|T_{\pi^- \Sigma^+}|^2 &=& \frac{1}{3}|t_0|^2 + \frac{1}{2}|t_1|^2  - \frac{2}{\sqrt{6}}|t_0||t_1|\cos{\phi_{01}} \label{eq:sigppim},\\
|T_{\pi^0 \Sigma^0}|^2 &=& \frac{1}{3}|t_0|^2 \label{eq:sig0pi0},\\
|T_{\pi^+ \Sigma^-}|^2 &=& \frac{1}{3}|t_0|^2 + \frac{1}{2}|t_1|^2  + \frac{2}{\sqrt{6}}|t_0||t_1|\cos{\phi_{01}}. \label{eq:sigmpip}
\end{eqnarray}
These relationships can be combined to show several things.  First,
the sum of the measured line shapes gives the sum of the $I=0$ and $I=
1$ amplitudes' squared magnitudes:
\begin{equation}
|T_{\pi^- \Sigma^+}|^2  + |T_{\pi^0 \Sigma^0}|^2  + |T_{\pi^+
 \Sigma^-}|^2 = |t_0|^2 + |t_1|^2,
\end{equation}
that is, the interference terms cancel and we see the incoherent sum of
the isospin channels.  The $I=0$ amplitude is proportional to the
$\Sigma^0\pi^0$ channel alone, as per Eq.~\eqref{eq:sig0pi0}. The
$I=1$ amplitude's magnitude squared is given by 
\begin{equation}
|t_1|^2 = 
|T_{\pi^- \Sigma^+}|^2  + |T_{\pi^+ \Sigma^-}|^2 - 2 |T_{\pi^0
  \Sigma^0}|^2,
\label{eq:averages}
\end{equation}
which implies that the average of the charged final states should be
greater than or equal to the neutral final state, depending on the
size of $|t_1|$. The interference between the
isospin states is accessed using
\begin{equation}
|T_{\pi^+ \Sigma^-}|^2  - |T_{\pi^- \Sigma^+}|^2 = 
\frac{4}{\sqrt{6}}|t_0||t_1|\cos{\phi_{01}}. 
\end{equation}
This equation shows how any difference between the charged decay modes
is directly related to the interference of the two isospin channels.
Note that $\phi_{01}$ is the mass-dependent phase between $t_0(m)$ and
$t_1(m)$.  Apart from that mass dependence, we allow an arbitrary
strong production phase for each of the amplitudes, called
$\Delta\phi_{I}$ below.

For the production reaction $\gamma + p \to K^+ + (\Sigma\pi)$ we
write the contribution from an amplitude of isospin $I$ at fixed
$\gamma p$ center-of-mass~energy $W$ and $\Sigma\pi$ mass $m$ as
\begin{equation}
t_I(m) = C_I(W) e^{i \Delta\phi_I} B_I(m), 
\label{eq:normform}
\end{equation}
where $C_I(W)$ is a real number representing the effective strength of
the excitation and $\Delta\phi_I$ is a corresponding
production phase.  The Breit-Wigner amplitude has the form
\begin{equation} 
B_I(m) = \sqrt{\frac{2}{\pi}} \left[ \frac{\sqrt{m_R m \Gamma_I^0 (q/q_R)^{2L}}}{m_R^2 - m^2 -
  i m_R \Gamma_{\mathrm{tot}}(q)} \right],
\label{eq:bwamp}
\end{equation} 
\noindent
where $m_R$ is the centroid of the resonance distribution, in this
case the $\Sigma\pi$ invariant mass, $\Gamma_I^0$ is the fixed decay
width to a given final state, and $\Gamma_{\mathrm{tot}}(q)$ is the total width
to all final states.  The available momentum in the decaying hyperon
center-of-mass system is called $q$, and in this frame $q_{R}$ is the
available decay momentum at $m = m_R$.  In this way of writing the
amplitude, the numerator has no phase space factor, but this will
be included below when we write the final expression for the line
shape.

We assume that the line shape for each isospin contribution to the
intermediate hyperon state is described by a relativistic Breit-Wigner
distribution with suitable phase space factors and normalization.
The total width of the resonance, $\Gamma_{\mathrm{tot}}(q)$, is the sum of 
partial decay widths, but for a single decay channel designated by a
``1'', let it be the partial decay width $\Gamma_{I,1}(q_1(m))$.   
The width of the resonance going into a single decay mode ``1'' 
is, in the relativistic formulation, dependent on the mass and is written as
\begin{equation} 
\Gamma_{\mathrm{tot}}(q) \to \Gamma_{I,1}(q_1) = \Gamma^0_{I,1} \frac{m_R}{m} \left(\frac{q_1(m)}{q_{R}}\right)^{2L+1},
\label{eq:q}
\end{equation} 
\noindent
where $\Gamma^0_{I,1}$ denotes a fixed decay width that will be
determined by the fit, and $q_1(m)$ is the available momentum in this
decay mode at mass $m$.  This expression accounts for the increasing
phase space available for the two-body decay across the resonance, and
it forces the width to zero at threshold.  (Later we will analytically
continue $q$ below threshold, however.)  We will consider only $L=0$
or $S$-wave decays, as required for the odd-parity $\Lambda(1405)$
decaying to a pseudo-scalar meson and an octet baryon.
%The factor containing the ratio of masses is not found in all
%representations of the relativistic Breit-Wigner in the literature,
%but we chose to use it.

The overall coupling strength of the resonance represented by
Eq.~\eqref{eq:bwamp} for the reaction $\gamma + p \rightarrow K^+ +
\Lambda(1405)$ is given by the parameters in Eq.~\eqref{eq:normform}.
We take these to be fixed (at a given value of $W$) over the whole
range of the mass distribution $m$.

For several of the fits to the data (discussed below) we used either
two $I=0$ or two $I=1$ Breit-Wigner amplitudes.  In all cases these amplitudes
were added coherently.  We selected $\Delta\phi_0$ for the `first' or
`dominant' $I=0$ amplitude to be zero, so the other strong phases were
determined relative to it.

%If we distinguish the two
%amplitudes within one isospin by $A$ and $B$, they can be expressed
%as
%\begin{equation}
%t_I(m) = C_I^A  e^{i \Delta\phi_I^A}B_I^A(m) + C_I^B  e^{i\Delta\phi_I^B} B_I^B(m),
%\end{equation}
%where the ``A'' and ``B'' superscripts indicate the first and second
%decay amplitudes of isospin $I$.  This is, they add coherently.  For
%example, the parameters in the most general fit to isospin $I=0$ are
%$C_0^A$, $C_0^B$, $\Delta\phi_0^A$ and $\Delta\phi_0^B$.  We select
%$\Delta\phi_0^A$ for the `first' or `dominant' $I=0$ amplitude to be
%zero.  Additional parameters used to characterize the centroid, width,
%and channel coupling of $B_I(m)$ are discussed below.

For hadronic reactions we must also consider the dynamical
consequences of the opening of thresholds to decay channels other than
the single channel denoted ``1''.  In the present situation
there is the $N\bar{K}$ channel that opens at $m_{\mathrm{thresh}} = m_K + m_N
\simeq 1434$~\mevcc, which is within the range of the mass distribution
of the $\Sigma\pi$ system under study.  This can significantly impact
the line shape of the resonance.  To preserve unitarity and the
analytic form of the decay amplitude as a mass threshold is crossed,
we modify the amplitude of Eq.~\eqref{eq:bwamp} in a specific way.  If
we denote the second decay mode as channel ``2'', then the total width
of the resonance is
\begin{equation}
\Gamma_{\mathrm{tot}}(m) = \Gamma_{I,1}(q_1(m)) + \Gamma_{I,2}(q_2(m)), 
\end{equation}
\noindent
where the second decay channel is described by the width
\begin{equation} 
\Gamma_{I,2}(q) = \Gamma^0_{I,2} \frac{m_R}{m} \left(\frac{q_2(m)}{q_R}\right)^{2L+1}.
\end{equation} 
\noindent
Here, $q_2(m)$ is the decay momentum available for decay mode ``2'' at
mass $m$, and $\Gamma^0_{I,2}$ is the constant factor for the width of
this partial decay mode.  Below threshold $m_{\mathrm{thresh}}$, the momentum
$q_2(m)$ is nominally zero.  However, in the Flatt\'e method~\cite{Flatte}  we
analytically continue the momentum to imaginary values, denoting it as
${q_2}^\prime = -i q_2$ for $m<m_{\mathrm{thresh}}$.  Furthermore, we introduce
a Flatt\'e parameter for the branching fraction of the decay modes as
\begin{equation} 
\gamma = \Gamma^0_{I,2}/\Gamma^0_{I,1}.
\end{equation} 
\noindent
Below threshold for decay mode 2, the total decay width is
\begin{equation}
\Gamma_{\mathrm{tot}}(m) = \Gamma_{I,1}(q_1(m)) +  i\gamma\Gamma_{I,1}({q_2}^\prime(m)), 
\end{equation}
\noindent
%and for the mass distribution the line shape is proportional to
%\begin{equation} 
%|B_I(m)|^2 = \frac{2}{\pi} \frac{m m_R  \Gamma^0_{I,1}}{(m_R^2 - m^2
% +\gamma m_R\Gamma_{I,1}(q_2^\prime))^2 + m_R^2 \Gamma_{I,1}(q_1)^2}.
%\end{equation} 
%Above the threshold the total decay width is
while above the threshold the total decay width is
\begin{equation}
\Gamma_{\mathrm{tot}}(m) = \Gamma_1(q_1(m)) +  \gamma\Gamma_1({q_2}(m)),
\end{equation}
\noindent
%and the mass distribution becomes proportional to
%\begin{equation} 
%|B_I(m)|^2 = \frac{2}{\pi} \frac{m m_R  \Gamma^0_{I,1}}{(m_R^2 -
% m^2)^2 + m_R^2 (\Gamma_{I,1}(q_1) + \gamma\Gamma_{I,1}(q_2))^2}.
%\end{equation} 
%Similar expressions occur for the line shape of decay mode ``2'' above
%the threshold mass $m_{\mathrm{thresh}}$, but these do not interest us since
%the experimental result is for the $\Sigma\pi$ mass distribution ``1''
%only, not for $N \bar{K} $ mode ``2''.
and these two expressions are used, respectively,  in Eq~\eqref{eq:bwamp}.

Apart from the overall strength $C_I$ and phase $\Delta\phi_I$, there
are two free parameters in these expressions for a single resonance:
the intrinsic width $\Gamma^0_{I,1}$, and the relative branching
fraction between decay modes $\gamma$.  The fits were made over the
whole range of energy $W$ (in nine bins from 2.0 to 2.8 GeV), and
these two parameters were fixed to the same value for all $W$.

The experimental results for kaon-angle integrated mass distributions
are in the form of differential cross sections $d\sigma_{ab}/dm$ with
$ab \in \{+-, 00, -+\}$; the expression for this cross section
includes relevant flux and phase space factors.
Fig.~\ref{fig:cartoon} illustrates how this reaction requires the use
of three-body phase space. To arrive at it we factorize this phase
space into two two-body pieces using standard
methods~\cite{Chung:1971ri}, the first for the $K^+ Y^*$ hyperon
intermediate state of mass $m$, and the second for the decay of this
state into $\Sigma\pi$.

%The fully differential cross section is 
%\begin{equation}
%d\sigma_{ab} = \frac{(\hbar c)^2}{4 p_{\gamma p} \sqrt{s}} | T_{\pi^a\Sigma^b}|^2 d\Phi_3,
%\label{eq:cross}
%\end{equation}
%where the denominator is the initial state flux at center-of-mass
%momentum $p_{\gamma p}$ and energy $\sqrt{s} = W$, $T_{\pi\Sigma}$ is
%the invariant production amplitude defined by Eqs.~\eqref{eq:sigppim},
%\eqref{eq:sig0pi0} and \eqref{eq:sigmpip}, and $d\Phi_3$ is an element of
%three-body phase space.

%The two-body phase space elements are
%\begin{equation}
%d\Phi_2(\gamma p\to K^+Y^*) = \frac{1}{(4\pi)^2} \frac{p_{K^+}}{\sqrt{s}}d\Omega_{K^+}
%\end{equation}
%and
%\begin{equation}
%d\Phi_2(Y^* \to\pi\Sigma) = \frac{1}{(4\pi)^2} \frac{q}{m}d\Omega_{\Sigma}
%\end{equation}
%for the respective two-body steps, each in their respective
%c.m. frames, where $Y^*$ denotes any intermediate hyperon state of
%mass $m$.  The momentum of the kaon in the overall c.m. system is
%$p_{K^+}$.  The momentum in the $\Sigma\pi$ final state is the
%afrementioned $q$.  The factorized three-body phase space is given by
%\begin{equation}
%d\Phi_3 =
%\left( d\Phi_2(\gamma p\to K^+Y^*) \right) 
%\frac{dm^2}{2\pi}
%\left( d\Phi_2(Y^*   \to\pi\Sigma) \right)
%\label{eq:phasespace}
%\end{equation}
%Combining Eqs.~\eqref{eq:cross} and ~\eqref{eq:phasespace} gives

The fully differential form of the cross section is
\begin{equation}
\frac{d\sigma_{ab}}{d\Omega_{K^+} d\Omega_\Sigma dm} = 
\frac{(\hbar c)^2}{(4\pi)^5} \frac{p_{K^+}q}{p_{\gamma p} s} | T_{\pi^a\Sigma^b}|^2,
\end{equation}
where the momentum of the kaon in the overall center-of-mass system is
$p_{K^+}$, the momentum in the $\Sigma\pi$ final state is the
aforementioned $q$, and $\sqrt{s} = W$. The invariant production
amplitude $T_{\pi\Sigma}$ is defined by Eqs.~\eqref{eq:sigppim},
\eqref{eq:sig0pi0} and \eqref{eq:sigmpip}.

In the experiment we measure the decay distribution of $Y^*
\to\Sigma\pi$ over the full solid angle $\Omega_\Sigma$, so the data
are automatically integrated over this variable. Formally, we take
$T_{\pi \Sigma}$ to be independent of this decay angle.  The reaction
is not ``flat'' with respect to kaon angle, as we know from
measurement of the differential cross section $d\sigma/d\Omega_{K^+}$.
But for studying the line shapes we are forced to integrate over kaon
angle to gain enough statistics for the analysis.  We therefore take
$T_{\pi\Sigma}$ to be the kaon-angle averaged matrix element and
integrate over $\Omega_{K^+}$.  Figure~\ref{fig:cartoon} shows that
there is a vertex involving the photon, and the strength at this
vertex must be proportional to $\sqrt{\alpha}$, where $\alpha$ is the
fine structure constant.  Factoring this out of the matrix element
means the previously-defined fit parameters $C_I$ become an effective
strong coupling with units of $\sqrt{\mathrm{GeV}}$.  The final
expression for the differential-in-mass cross sections is then
\begin{equation}
\frac{d\sigma_{ab}}{dm} = 
\frac{(\hbar c)^2\alpha}{64\pi^3} \frac{p_{K^+}q}{p_{\gamma p} W^2} | T_{\pi^a\Sigma^b}|^2.
\end{equation}
There is an interplay among the phase space factors in front of the
matrix element.  For a given invariant energy $W$, $p_{\gamma p}$ is
determined.  But the possible ranges of $p_{K^+}$ and $m$ are also
limited, so the larger $p_{K^+}$ becomes, the smaller $m$, and
therefore the smaller $q$ must be.

In addition to the coherent sum of the isospin components of the line
shapes, it was necessary to include a linear background function under
each of the $\Sigma\pi$ mass distributions.  This sloping background
was introduced to represent less-than-perfect subtraction of the
backgrounds due, for example, to $K^*$ production or tails of higher
mass hyperons.  The need for such a background parametrization is seen
in the data, which show that in several mass distributions the trend
at the high-mass end of the scales is not toward zero, but rather to a
constant or even a rising slope.  The problem was mainly with the
$\Sigma^0\pi^0$ final state, the one for which it was not possible to
make a direct experimental measurement of the $K^*$ background, and we
had to rely on the Monte Carlo model alone.  The slopes of the
backgrounds were not fit parameters, but were matched to the
differential mass distributions at $1.6$ GeV.

%%%%%%%%%%%%%%%%%%%%%%%%% end of file %%%%%%%%%%%%%%%%%%%%%%%%%%%%%%%%%%%%

%*************************************************************************************
%                                                                                    *
%                               Isospin Decomposition                                *
%                                                                                    *
%*************************************************************************************
% Revisions
% 08-29-12 Add text re Sigma(1385)
% 09-10-12 RS big edits...
% 09-13-12 RS wording...
% 09-17-12 RS wording...
% 11-16-12 RS edits after Ad Hoc comments
% 01-11-13 RS edits after CLAS comments

\section{Isospin Decomposition}
\label{section:isospin}
We can now take the mass distributions found in this analysis and
separate the information from the three charge combinations in the
$\Sigma\pi$ final states according to $I=0$ and $I=1$ components.
This is crucial toward the goal of understanding the contribution from
the true $\Lambda(1405)$, which is by definition $I=0$, and anything
else happening in the reaction mechanism.

We found that fitting two $I=0$ amplitudes to just the $\Sigma^0\pi^0$
data led to a very good fit after including the Flatt\'e channel
coupling~\cite{isospinfits}.  A ``two-pole'' explanation of the
$\Lambda(1405)$ would favor such a result.  The centroids and widths
of the $I=0$ states remained stable when an $I=1$ amplitude was added
to include the $\Sigma^+\pi^-$ and $\Sigma^-\pi^+$ final state
combinations.  However, it was found that a much better fit could be
obtained with a {\it single} $I=0$ amplitude and two separate coherent
$I=1$ amplitudes.  This is the result we show here.  More complete
details of the fits will be given in the separate
paper~\cite{isospinfits}, but here we present the ``best fit''
results.

The fits were made to a reduced data set in order to exactly match the
kaon angular coverage of the three decay modes, and to remove data
points in the vicinity of the $\Lambda(1520)$ where there was evidence
(Fig.~\ref{fig:lineshape}) of less than perfect Monte Carlo matching.
There were a total of 34 free parameters and 1128 data points.  The
reduced $\chi^2$ of the fit was 2.15, the best we achieved with any
amplitude combination.  Most of the parameters were taken up with the
overall strength of each amplitude, $C_I$, in each $W$ bin.  The
centroid, width, and Flatt\'e parameters of the fitted amplitudes as
per Eq.~\eqref{eq:normform} are given in Table~\ref{resultstable4}.

%%%%%%%%%%%%%%%%%%%%%%%%%%%%%%%%% TABLE III %%%%%%%%%%%%%%%%%%%%%%%%%%%%%%%%%%%%%%%%%%%%
\begin{table}[h]
\caption{ Results of the fit using one $I=0$ and two $I=1$
  Breit-Wigner line shapes. }
\begin{center}
\begin{tabular}{c|c|c|c|c}
\hline
\hline
Amplitude & Centroid     &   Width         & Phase                 & Flatt\'e \\ 
          & $m_R$        &$\Gamma^0_{I,1}$ & $\Delta\Phi_I$        & Factor   \\ 
          & (\mevcc)        &   (\mevcc)         & (radians)             & $\gamma$ \\ 
\hline 
$I=0$             & $1338\pm10$ & $ 85\pm 10$  & N/A               & $0.91\pm 0.20$\\
$I=1$ (narrow)    & $1413\pm10$ & $ 52\pm 10$  & $2.0 \pm 0.2$     & $0.41\pm 0.20$\\
$I=1$  (broad)    & $1394\pm20$ & $149\pm 40$  & $0.1 \pm 0.3$     & N/A \\

\hline
\hline
\end{tabular}
\end{center}
\label{resultstable4}
\end{table}
%%%%%%%%%%%%%%%%%%%%%%%%%%%%%%%%%%%%%%%%%%%%%%%%%%%%%%%%%%%%%%%%%%%%%%%%%%%%%%%%%%%%%%%

The $I=0$ piece of the reaction was found in the fit to be at the
$\Sigma\pi$ threshold.  The fit was flexible enough to let this
centroid move smoothly below threshold if necessary, but the fit was
optimal with the centroid of the $\Lambda(1405)$, nominally at $1405$
\mevcc{}, pushed down to $1338$ \mevcc. The rising and falling of the
line shape is controlled by the opening of phase space from threshold
on the low mass side, and by the inflection caused by the opening of
the $N \bar{K} $ threshold on the high-mass side. The intrinsic width of
the $I=0$ resonance was fitted to $85$ \mevcc. However, we expect this
width to be poorly determined due to the dominance of the thresholds
above and below the centroid.  The Flatt\'e coupling parameter is close
to unity.  A value of 0.91 means there is a strong switch-over to the
$N \bar{K}$ decay mode as the available energy exceeds this threshold.
This switch-over is consistent with theoretical
expectations~\cite{Jido:2003cb}.

\begin{figure*}[htpb]
%\begin{figure}[htpb]
\resizebox{1.0\textwidth}{!}{\includegraphics[angle=0.0]{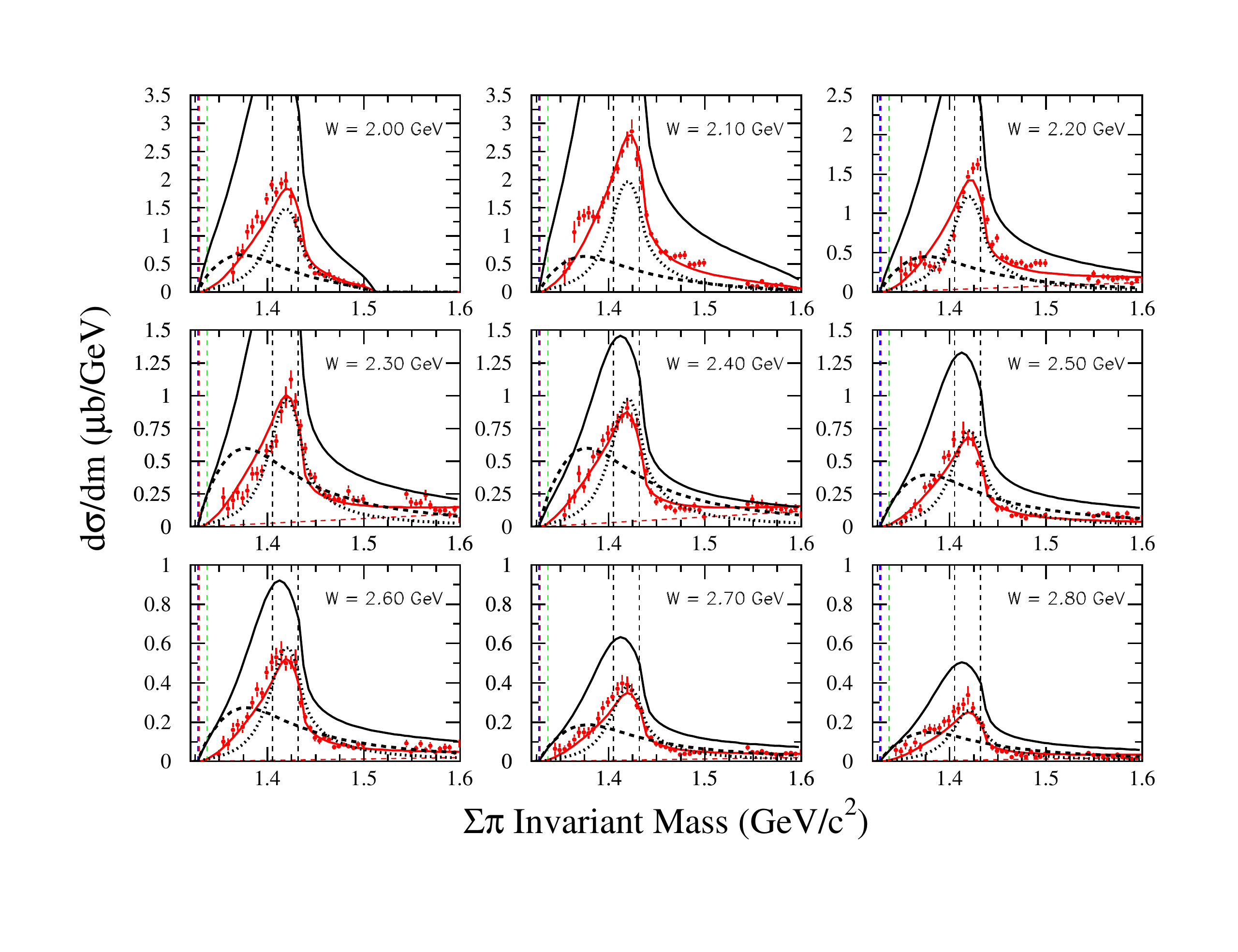}}
\vspace{-2cm}
\caption{(Color online) Data and fits for $\Sigma^+ \pi^-$, with each
  panel showing a different value of $W$.  Data and fitted shapes are
  in red.  The isospin contributions are $I=0$ (solid black), narrow
  $I=1$ (dotted black), and wide $I=1$ (dashed black).  The black
  curves are the same in all panels except for normalization.  The
  vertical dashed lines show the $\Sigma\pi$ thresholds on the left,
  the nominal 1.405 GeV location, and the $N\bar{K}$ threshold. 
  The incoherent background is shown as a thin dashed line (red).
 }
\label{fig:page12_4}       % Give a unique label
%\end{figure}
\end{figure*}

Fig.~\ref{fig:page12_4} shows only the $\Sigma^+ \pi^-$ data and the
corresponding fit, including the underlying separate isospin curves.
The black solid curve shows the dominant $I=0$ line shape which is the
same for all $W$ bins.  It exhibits a distinct edge and change in
curvature at the $N \bar{K}$ mass $m_{\mathrm{thresh}}$ due to
operation of the Flatt\'e effect.  It is evident that the data demand
this sort of slope discontinuity in the $\Sigma \pi$ distributions.
The fit has some problems for $W=2.1, 2.2$ and 2.3 GeV, where the
prominent narrowing around $1400$ \mevcc{} is not reproduced.  We have
been unable to find a combination of fit parameters and amplitudes
that would improve this situation.

One sees in this and the next figures the ``narrow'' $I=1$
contribution (dotted lines) plus a second quite ``wide'' contribution
(dashed lines).  Only the narrow line was allowed to have a Flatt\'e
break at the $N \bar{K}$ threshold, but not the very wide
contribution.  The centroid, width, and Flatt\'e parameter for this
and the other curves are given in Table~\ref{resultstable4}.

%%%%%%%%%%%%%%%%%%%%%%%%%% FIGURE 20 %%%%%%%%%%%%%%%%%%%%%%%%%%%%%%%%%%%%
%remove the ``*'' to put the figures in-line with text.  
%%%%%%%%%%%%%%%%%%%%%%%%%%Fig %%%%%%%%%%%%%%%%%%%%%%%%%%%%%%%%%%%%

\begin{figure*}
%\begin{figure}[htpb]
\resizebox{1.0\textwidth}{!}{\includegraphics[angle=0.0]{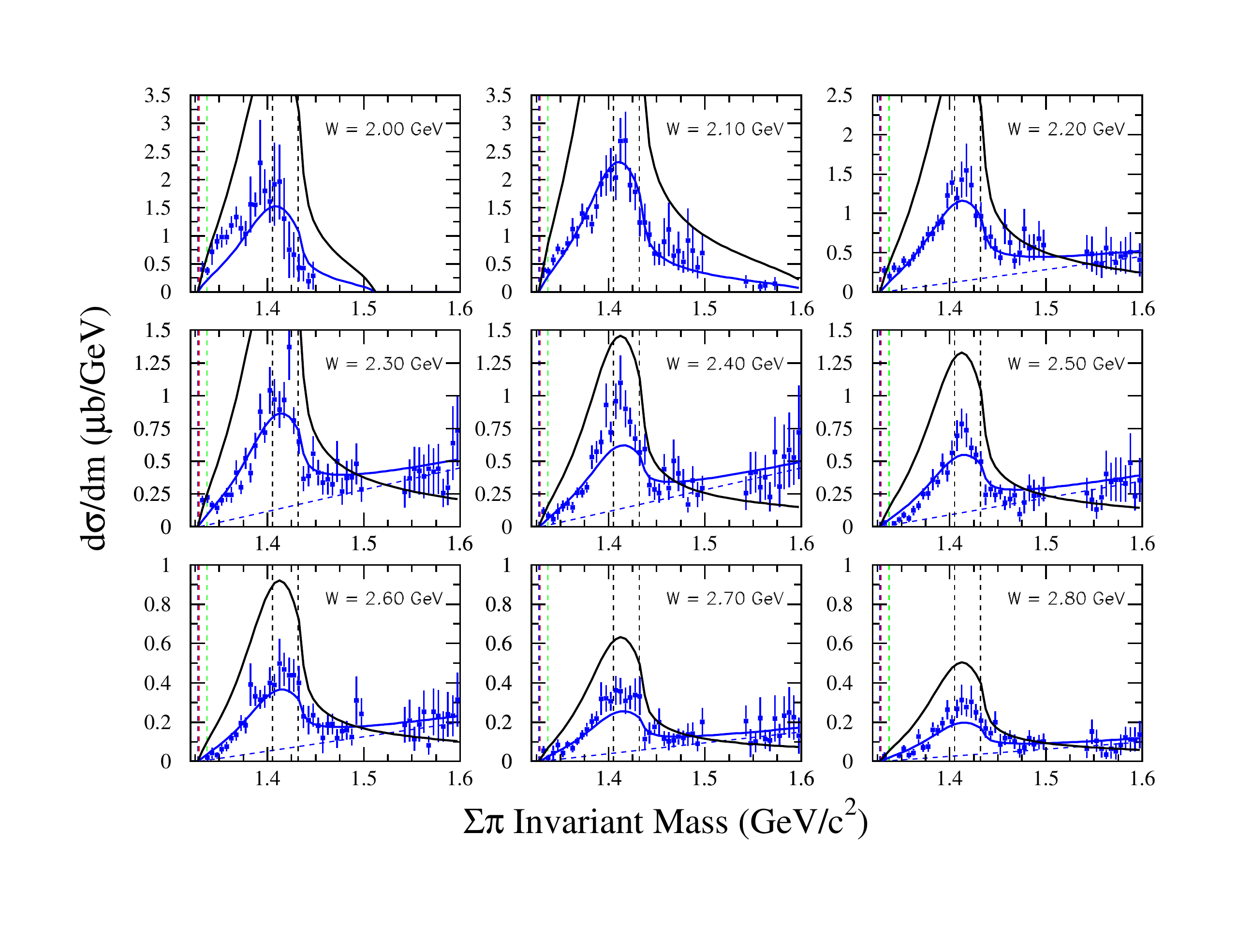}}
\vspace{-2cm}
\caption{ (Color online) 
  Data and fits for $\Sigma^0 \pi^0$, with each panel
  showing a different value of $W$.
  Data and fitted shapes are in blue.
  All the other lines and curves are exactly the same as in 
  Fig.~\ref{fig:page12_4}.
  The $I=1$ curves are not included here.
  The incoherent background is shown as a thin dashed line (blue).
  }
\label{fig:page13_4}       % Give a unique label
%\end{figure}
\end{figure*}

Analogous to Fig.~\ref{fig:page12_4}, Fig.~\ref{fig:page13_4} shows
only the $\Sigma^0 \pi^0$ data and corresponding fit, including the
underlying separate isospin curves.  The $I=0$ line shape (solid
black) is three times the $\Sigma^0\pi^0$ curves (solid blue), as
given in Eq.~\eqref{eq:sig0pi0}, apart from the incoherent background.
Here the effect of using two $I=1$ amplitudes can be considered.  This
channel is all $I=0$, but in accommodating the global fit to all
channels, the position, strength, and width of the single $I=0$ piece
is affected.  The fit is less good than when fitting the
$\Sigma^0\pi^0$ final state alone, and of about equal qualitative
goodness as when using two $I=0$ amplitudes and one $I=1$
amplitude~\cite{isospinfits}.  The $\Sigma^0\pi^0$ channel did not
help us discriminate which amplitude combination is superior.

%%%%%%%%%%%%%%%%%%%%%%%%%% FIGURE 21 %%%%%%%%%%%%%%%%%%%%%%%%%%%%%%%%%%%%
%remove the ``*'' to put the figures in-line with text.  
%%%%%%%%%%%%%%%%%%%%%%%%%%Fig %%%%%%%%%%%%%%%%%%%%%%%%%%%%%%%%%%%%

\begin{figure*}
%\begin{figure}[htpb]
\resizebox{1.0\textwidth}{!}{\includegraphics[angle=0.0]{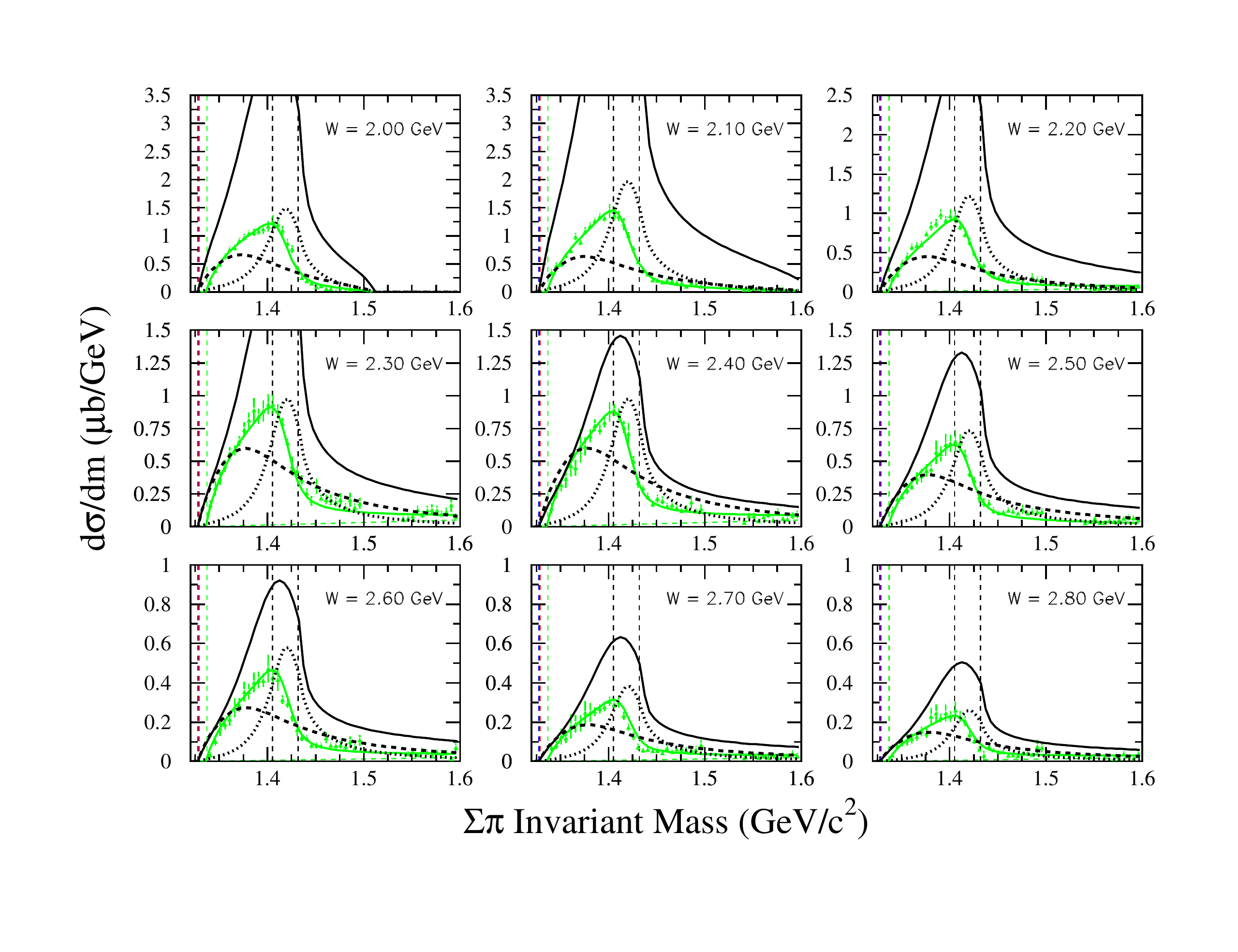}}
\vspace{-2cm}
\caption{ (Color online) 
  Data and fits for $\Sigma^- \pi^+$, with each panel
  showing a different value of $W$.
  Data and fitted shapes are in green.
  All the other lines and curves are exactly the same as in 
  Fig.~\ref{fig:page12_4}.
  The incoherent background is shown as a thin dashed line (green).
  }
\label{fig:page14_4}       % Give a unique label
%\end{figure}
\end{figure*}

Figure~\ref{fig:page14_4} shows only the $\Sigma^- \pi^+$ data with
the corresponding fit (solid green), including the underlying separate
isospin contributions.  In this case the fits are uniformly good
across all values of $W$.  The black curves are the same in each panel
except for their fitted magnitudes, which are the same in
Figs.~\ref{fig:page12_4},~\ref{fig:page13_4}, and ~\ref{fig:page14_4}
at each $W$.

%%%%%%%%%%%%%%%%%%%%%%%%%% FIGURE 22 %%%%%%%%%%%%%%%%%%%%%%%%%%%%%%%%%%%%
%remove the ``*'' to put the figures in-line with text.  
%%%%%%%%%%%%%%%%%%%%%%%%%%Fig %%%%%%%%%%%%%%%%%%%%%%%%%%%%%%%%%%%%

The fit comfortably accommodates Breit-Wigner-like $I=1$ structures centered
near $1394$ and $1413$ \mevcc.  There are no standard quark-model
$\Sigma$ states that would fit this description.  The observation at
least tentatively suggests evidence for the $I=1$, $J^P=1/2^-$,
$\Sigma^*$ state predicted in some extensions of the basic quark
model~\cite{Zou:2010tc}.  However, our fit is a phenomenological
parametrization of the $I=1$ amplitude, and not a direct
identification of resonant states.

The broad $I=1$ structure is hard to interpret because it is so wide.
It could result from a non-resonant coherent 3-body amplitude present
in the reaction mechanism.  The fit is substantially better when
including this second $I=1$ amplitude, in fact, it is crucial for
providing the separation between the mass distributions in the
threshold region of the three charge states.

\begin{figure}[htpb]
\resizebox{0.5\textwidth}{!}{\includegraphics[angle=0.0]{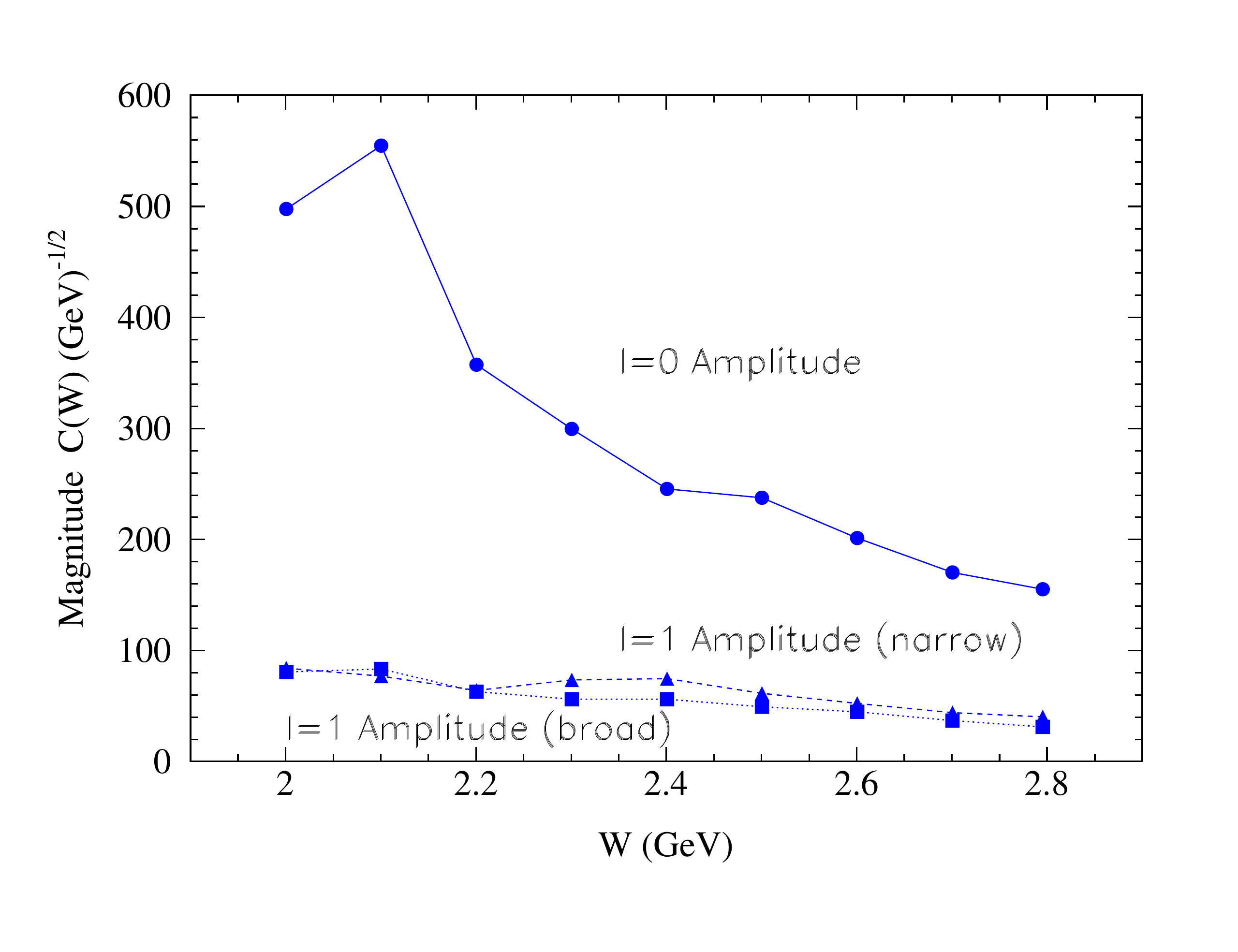}}
\caption{(Color online) Strength of each of the isospin amplitudes as
  a function of $W$.  These are the real coefficients of the
  amplitudes, of which the magnitudes give the contributions of each
  isospin component.  }
\label{fig:page15_4}       % Give a unique label
\end{figure}

The component curves for the one $I=0$ and two $I=1$ amplitudes
contributing to $d\sigma/dm$ are the same in shape, but differ in
magnitude, on each panel.  It is evident that the $I=0$ strength is the
largest contribution to the reaction, but the two $I=1$ contributions
are far from small in comparison.  The magnitudes of the isospin
components as a function of $W$ are shown in Fig.~\ref{fig:page15_4}.
These are the real coefficients as per Eq.~\eqref{eq:normform} that
enter each Breit-Wigner amplitude (in magnitude).  Above 2.2 GeV the
$I=1$ strengths combined are as large as half of the $I=0$ strength.
The relative phase angle of the broad $I=1$ amplitude is close to zero
with respect to the $I=0$ amplitude.  This means there is no
interference between them apart from the Breit-Wigner phase dependence.  However
the two $I=1$ amplitudes have a large phase with respect to each
other, as given in Table~\ref{resultstable4}, and for this we have is
no simple explanation.

%%%%%%%%%%%%%%%%%%%%%%%%%% FIGURE 23 %%%%%%%%%%%%%%%%%%%%%%%%%%%%%%%%%%%%
%remove the ``*'' to put the figures in-line with text.  
%\begin{figure*}
%\end{figure*} 
%%%%%%%%%%%%%%%%%%%%%%%%%% End Fig %%%%%%%%%%%%%%%%%%%%%%%%%%%%%%%%%%%%

We think the work discussed above makes the case that the
$\Lambda(1405)$, as seen experimentally in photoproduction on the
proton, is not an isolated $I=0$ resonance centered near $1405$
\mevcc.  The observed line shape (or mass distribution) differs in
each of the three $\Sigma\pi$ decay modes, which shows that there is
substantial $I=1$ strength in the system.  We found it necessary to
carefully consider the opening of the $N \bar{K}$ decay mode.  We have
interpreted the $I=1$ strength in terms of two Breit-Wigner resonances
that interfere with the pure $I=0$ state $\Lambda(1405)$.  After this
was done, we arrived at a satisfactory representation of the
experimental results.  Even our best fit does not reproduce the data
fully, and it is difficult to tell whether the remaining discrepancies
indicate unresolved systematic issues with the data or additional
physics content that we have not identified.

According to our best fit, a narrow $I=1$ amplitude is a substantial
piece of the overall production strength of what has loosely been
called the ``$\Lambda(1405)$''.  A wide contribution also appears to
be needed.  The extra $I=1$ strength must have $J^P = 1/2^-$ in order
to interfere as it does with the $I=0$ amplitude, the true
$\Lambda(1405)$.  It must be emphasized that this $I=1$ strength has
nothing to do with the standard $\Sigma^0(1385)$ $J^P = 3/2^+$ since
that state was carefully excluded much earlier in the analysis
process, both by explicit subtraction and by recognition that it
cannot interfere in the present angle-integrated spectra.  Although
our angular coverage of the hyperon decays is not complete, a majority
of the range has been measured.

Assuming we are correct in the identification and assignment of
quantum numbers of the Breit-Wigner amplitudes we see, we can discuss them in
light of recent theoretical models.  First, the low mass of the $I=0$
amplitude is consistent with predictions of a two-pole structure for
the $\Lambda(1405)$, wherein the lower of the two poles is more likely
to couple to the $\Sigma\pi$ final state.  For example, in the
chiral-unitary model of Ramos, Oset and Bennhold~\cite{Ramos:2003mu},
the lower-mass pole is at $1390+i66$~\mevcc.  However, the same
analysis predicts a $\Sigma$ with $1/2^-$ at $1579+i274$ \mevcc, which
is not consistent with the structure we see.  In the model of Oller
and Meissner~\cite{Oller:2000fj} the $I=0$ lower-mass pole is on two
Riemann sheets at $1379-i28$ and $1433-i11$~\mevcc, whereas the $I=1$
pole is at $1444-i69$ and $1419+i42$~\mevcc.  Hence this latter model
is somewhat closer to our results.  The meson-exchange model of
Haidenbauer \etal~\cite{Haidenbauer:2010ch} also predicts a two-pole
structure for the \LambdaOne{} with positions at $1334.3 + i62.3$
\mevcc{} and $1435.8 + i25.6$ \mevcc.  The lower of these is at the
$\Sigma\pi$ threshold, as found by us in the present fit.  When we fit
with two $I=0$ line shapes~\cite{isospinfits}, however, the higher
mass centroid does not match the predicted pole position.  The
resonance pole positions in the various models do not correspond
directly to the centroids of Breit-Wigner mass distributions, so these
numerical comparisons are only qualitative.

To make a further connection to previous theoretical work we can make
some remarks about previous efforts to identify a $\Sigma^*(1/2^-)$
state near $1380$ \mevcc.  As mentioned in the Introduction, positing
such a state was a consequence of examining several open issues in
hadron structure using a 5-quark baryon ansatz~\cite{Zou:2010tc}.  In
that class of models, the dominant configuration of some excited
baryons consists of two diquarks and an anti-quark in a mutual $L=0$ or $L=1$
state.  This in turn can lead to low mass, negative parity, isovector
states such as the one under discussion here.  The results we obtained
here may relate to the observation that the line shape of the
$\Sigma(1385)$ does not conform to its expected $P$-wave character,
as discussed in Section~\ref{subsection:YieldExtraction:chLambda}.  If
there is indeed an admixture of an $I=1$ amplitude with $J^P=1/2^-$ at
nearly the same mass, one can in principle have interferences that
modify the line shape of the experimentally-seen $\Lambda\pi$ final
state.  We have not pursued this question further at this time.  We
emphasize once again that a $P$-wave decay cannot be biasing our results
for the $S$-wave $\Sigma\pi$ data because we integrate over the hyperon
decay angles, canceling any interference.

The CLAS results for the $\Sigma\pi$ mass distributions
in the vicinity of $1405$ \mevcc{} are compelling in the following sense.
The mass distribution differences between the charge states are large
and systematic across our measured kinematic space.  The need for
$I=1$ strength is inescapable.  Furthermore, we have shown that the
$\Sigma(1385)(3/2^+)$ is not a player in this phenomenology, and we
have taken care to show that the $K^*$ production background also does 
not play a role.  Finally, the line shape fits that we have made
show that the $I=1$ strength is described at least in part by 
Breit-Wigner $I=1$ amplitudes with the masses and widths given in
Table~\ref{resultstable4}.

%%%%%%%%%%%%%%%%%%%%%%%% end of file %%%%%%%%%%%%%%%%%%%%%%%%%%%%%%%%%%%%

%*************************************************************************************
%                                                                                    *
%                                    Conclusions                                     *
%                                                                                    *
%*************************************************************************************
% Revisions:
% 09-10-12  RS wordsmithing
% 09-14-12  RS wordsmithing
% 11-20-12 RS changes after Ad Hoc comments
% 01-11-13 RS changes after CLAS comments

\section{Conclusions}
\label{section:Conclusions}

The mass distributions or line shapes of the invariant $\Sigma \pi$
mass have been measured in the region of the \LambdaOne{} using CLAS
at Jefferson Lab.  All three charge combinations were measured, and
the main qualitative conclusion is that they are significantly
different from each other and none is well represented by a simple
Breit-Wigner line shape.  We have shown that the background from the
$\Sigma(1385) \to \Sigma^\pm\pi^\mp$ states is small and
well-controlled by scaling the dominant $\Sigma(1385) \to
\Lambda\pi^0$ decay.  We have shown that the interference with
$K^*\Sigma$ final states is unimportant in the sense that the line
shape results are unaffected.

Interference of $I=0$ and $I=1$ isospin channels appears to lie at the
root of the differing line shapes for the three $\Sigma\pi$ final
states.  That is to say, there is $I=1$ strength present with the same
$J^P=1/2^-$ quantum numbers as the $\Lambda(1405)$.  Amplitude-level
fits suggest that there may be a $\Sigma$-like state in this mass
range, and that the centroid of the $I=0$ $\Lambda(1405)$ state lies
essentially at the $\Sigma\pi$ threshold.  This places the
\LambdaOne{} far from the nominal PDG mass value, in a place where
$\Sigma\pi$ threshold effects will have to be understood
quantitatively to obtain an accurate picture of this state.  From the
same analysis, even the $\Sigma^0\pi^0$ channel, which is purely
$I=0$, cannot be represented by a relativistic Breit-Wigner line shape
alone.  We find that a channel-coupling to the unmeasured $N \bar{K}$
final state via a Flatt\'e-style unitarization can lead to a
satisfactory shape, and that indeed this channel-coupling dominates
the observed mass distribution.  Thus, we find some signature effect
for a two-pole picture of the $I=0$ ~$\Lambda(1405)$, in which the
reaction amplitude couples significantly to both final states.
However, we see also how the $I=1$ amplitude adds one more layer of
complexity to the experimental picture by influencing the charged
final states.  The choice of one $I=0$ and two $I=1$ amplitudes
presented in this paper led to the best fit among several choices.
Similar results were obtained using two $I=0$ and one $I=1$ amplitude,
which may correspond more closely to current theoretical ideas, but
these are described elsewhere~\cite{isospinfits}.

In addition to the results shown in this paper, the photoproduction
differential cross sections of the \LambdaOne, \LambdaTwo, and
$\Sigma(1385)$, will be presented in a separate
paper~\cite{crosssectionpaper}.  Also, the same data have been used to
directly measure the spin and parity of the $\Lambda(1405)$, and this
result will also be presented separately~\cite{paritypaper}.

Clearly, both more theoretical modeling of the present results and
additional experimental data are needed.  The present work has, we
conclude, provided detailed line shape results, to which a
parametrization with a set of Breit-Wigner amplitudes shows the
importance of $I=1$, $J^P=1/2^-$ strength centered near $1394$ and
$1413$ \mevcc, with a dominant $I=0$ piece very near the $\Sigma\pi$
threshold.

%%%%%%%%%%%%%%%%%%%%%%%%%%%%%%%%%%%%%%%%%%%%%%%%%%%%%%%%%%%%%%%%%%%

\begin{acknowledgments}
We acknowledge the outstanding efforts of the staff of the Accelerator
and Physics Divisions at Jefferson Lab that made this experiment
possible. The work of the Medium Energy Physics group at Carnegie
Mellon University was supported by DOE grant DE-FG02-87ER40315.  The
Southeastern Universities Research Association (SURA) operated the
Thomas Jefferson National Accelerator Facility for the United States
Department of Energy under contract DE-AC05-84ER40150.  Further
support was provided by the National Science Foundation and the United
Kingdom's Science and Technology Facilities Council.

\end{acknowledgments}
\vfill

\bibliography{lineshape}

\begin{thebibliography}{55}%
\makeatletter
\providecommand \@ifxundefined [1]{%
 \@ifx{#1\undefined}
}%
\providecommand \@ifnum [1]{%
 \ifnum #1\expandafter \@firstoftwo
 \else \expandafter \@secondoftwo
 \fi
}%
\providecommand \@ifx [1]{%
 \ifx #1\expandafter \@firstoftwo
 \else \expandafter \@secondoftwo
 \fi
}%
\providecommand \natexlab [1]{#1}%
\providecommand \enquote  [1]{``#1''}%
\providecommand \bibnamefont  [1]{#1}%
\providecommand \bibfnamefont [1]{#1}%
\providecommand \citenamefont [1]{#1}%
\providecommand \href@noop [0]{\@secondoftwo}%
\providecommand \href [0]{\begingroup \@sanitize@url \@href}%
\providecommand \@href[1]{\@@startlink{#1}\@@href}%
\providecommand \@@href[1]{\endgroup#1\@@endlink}%
\providecommand \@sanitize@url [0]{\catcode `\\12\catcode `\$12\catcode
  `\&12\catcode `\#12\catcode `\^12\catcode `\_12\catcode `\%12\relax}%
\providecommand \@@startlink[1]{}%
\providecommand \@@endlink[0]{}%
\providecommand \url  [0]{\begingroup\@sanitize@url \@url }%
\providecommand \@url [1]{\endgroup\@href {#1}{\urlprefix }}%
\providecommand \urlprefix  [0]{URL }%
\providecommand \Eprint [0]{\href }%
\providecommand \doibase [0]{http://dx.doi.org/}%
\providecommand \selectlanguage [0]{\@gobble}%
\providecommand \bibinfo  [0]{\@secondoftwo}%
\providecommand \bibfield  [0]{\@secondoftwo}%
\providecommand \translation [1]{[#1]}%
\providecommand \BibitemOpen [0]{}%
\providecommand \bibitemStop [0]{}%
\providecommand \bibitemNoStop [0]{.\EOS\space}%
\providecommand \EOS [0]{\spacefactor3000\relax}%
\providecommand \BibitemShut  [1]{\csname bibitem#1\endcsname}%
\let\auto@bib@innerbib\@empty
%</preamble>
\bibitem [{\citenamefont {Alston}\ \emph {et~al.}(1961)\citenamefont {Alston}
  \emph {et~al.}}]{Alston_1405}%
  \BibitemOpen
  \bibfield  {author} {\bibinfo {author} {\bibfnamefont {M.~H.}\ \bibnamefont
  {Alston}} \emph {et~al.},\ }\href {\doibase 10.1103/PhysRevLett.6.698}
  {\bibfield  {journal} {\bibinfo  {journal} {Phys. Rev. Lett.}\ }\textbf
  {\bibinfo {volume} {6}},\ \bibinfo {pages} {698} (\bibinfo {year}
  {1961})}\BibitemShut {NoStop}%
%%CITATION = PRLTA,6,698;%%
\bibitem [{\citenamefont {Isgur}\ and\ \citenamefont
  {Karl}(1978)}]{Isgur-Karl_PRD18}%
  \BibitemOpen
  \bibfield  {author} {\bibinfo {author} {\bibfnamefont {N.}~\bibnamefont
  {Isgur}}\ and\ \bibinfo {author} {\bibfnamefont {G.}~\bibnamefont {Karl}},\
  }\href {\doibase 10.1103/PhysRevD.18.4187} {\bibfield  {journal} {\bibinfo
  {journal} {Phys. Rev.}\ }\textbf {\bibinfo {volume} {D18}},\ \bibinfo {pages}
  {4187} (\bibinfo {year} {1978})}\BibitemShut {NoStop}%
%%CITATION = PHRVA,D18,4187;%%
\bibitem [{\citenamefont {Capstick}\ and\ \citenamefont
  {Isgur}(1986)}]{Capstick-Isgur}%
  \BibitemOpen
  \bibfield  {author} {\bibinfo {author} {\bibfnamefont {S.}~\bibnamefont
  {Capstick}}\ and\ \bibinfo {author} {\bibfnamefont {N.}~\bibnamefont
  {Isgur}},\ }\href {\doibase 10.1103/PhysRevD.34.2809} {\bibfield  {journal}
  {\bibinfo  {journal} {Phys. Rev.}\ }\textbf {\bibinfo {volume} {D34}},\
  \bibinfo {pages} {2809} (\bibinfo {year} {1986})}\BibitemShut {NoStop}%
%%CITATION = PHRVA,D34,2809;%%
\bibitem [{\citenamefont {Capstick}\ and\ \citenamefont
  {Roberts}(2000)}]{Capstick:2000qj}%
  \BibitemOpen
  \bibfield  {author} {\bibinfo {author} {\bibfnamefont {S.}~\bibnamefont
  {Capstick}}\ and\ \bibinfo {author} {\bibfnamefont {W.}~\bibnamefont
  {Roberts}},\ }\href@noop {} {\bibfield  {journal} {\bibinfo  {journal}
  {Prog.Part.Nucl.Phys.}\ }\textbf {\bibinfo {volume} {45}},\ \bibinfo {pages}
  {S241} (\bibinfo {year} {2000})},\ \Eprint
  {http://arxiv.org/abs/nucl-th/0008028} {arXiv:nucl-th/0008028 [nucl-th]}
  \BibitemShut {NoStop}%
%%CITATION = NUCL-TH/0008028;%%
\bibitem [{\citenamefont {Dalitz}\ and\ \citenamefont
  {Tuan}(1959{\natexlab{a}})}]{Dalitz_Tuan:AnnPhys8}%
  \BibitemOpen
  \bibfield  {author} {\bibinfo {author} {\bibfnamefont {R.~H.}\ \bibnamefont
  {Dalitz}}\ and\ \bibinfo {author} {\bibfnamefont {S.~F.}\ \bibnamefont
  {Tuan}},\ }\href@noop {} {\bibfield  {journal} {\bibinfo  {journal} {Annals
  Phys.}\ }\textbf {\bibinfo {volume} {8}},\ \bibinfo {pages} {100} (\bibinfo
  {year} {1959}{\natexlab{a}})}\BibitemShut {NoStop}%
%%CITATION = APNYA,8,100;%%
\bibitem [{\citenamefont {Dalitz}\ and\ \citenamefont
  {Tuan}(1959{\natexlab{b}})}]{Dalitz_Tuan:PRL}%
  \BibitemOpen
  \bibfield  {author} {\bibinfo {author} {\bibfnamefont {R.~H.}\ \bibnamefont
  {Dalitz}}\ and\ \bibinfo {author} {\bibfnamefont {S.~F.}\ \bibnamefont
  {Tuan}},\ }\href {\doibase 10.1103/PhysRevLett.2.425} {\bibfield  {journal}
  {\bibinfo  {journal} {Phys. Rev. Lett.}\ }\textbf {\bibinfo {volume} {2}},\
  \bibinfo {pages} {425} (\bibinfo {year} {1959}{\natexlab{b}})}\BibitemShut
  {NoStop}%
%%CITATION = PRLTA,2,425;%%
\bibitem [{\citenamefont {Dalitz}\ and\ \citenamefont
  {Tuan}(1960)}]{Dalitz_Tuan:AnnPhys10}%
  \BibitemOpen
  \bibfield  {author} {\bibinfo {author} {\bibfnamefont {R.~H.}\ \bibnamefont
  {Dalitz}}\ and\ \bibinfo {author} {\bibfnamefont {S.~F.}\ \bibnamefont
  {Tuan}},\ }\href@noop {} {\bibfield  {journal} {\bibinfo  {journal} {Annals
  Phys.}\ }\textbf {\bibinfo {volume} {10}},\ \bibinfo {pages} {307} (\bibinfo
  {year} {1960})}\BibitemShut {NoStop}%
%%CITATION = APNYA,10,307;%%
\bibitem [{\citenamefont {Oset}\ and\ \citenamefont
  {Ramos}(1998)}]{Oset-Ramos}%
  \BibitemOpen
  \bibfield  {author} {\bibinfo {author} {\bibfnamefont {E.}~\bibnamefont
  {Oset}}\ and\ \bibinfo {author} {\bibfnamefont {A.}~\bibnamefont {Ramos}},\
  }\href {\doibase 10.1016/S0375-9474(98)00170-5} {\bibfield  {journal}
  {\bibinfo  {journal} {Nucl. Phys.}\ }\textbf {\bibinfo {volume} {A635}},\
  \bibinfo {pages} {99} (\bibinfo {year} {1998})},\ \Eprint
  {http://arxiv.org/abs/9711022} {arXiv:9711022} \BibitemShut {NoStop}%
%%CITATION = NUCL-TH/9711022;%%
\bibitem [{\citenamefont {Oller}\ and\ \citenamefont
  {Meissner}(2001)}]{Oller:2000fj}%
  \BibitemOpen
  \bibfield  {author} {\bibinfo {author} {\bibfnamefont {J.~A.}\ \bibnamefont
  {Oller}}\ and\ \bibinfo {author} {\bibfnamefont {U.~G.}\ \bibnamefont
  {Meissner}},\ }\href {\doibase 10.1016/S0370-2693(01)00078-8} {\bibfield
  {journal} {\bibinfo  {journal} {Phys. Lett.}\ }\textbf {\bibinfo {volume}
  {B500}},\ \bibinfo {pages} {263} (\bibinfo {year} {2001})},\ \Eprint
  {http://arxiv.org/abs/0011146} {arXiv:0011146} \BibitemShut {NoStop}%
%%CITATION = HEP-PH/0011146;%%
\bibitem [{\citenamefont {Hyodo}\ and\ \citenamefont
  {Jido}(2012)}]{Hyodo:2011ur}%
  \BibitemOpen
  \bibfield  {author} {\bibinfo {author} {\bibfnamefont {T.}~\bibnamefont
  {Hyodo}}\ and\ \bibinfo {author} {\bibfnamefont {D.}~\bibnamefont {Jido}},\
  }\href {\doibase 10.1016/j.ppnp.2011.07.002} {\bibfield  {journal} {\bibinfo
  {journal} {Prog.Part.Nucl.Phys.}\ }\textbf {\bibinfo {volume} {67}},\
  \bibinfo {pages} {55} (\bibinfo {year} {2012})},\ \Eprint
  {http://arxiv.org/abs/1104.4474} {arXiv:1104.4474 [nucl-th]} \BibitemShut
  {NoStop}%
%%CITATION = ARXIV:1104.4474;%%
\bibitem [{\citenamefont {Nacher}\ \emph {et~al.}(1999)\citenamefont {Nacher},
  \citenamefont {Oset}, \citenamefont {Toki},\ and\ \citenamefont
  {Ramos}}]{Nacher:1998mi}%
  \BibitemOpen
  \bibfield  {author} {\bibinfo {author} {\bibfnamefont {J.~C.}\ \bibnamefont
  {Nacher}}, \bibinfo {author} {\bibfnamefont {E.}~\bibnamefont {Oset}},
  \bibinfo {author} {\bibfnamefont {H.}~\bibnamefont {Toki}}, \ and\ \bibinfo
  {author} {\bibfnamefont {A.}~\bibnamefont {Ramos}},\ }\href {\doibase
  10.1016/S0370-2693(99)00380-9} {\bibfield  {journal} {\bibinfo  {journal}
  {Phys. Lett.}\ }\textbf {\bibinfo {volume} {B455}},\ \bibinfo {pages} {55}
  (\bibinfo {year} {1999})},\ \Eprint {http://arxiv.org/abs/9812055}
  {arXiv:9812055} \BibitemShut {NoStop}%
%%CITATION = NUCL-TH/9812055;%%
\bibitem [{\citenamefont {Kaiser}\ \emph {et~al.}(1995)\citenamefont {Kaiser},
  \citenamefont {Siegel},\ and\ \citenamefont {Weise}}]{Kaiser-Siegel-Weise}%
  \BibitemOpen
  \bibfield  {author} {\bibinfo {author} {\bibfnamefont {N.}~\bibnamefont
  {Kaiser}}, \bibinfo {author} {\bibfnamefont {P.~B.}\ \bibnamefont {Siegel}},
  \ and\ \bibinfo {author} {\bibfnamefont {W.}~\bibnamefont {Weise}},\ }\href
  {\doibase 10.1016/0375-9474(95)00362-5} {\bibfield  {journal} {\bibinfo
  {journal} {Nucl. Phys.}\ }\textbf {\bibinfo {volume} {A594}},\ \bibinfo
  {pages} {325} (\bibinfo {year} {1995})},\ \Eprint
  {http://arxiv.org/abs/9505043} {arXiv:9505043} \BibitemShut {NoStop}%
%%CITATION = NUCL-TH/9505043;%%
\bibitem [{\citenamefont {Jido}\ \emph {et~al.}(2003)\citenamefont {Jido},
  \citenamefont {Oller}, \citenamefont {Oset}, \citenamefont {Ramos},\ and\
  \citenamefont {Meissner}}]{Jido:2003cb}%
  \BibitemOpen
  \bibfield  {author} {\bibinfo {author} {\bibfnamefont {D.}~\bibnamefont
  {Jido}}, \bibinfo {author} {\bibfnamefont {J.~A.}\ \bibnamefont {Oller}},
  \bibinfo {author} {\bibfnamefont {E.}~\bibnamefont {Oset}}, \bibinfo {author}
  {\bibfnamefont {A.}~\bibnamefont {Ramos}}, \ and\ \bibinfo {author}
  {\bibfnamefont {U.~G.}\ \bibnamefont {Meissner}},\ }\href {\doibase
  10.1016/S0375-9474(03)01598-7} {\bibfield  {journal} {\bibinfo  {journal}
  {Nucl. Phys.}\ }\textbf {\bibinfo {volume} {A725}},\ \bibinfo {pages} {181}
  (\bibinfo {year} {2003})},\ \Eprint {http://arxiv.org/abs/0303062}
  {arXiv:0303062} \BibitemShut {NoStop}%
%%CITATION = NUCL-TH/0303062;%%
\bibitem [{\citenamefont {Borasoy}\ \emph {et~al.}(2005)\citenamefont
  {Borasoy}, \citenamefont {Nissler},\ and\ \citenamefont {Weise}}]{Borasoy}%
  \BibitemOpen
  \bibfield  {author} {\bibinfo {author} {\bibfnamefont {B.}~\bibnamefont
  {Borasoy}}, \bibinfo {author} {\bibfnamefont {R.}~\bibnamefont {Nissler}}, \
  and\ \bibinfo {author} {\bibfnamefont {W.}~\bibnamefont {Weise}},\ }\href
  {\doibase 10.1140/epja/i2005-10079-1} {\bibfield  {journal} {\bibinfo
  {journal} {Eur. Phys. J.}\ }\textbf {\bibinfo {volume} {A25}},\ \bibinfo
  {pages} {79} (\bibinfo {year} {2005})},\ \Eprint
  {http://arxiv.org/abs/0505239} {arXiv:0505239} \BibitemShut {NoStop}%
%%CITATION = HEP-PH/0505239;%%
\bibitem [{\citenamefont {Lutz}\ and\ \citenamefont
  {Soyeur}(2005)}]{Lutz:2004sg}%
  \BibitemOpen
  \bibfield  {author} {\bibinfo {author} {\bibfnamefont {M.~F.}\ \bibnamefont
  {Lutz}}\ and\ \bibinfo {author} {\bibfnamefont {M.}~\bibnamefont {Soyeur}},\
  }\href {\doibase 10.1016/j.nuclphysa.2004.11.007} {\bibfield  {journal}
  {\bibinfo  {journal} {Nucl.Phys.}\ }\textbf {\bibinfo {volume} {A748}},\
  \bibinfo {pages} {499} (\bibinfo {year} {2005})},\ \Eprint
  {http://arxiv.org/abs/nucl-th/0407115} {arXiv:nucl-th/0407115 [nucl-th]}
  \BibitemShut {NoStop}%
%%CITATION = NUCL-TH/0407115;%%
\bibitem [{\citenamefont {Haidenbauer}\ \emph {et~al.}(2011)\citenamefont
  {Haidenbauer}, \citenamefont {Krein}, \citenamefont {Meissner},\ and\
  \citenamefont {Tolos}}]{Haidenbauer:2010ch}%
  \BibitemOpen
  \bibfield  {author} {\bibinfo {author} {\bibfnamefont {J.}~\bibnamefont
  {Haidenbauer}}, \bibinfo {author} {\bibfnamefont {G.}~\bibnamefont {Krein}},
  \bibinfo {author} {\bibfnamefont {U.-G.}\ \bibnamefont {Meissner}}, \ and\
  \bibinfo {author} {\bibfnamefont {L.}~\bibnamefont {Tolos}},\ }\href
  {\doibase 10.1140/epja/i2011-11018-3} {\bibfield  {journal} {\bibinfo
  {journal} {Eur.Phys.J.}\ }\textbf {\bibinfo {volume} {A47}},\ \bibinfo
  {pages} {18} (\bibinfo {year} {2011})},\ \Eprint
  {http://arxiv.org/abs/1008.3794} {arXiv:1008.3794 [nucl-th]} \BibitemShut
  {NoStop}%
%%CITATION = ARXIV:1008.3794;%%
\bibitem [{\citenamefont {Akaishi}\ and\ \citenamefont
  {Yamazaki}(2002)}]{Akaishi}%
  \BibitemOpen
  \bibfield  {author} {\bibinfo {author} {\bibfnamefont {Y.}~\bibnamefont
  {Akaishi}}\ and\ \bibinfo {author} {\bibfnamefont {T.}~\bibnamefont
  {Yamazaki}},\ }\href {\doibase 10.1103/PhysRevC.65.044005} {\bibfield
  {journal} {\bibinfo  {journal} {Phys. Rev.}\ }\textbf {\bibinfo {volume}
  {C65}},\ \bibinfo {pages} {044005} (\bibinfo {year} {2002})}\BibitemShut
  {NoStop}%
%%CITATION = PHRVA,C65,044005;%%
\bibitem [{\citenamefont {Akaishi}\ \emph {et~al.}(2010)\citenamefont
  {Akaishi}, \citenamefont {Yamazaki}, \citenamefont {Obu},\ and\ \citenamefont
  {Wada}}]{Akaishi:2010wt}%
  \BibitemOpen
  \bibfield  {author} {\bibinfo {author} {\bibfnamefont {Y.}~\bibnamefont
  {Akaishi}}, \bibinfo {author} {\bibfnamefont {T.}~\bibnamefont {Yamazaki}},
  \bibinfo {author} {\bibfnamefont {M.}~\bibnamefont {Obu}}, \ and\ \bibinfo
  {author} {\bibfnamefont {M.}~\bibnamefont {Wada}},\ }\href {\doibase
  10.1016/j.nuclphysa.2010.01.176} {\bibfield  {journal} {\bibinfo  {journal}
  {Nucl.Phys.}\ }\textbf {\bibinfo {volume} {A835}},\ \bibinfo {pages} {67}
  (\bibinfo {year} {2010})},\ \Eprint {http://arxiv.org/abs/1002.2560}
  {arXiv:1002.2560 [nucl-th]} \BibitemShut {NoStop}%
%%CITATION = ARXIV:1002.2560;%%
\bibitem [{\citenamefont {Wohl}(1998)}]{Wohl}%
  \BibitemOpen
  \bibfield  {author} {\bibinfo {author} {\bibfnamefont {C.~G.}\ \bibnamefont
  {Wohl}} (\bibinfo {collaboration} {Particle Data Group}),\ }\href {\doibase
  10.1103/PhysRevD.86.010001} {\bibfield  {journal} {\bibinfo  {journal}
  {Phys.Rev.}\ }\textbf {\bibinfo {volume} {D86}},\ \bibinfo {pages} {pp 1383}
  (\bibinfo {year} {1998})},\ \bibinfo {note} {article ``Charmed Baryons'' in
  Ref.~\cite{Beringer:1900zz}}\BibitemShut {NoStop}%
\bibitem [{\citenamefont {Kisslinger}\ and\ \citenamefont
  {Henley}(2011)}]{Kisslinger:2009dr}%
  \BibitemOpen
  \bibfield  {author} {\bibinfo {author} {\bibfnamefont {L.~S.}\ \bibnamefont
  {Kisslinger}}\ and\ \bibinfo {author} {\bibfnamefont {E.~M.}\ \bibnamefont
  {Henley}},\ }\href {\doibase 10.1140/epja/i2011-11008-5} {\bibfield
  {journal} {\bibinfo  {journal} {Eur.Phys.J.}\ }\textbf {\bibinfo {volume}
  {A47}},\ \bibinfo {pages} {8} (\bibinfo {year} {2011})},\ \Eprint
  {http://arxiv.org/abs/0911.1179} {arXiv:0911.1179 [hep-ph]} \BibitemShut
  {NoStop}%
%%CITATION = ARXIV:0911.1179;%%
\bibitem [{\citenamefont {{Kittel, Olaf and Farrar, Glennys
  R.}}(2000)}]{Kittel-Farrar1}%
  \BibitemOpen
  \bibfield  {author} {\bibinfo {author} {\bibnamefont {{Kittel, Olaf and
  Farrar, Glennys R.}}},\ }\href@noop {} {\  (\bibinfo {year} {2000})},\
  \Eprint {http://arxiv.org/abs/0010186} {arXiv:0010186 [hep-ph]} \BibitemShut
  {NoStop}%
%%CITATION = HEP-PH/0010186;%%
\bibitem [{\citenamefont {{Kittel, Olaf and Farrar, Glennys
  R.}}(2005)}]{Kittel-Farrar2}%
  \BibitemOpen
  \bibfield  {author} {\bibinfo {author} {\bibnamefont {{Kittel, Olaf and
  Farrar, Glennys R.}}},\ }\href@noop {} {\  (\bibinfo {year} {2005})},\
  \Eprint {http://arxiv.org/abs/0508150} {arXiv:0508150 [hep-ph]} \BibitemShut
  {NoStop}%
%%CITATION = HEP-PH/0508150;%%
\bibitem [{\citenamefont {Gao}\ \emph {et~al.}(2012)\citenamefont {Gao},
  \citenamefont {Shi},\ and\ \citenamefont {Zou}}]{Gao:2012zh}%
  \BibitemOpen
  \bibfield  {author} {\bibinfo {author} {\bibfnamefont {P.}~\bibnamefont
  {Gao}}, \bibinfo {author} {\bibfnamefont {J.}~\bibnamefont {Shi}}, \ and\
  \bibinfo {author} {\bibfnamefont {B.}~\bibnamefont {Zou}},\ }\href {\doibase
  10.1103/PhysRevC.86.025201} {\bibfield  {journal} {\bibinfo  {journal}
  {Phys.Rev.}\ }\textbf {\bibinfo {volume} {C86}},\ \bibinfo {pages} {025201}
  (\bibinfo {year} {2012})},\ \Eprint {http://arxiv.org/abs/1206.4210}
  {arXiv:1206.4210 [nucl-th]} \BibitemShut {NoStop}%
%%CITATION = ARXIV:1206.4210;%%
\bibitem [{\citenamefont {Zou}(2010)}]{Zou:2010tc}%
  \BibitemOpen
  \bibfield  {author} {\bibinfo {author} {\bibfnamefont {B.-S.}\ \bibnamefont
  {Zou}},\ }\href {\doibase 10.1016/j.nuclphysa.2010.01.194} {\bibfield
  {journal} {\bibinfo  {journal} {Nucl. Phys.}\ }\textbf {\bibinfo {volume}
  {A835}},\ \bibinfo {pages} {199} (\bibinfo {year} {2010})},\ \Eprint
  {http://arxiv.org/abs/1001.1084} {arXiv:1001.1084 [nucl-th]} \BibitemShut
  {NoStop}%
%%CITATION = 1001.1084;%%
\bibitem [{\citenamefont {Wu}\ \emph {et~al.}(2009)\citenamefont {Wu},
  \citenamefont {Dulat},\ and\ \citenamefont {Zou}}]{Wu:2009tu}%
  \BibitemOpen
  \bibfield  {author} {\bibinfo {author} {\bibfnamefont {J.-J.}\ \bibnamefont
  {Wu}}, \bibinfo {author} {\bibfnamefont {S.}~\bibnamefont {Dulat}}, \ and\
  \bibinfo {author} {\bibfnamefont {B.~S.}\ \bibnamefont {Zou}},\ }\href
  {\doibase 10.1103/PhysRevD.80.017503} {\bibfield  {journal} {\bibinfo
  {journal} {Phys. Rev.}\ }\textbf {\bibinfo {volume} {D80}},\ \bibinfo {pages}
  {017503} (\bibinfo {year} {2009})},\ \Eprint {http://arxiv.org/abs/0906.3950}
  {arXiv:0906.3950 [hep-ph]} \BibitemShut {NoStop}%
%%CITATION = 0906.3950;%%
\bibitem [{\citenamefont {Wu}\ \emph {et~al.}(2010)\citenamefont {Wu},
  \citenamefont {Dulat},\ and\ \citenamefont {Zou}}]{Wu:2009nw}%
  \BibitemOpen
  \bibfield  {author} {\bibinfo {author} {\bibfnamefont {J.-J.}\ \bibnamefont
  {Wu}}, \bibinfo {author} {\bibfnamefont {S.}~\bibnamefont {Dulat}}, \ and\
  \bibinfo {author} {\bibfnamefont {B.}~\bibnamefont {Zou}},\ }\href {\doibase
  10.1103/PhysRevC.81.045210} {\bibfield  {journal} {\bibinfo  {journal}
  {Phys.Rev.}\ }\textbf {\bibinfo {volume} {C81}},\ \bibinfo {pages} {045210}
  (\bibinfo {year} {2010})},\ \Eprint {http://arxiv.org/abs/0909.1380}
  {arXiv:0909.1380 [hep-ph]} \BibitemShut {NoStop}%
%%CITATION = ARXIV:0909.1380;%%
\bibitem [{\citenamefont {Gao}\ \emph {et~al.}(2010)\citenamefont {Gao},
  \citenamefont {Wu},\ and\ \citenamefont {Zou}}]{Gao:2010hy}%
  \BibitemOpen
  \bibfield  {author} {\bibinfo {author} {\bibfnamefont {P.}~\bibnamefont
  {Gao}}, \bibinfo {author} {\bibfnamefont {J.-J.}\ \bibnamefont {Wu}}, \ and\
  \bibinfo {author} {\bibfnamefont {B.}~\bibnamefont {Zou}},\ }\href {\doibase
  10.1103/PhysRevC.81.055203} {\bibfield  {journal} {\bibinfo  {journal}
  {Phys.Rev.}\ }\textbf {\bibinfo {volume} {C81}},\ \bibinfo {pages} {055203}
  (\bibinfo {year} {2010})},\ \Eprint {http://arxiv.org/abs/1001.0805}
  {arXiv:1001.0805 [nucl-th]} \BibitemShut {NoStop}%
%%CITATION = ARXIV:1001.0805;%%
\bibitem [{\citenamefont {Khemchandani}\ \emph {et~al.}(2011)\citenamefont
  {Khemchandani}, \citenamefont {Martinez~Torres}, \citenamefont {Kaneko},
  \citenamefont {Nagahiro},\ and\ \citenamefont
  {Hosaka}}]{Khemchandari:PhysRevD.84.094018}%
  \BibitemOpen
  \bibfield  {author} {\bibinfo {author} {\bibfnamefont {K.~P.}\ \bibnamefont
  {Khemchandani}}, \bibinfo {author} {\bibfnamefont {A.}~\bibnamefont
  {Martinez~Torres}}, \bibinfo {author} {\bibfnamefont {H.}~\bibnamefont
  {Kaneko}}, \bibinfo {author} {\bibfnamefont {H.}~\bibnamefont {Nagahiro}}, \
  and\ \bibinfo {author} {\bibfnamefont {A.}~\bibnamefont {Hosaka}},\ }\href
  {\doibase 10.1103/PhysRevD.84.094018} {\bibfield  {journal} {\bibinfo
  {journal} {Phys. Rev. D}\ }\textbf {\bibinfo {volume} {84}},\ \bibinfo
  {pages} {094018} (\bibinfo {year} {2011})}\BibitemShut {NoStop}%
\bibitem [{\citenamefont {Oh}(2007)}]{Oh:PhysRevD.75.074002}%
  \BibitemOpen
  \bibfield  {author} {\bibinfo {author} {\bibfnamefont {Y.}~\bibnamefont
  {Oh}},\ }\href {\doibase 10.1103/PhysRevD.75.074002} {\bibfield  {journal}
  {\bibinfo  {journal} {Phys. Rev. D}\ }\textbf {\bibinfo {volume} {75}},\
  \bibinfo {pages} {074002} (\bibinfo {year} {2007})}\BibitemShut {NoStop}%
\bibitem [{\citenamefont {Oset}\ \emph {et~al.}(2002)\citenamefont {Oset},
  \citenamefont {Ramos},\ and\ \citenamefont {Bennhold}}]{Oset:2001cn}%
  \BibitemOpen
  \bibfield  {author} {\bibinfo {author} {\bibfnamefont {E.}~\bibnamefont
  {Oset}}, \bibinfo {author} {\bibfnamefont {A.}~\bibnamefont {Ramos}}, \ and\
  \bibinfo {author} {\bibfnamefont {C.}~\bibnamefont {Bennhold}},\ }\href
  {\doibase 10.1016/S0370-2693(01)01523-4} {\bibfield  {journal} {\bibinfo
  {journal} {Phys.Lett.}\ }\textbf {\bibinfo {volume} {B527}},\ \bibinfo
  {pages} {99} (\bibinfo {year} {2002})},\ \Eprint
  {http://arxiv.org/abs/nucl-th/0109006} {arXiv:nucl-th/0109006 [nucl-th]}
  \BibitemShut {NoStop}%
%%CITATION = NUCL-TH/0109006;%%
\bibitem [{\citenamefont {Beringer}\ \emph {et~al.}(2012)\citenamefont
  {Beringer} \emph {et~al.}}]{Beringer:1900zz}%
  \BibitemOpen
  \bibfield  {author} {\bibinfo {author} {\bibfnamefont {J.}~\bibnamefont
  {Beringer}} \emph {et~al.} (\bibinfo {collaboration} {Particle Data Group}),\
  }\href {\doibase 10.1103/PhysRevD.86.010001} {\bibfield  {journal} {\bibinfo
  {journal} {Phys.Rev.}\ }\textbf {\bibinfo {volume} {D86}},\ \bibinfo {pages}
  {010001} (\bibinfo {year} {2012})}\BibitemShut {NoStop}%
%%CITATION = PHRVA,D86,010001;%%
\bibitem [{\citenamefont {Thomas}\ \emph {et~al.}(1973)\citenamefont {Thomas},
  \citenamefont {Engler}, \citenamefont {Fisk},\ and\ \citenamefont
  {Kraemer}}]{Thomas}%
  \BibitemOpen
  \bibfield  {author} {\bibinfo {author} {\bibfnamefont {D.~W.}\ \bibnamefont
  {Thomas}}, \bibinfo {author} {\bibfnamefont {A.}~\bibnamefont {Engler}},
  \bibinfo {author} {\bibfnamefont {H.~E.}\ \bibnamefont {Fisk}}, \ and\
  \bibinfo {author} {\bibfnamefont {R.~W.}\ \bibnamefont {Kraemer}},\ }\href
  {\doibase 10.1016/0550-3213(73)90217-4} {\bibfield  {journal} {\bibinfo
  {journal} {Nucl. Phys.}\ }\textbf {\bibinfo {volume} {B56}},\ \bibinfo
  {pages} {15} (\bibinfo {year} {1973})}\BibitemShut {NoStop}%
%%CITATION = NUPHA,B56,15;%%
\bibitem [{\citenamefont {Hemingway}(1985)}]{Hemingway}%
  \BibitemOpen
  \bibfield  {author} {\bibinfo {author} {\bibfnamefont {R.~J.}\ \bibnamefont
  {Hemingway}},\ }\href {\doibase 10.1016/0550-3213(85)90556-5} {\bibfield
  {journal} {\bibinfo  {journal} {Nucl. Phys.}\ }\textbf {\bibinfo {volume}
  {B253}},\ \bibinfo {pages} {742} (\bibinfo {year} {1985})}\BibitemShut
  {NoStop}%
%%CITATION = NUPHA,B253,742;%%
\bibitem [{\citenamefont {Zychor}\ \emph {et~al.}(2008)\citenamefont {Zychor}
  \emph {et~al.}}]{Zychor}%
  \BibitemOpen
  \bibfield  {author} {\bibinfo {author} {\bibfnamefont {I.}~\bibnamefont
  {Zychor}} \emph {et~al.},\ }\href {\doibase 10.1016/j.physletb.2008.01.002}
  {\bibfield  {journal} {\bibinfo  {journal} {Phys. Lett.}\ }\textbf {\bibinfo
  {volume} {B660}},\ \bibinfo {pages} {167} (\bibinfo {year} {2008})},\ \Eprint
  {http://arxiv.org/abs/0705.1039} {arXiv:0705.1039 [nucl-ex]} \BibitemShut
  {NoStop}%
%%CITATION = 0705.1039;%%
\bibitem [{\citenamefont {Agakishiev}\ \emph {et~al.}(2012)\citenamefont
  {Agakishiev} \emph {et~al.}}]{Hades}%
  \BibitemOpen
  \bibfield  {author} {\bibinfo {author} {\bibfnamefont {G.}~\bibnamefont
  {Agakishiev}} \emph {et~al.} (\bibinfo {collaboration} {HADES}),\ }\href
  {\doibase 10.1103/PhysRevC.85.035203} {\bibfield  {journal} {\bibinfo
  {journal} {Phys.Rev.}\ }\textbf {\bibinfo {volume} {C85}},\ \bibinfo {pages}
  {035203} (\bibinfo {year} {2012})},\ \Eprint {http://arxiv.org/abs/1109.6806}
  {arXiv:1109.6806 [nucl-ex]} \BibitemShut {NoStop}%
%%CITATION = ARXIV:1109.6806;%%
\bibitem [{\citenamefont {Ahn}\ \emph {et~al.}(2003)\citenamefont {Ahn} \emph
  {et~al.}}]{Ahn}%
  \BibitemOpen
  \bibfield  {author} {\bibinfo {author} {\bibfnamefont {J.~K.}\ \bibnamefont
  {Ahn}} \emph {et~al.} (\bibinfo {collaboration} {LEPS}),\ }\href {\doibase
  10.1016/S0375-9474(03)01164-3} {\bibfield  {journal} {\bibinfo  {journal}
  {Nucl. Phys.}\ }\textbf {\bibinfo {volume} {A721}},\ \bibinfo {pages} {715}
  (\bibinfo {year} {2003})}\BibitemShut {NoStop}%
%%CITATION = NUPHA,A721,715;%%
\bibitem [{\citenamefont {Niiyama}\ \emph {et~al.}(2008)\citenamefont {Niiyama}
  \emph {et~al.}}]{Niiyama}%
  \BibitemOpen
  \bibfield  {author} {\bibinfo {author} {\bibfnamefont {M.}~\bibnamefont
  {Niiyama}} \emph {et~al.} (\bibinfo {collaboration} {LEPS}),\ }\href
  {\doibase 10.1103/PhysRevC.78.035202} {\bibfield  {journal} {\bibinfo
  {journal} {Phys. Rev.}\ }\textbf {\bibinfo {volume} {C78}},\ \bibinfo {pages}
  {035202} (\bibinfo {year} {2008})},\ \Eprint {http://arxiv.org/abs/0805.4051}
  {arXiv:0805.4051 [hep-ex]} \BibitemShut {NoStop}%
%%CITATION = 0805.4051;%%
\bibitem [{\citenamefont {Moriya}\ and\ \citenamefont
  {Schumacher}(2013{\natexlab{a}})}]{crosssectionpaper}%
  \BibitemOpen
  \bibfield  {author} {\bibinfo {author} {\bibfnamefont {K.}~\bibnamefont
  {Moriya}}\ and\ \bibinfo {author} {\bibfnamefont {R.~A.}\ \bibnamefont
  {Schumacher}} (\bibinfo {collaboration} {CLAS}),\ }\href@noop {} {\
  (\bibinfo {year} {2013}{\natexlab{a}})},\ \bibinfo {note} {to be
  submitted}\BibitemShut {NoStop}%
\bibitem [{\citenamefont {Sober}\ \emph {et~al.}(2000)\citenamefont {Sober}
  \emph {et~al.}}]{Sober}%
  \BibitemOpen
  \bibfield  {author} {\bibinfo {author} {\bibfnamefont {D.~I.}\ \bibnamefont
  {Sober}} \emph {et~al.},\ }\href {\doibase 10.1016/S0168-9002(99)00784-6}
  {\bibfield  {journal} {\bibinfo  {journal} {Nucl. Instrum. Meth.}\ }\textbf
  {\bibinfo {volume} {A440}},\ \bibinfo {pages} {263} (\bibinfo {year}
  {2000})}\BibitemShut {NoStop}%
%%CITATION = NUIMA,A440,263;%%
\bibitem [{\citenamefont {Mecking}\ \emph {et~al.}(2003)\citenamefont {Mecking}
  \emph {et~al.}}]{CLAS-NIM}%
  \BibitemOpen
  \bibfield  {author} {\bibinfo {author} {\bibfnamefont {B.~A.}\ \bibnamefont
  {Mecking}} \emph {et~al.} (\bibinfo {collaboration} {CLAS}),\ }\href
  {\doibase 10.1016/S0168-9002(03)01001-5} {\bibfield  {journal} {\bibinfo
  {journal} {Nucl. Instrum. Meth.}\ }\textbf {\bibinfo {volume} {A503}},\
  \bibinfo {pages} {513} (\bibinfo {year} {2003})}\BibitemShut {NoStop}%
%%CITATION = NUIMA,A503,513;%%
\bibitem [{\citenamefont {Moriya}(2010)}]{Moriya-thesis}%
  \BibitemOpen
  \bibfield  {author} {\bibinfo {author} {\bibfnamefont {K.}~\bibnamefont
  {Moriya}},\ }\href@noop {} {\bibfield  {journal} {\bibinfo  {journal} {Ph.D.
  Thesis, Carnegie Mellon University}\ } (\bibinfo {year} {2010})},\ \bibinfo
  {note} {available online at
  \url{http://www.jlab.org/Hall-B/general/clas_thesis.html}}\BibitemShut
  {NoStop}%
\bibitem [{\citenamefont {Williams}\ \emph
  {et~al.}(2009{\natexlab{a}})\citenamefont {Williams} \emph
  {et~al.}}]{Williams1}%
  \BibitemOpen
  \bibfield  {author} {\bibinfo {author} {\bibfnamefont {M.}~\bibnamefont
  {Williams}} \emph {et~al.} (\bibinfo {collaboration} {CLAS}),\ }\href
  {\doibase 10.1103/PhysRevC.80.045213} {\bibfield  {journal} {\bibinfo
  {journal} {Phys. Rev.}\ }\textbf {\bibinfo {volume} {C80}},\ \bibinfo {pages}
  {045213} (\bibinfo {year} {2009}{\natexlab{a}})},\ \Eprint
  {http://arxiv.org/abs/0909.0616} {arXiv:0909.0616 [nucl-ex]} \BibitemShut
  {NoStop}%
%%CITATION = 0909.0616;%%
\bibitem [{\citenamefont {Williams}\ \emph
  {et~al.}(2009{\natexlab{b}})\citenamefont {Williams} \emph
  {et~al.}}]{Williams2}%
  \BibitemOpen
  \bibfield  {author} {\bibinfo {author} {\bibfnamefont {M.}~\bibnamefont
  {Williams}} \emph {et~al.} (\bibinfo {collaboration} {CLAS}),\ }\href
  {\doibase 10.1103/PhysRevC.80.065208} {\bibfield  {journal} {\bibinfo
  {journal} {Phys. Rev.}\ }\textbf {\bibinfo {volume} {C80}},\ \bibinfo {pages}
  {065208} (\bibinfo {year} {2009}{\natexlab{b}})},\ \Eprint
  {http://arxiv.org/abs/0908.2910} {arXiv:0908.2910 [nucl-ex]} \BibitemShut
  {NoStop}%
%%CITATION = 0908.2910;%%
\bibitem [{\citenamefont {McCracken}\ \emph {et~al.}(2010)\citenamefont
  {McCracken} \emph {et~al.}}]{McCracken:2009ra}%
  \BibitemOpen
  \bibfield  {author} {\bibinfo {author} {\bibfnamefont {M.}~\bibnamefont
  {McCracken}} \emph {et~al.} (\bibinfo {collaboration} {CLAS}),\ }\href
  {\doibase 10.1103/PhysRevC.81.025201} {\bibfield  {journal} {\bibinfo
  {journal} {Phys.Rev.}\ }\textbf {\bibinfo {volume} {C81}},\ \bibinfo {pages}
  {025201} (\bibinfo {year} {2010})},\ \Eprint {http://arxiv.org/abs/0912.4274}
  {arXiv:0912.4274 [nucl-ex]} \BibitemShut {NoStop}%
%%CITATION = ARXIV:0912.4274;%%
\bibitem [{\citenamefont {Dey}\ \emph {et~al.}(2010)\citenamefont {Dey} \emph
  {et~al.}}]{Dey}%
  \BibitemOpen
  \bibfield  {author} {\bibinfo {author} {\bibfnamefont {B.}~\bibnamefont
  {Dey}} \emph {et~al.} (\bibinfo {collaboration} {CLAS}),\ }\href@noop {}
  {\bibfield  {journal} {\bibinfo  {journal} {Phys. Rev.}\ }\textbf {\bibinfo
  {volume} {C82}},\ \bibinfo {pages} {025202} (\bibinfo {year} {2010})},\
  \Eprint {http://arxiv.org/abs/1006.0374} {arXiv:1006.0374 [nucl-ex]}
  \BibitemShut {NoStop}%
%%CITATION = 1006.0374;%%
\bibitem [{\citenamefont {Williams}(2007)}]{MW-thesis}%
  \BibitemOpen
  \bibfield  {author} {\bibinfo {author} {\bibfnamefont {M.}~\bibnamefont
  {Williams}},\ }\href@noop {} {\bibfield  {journal} {\bibinfo  {journal}
  {Ph.D. Thesis, Carnegie Mellon University}\ } (\bibinfo {year} {2007})},\
  \bibinfo {note} {available online at
  \url{http://www.jlab.org/Hall-B/general/clas_thesis.html}}\BibitemShut
  {NoStop}%
\bibitem [{\citenamefont {Borenstein}\ \emph {et~al.}(1974)\citenamefont
  {Borenstein}, \citenamefont {Kalbfleisch}, \citenamefont {Strand},
  \citenamefont {VanderBurg},\ and\ \citenamefont {Chapman}}]{Borenstein}%
  \BibitemOpen
  \bibfield  {author} {\bibinfo {author} {\bibfnamefont {S.~R.}\ \bibnamefont
  {Borenstein}}, \bibinfo {author} {\bibfnamefont {G.~R.}\ \bibnamefont
  {Kalbfleisch}}, \bibinfo {author} {\bibfnamefont {R.~C.}\ \bibnamefont
  {Strand}}, \bibinfo {author} {\bibfnamefont {V.}~\bibnamefont {VanderBurg}},
  \ and\ \bibinfo {author} {\bibfnamefont {J.~W.}\ \bibnamefont {Chapman}},\
  }\href {\doibase 10.1103/PhysRevD.9.3006} {\bibfield  {journal} {\bibinfo
  {journal} {Phys. Rev.}\ }\textbf {\bibinfo {volume} {D9}},\ \bibinfo {pages}
  {3006} (\bibinfo {year} {1974})}\BibitemShut {NoStop}%
%%CITATION = PHRVA,D9,3006;%%
\bibitem [{\citenamefont {Cameron}\ \emph {et~al.}(1978)\citenamefont {Cameron}
  \emph {et~al.}}]{Cameron_1385}%
  \BibitemOpen
  \bibfield  {author} {\bibinfo {author} {\bibfnamefont {W.}~\bibnamefont
  {Cameron}} \emph {et~al.} (\bibinfo {collaboration} {Rutherford-London}),\
  }\href {\doibase 10.1016/0550-3213(78)90022-6} {\bibfield  {journal}
  {\bibinfo  {journal} {Nucl. Phys.}\ }\textbf {\bibinfo {volume} {B143}},\
  \bibinfo {pages} {189} (\bibinfo {year} {1978})}\BibitemShut {NoStop}%
%%CITATION = NUPHA,B143,189;%%
\bibitem [{\citenamefont {Holmgren}\ \emph {et~al.}(1977)\citenamefont
  {Holmgren} \emph {et~al.}}]{Holmgren}%
  \BibitemOpen
  \bibfield  {author} {\bibinfo {author} {\bibfnamefont {S.~O.}\ \bibnamefont
  {Holmgren}} \emph {et~al.} (\bibinfo {collaboration}
  {Amsterdam-CERN-Nijmegen-Oxford}),\ }\href {\doibase
  10.1016/0550-3213(77)90064-5} {\bibfield  {journal} {\bibinfo  {journal}
  {Nucl. Phys.}\ }\textbf {\bibinfo {volume} {B119}},\ \bibinfo {pages} {261}
  (\bibinfo {year} {1977})}\BibitemShut {NoStop}%
%%CITATION = NUPHA,B119,261;%%
\bibitem [{CLA()}]{CLASdatabase}%
  \BibitemOpen
  \href@noop {} {\ }\bibinfo {note} {Full results are available upon request
  from the lead authors or from the CLAS database at
  \url{http://clasweb.jlab.org/physicsdb/}}\BibitemShut {NoStop}%
\bibitem [{\citenamefont {Flatt\'e}(1976)}]{Flatte}%
  \BibitemOpen
  \bibfield  {author} {\bibinfo {author} {\bibfnamefont {S.~M.}\ \bibnamefont
  {Flatt\'e}},\ }\href {\doibase 10.1016/0370-2693(76)90654-7} {\bibfield
  {journal} {\bibinfo  {journal} {Phys. Lett.}\ }\textbf {\bibinfo {volume}
  {B63}},\ \bibinfo {pages} {224} (\bibinfo {year} {1976})}\BibitemShut
  {NoStop}%
%%CITATION = PHLTA,B63,224;%%
\bibitem [{\citenamefont {Chung}(1971)}]{Chung:1971ri}%
  \BibitemOpen
  \bibfield  {author} {\bibinfo {author} {\bibfnamefont {S.~U.}\ \bibnamefont
  {Chung}},\ }\href@noop {} {\  (\bibinfo {year} {1971})},\ \bibinfo {note}
  {{CERN-71-08}, Updated version available at
  \url{http://suchung.web.cern.ch/suchung/spinfm1.pdf}}\BibitemShut {NoStop}%
\bibitem [{\citenamefont {Schumacher}\ and\ \citenamefont
  {Moriya}(2012)}]{isospinfits}%
  \BibitemOpen
  \bibfield  {author} {\bibinfo {author} {\bibfnamefont {R.~A.}\ \bibnamefont
  {Schumacher}}\ and\ \bibinfo {author} {\bibfnamefont {K.}~\bibnamefont
  {Moriya}},\ }\href@noop {} {\  (\bibinfo {year} {2012})},\ \bibinfo {note}
  {proceedings of the International Conference on Hypernuclear and Strange
  Particle Physics XI, Barcelona, Spain, (Submitted to Nucl. Phys.
  A)}\BibitemShut {NoStop}%
\bibitem [{\citenamefont {Ramos}\ \emph {et~al.}(2003)\citenamefont {Ramos},
  \citenamefont {Oset},\ and\ \citenamefont {Bennhold}}]{Ramos:2003mu}%
  \BibitemOpen
  \bibfield  {author} {\bibinfo {author} {\bibfnamefont {A.}~\bibnamefont
  {Ramos}}, \bibinfo {author} {\bibfnamefont {E.}~\bibnamefont {Oset}}, \ and\
  \bibinfo {author} {\bibfnamefont {C.}~\bibnamefont {Bennhold}},\ }\href
  {\doibase 10.1016/S0375-9474(03)01163-1} {\bibfield  {journal} {\bibinfo
  {journal} {Nucl.Phys.}\ }\textbf {\bibinfo {volume} {A721}},\ \bibinfo
  {pages} {711} (\bibinfo {year} {2003})}\BibitemShut {NoStop}%
%%CITATION = NUPHA,A721,711;%%
\bibitem [{\citenamefont {Moriya}\ and\ \citenamefont
  {Schumacher}(2013{\natexlab{b}})}]{paritypaper}%
  \BibitemOpen
  \bibfield  {author} {\bibinfo {author} {\bibfnamefont {K.}~\bibnamefont
  {Moriya}}\ and\ \bibinfo {author} {\bibfnamefont {R.~A.}\ \bibnamefont
  {Schumacher}} (\bibinfo {collaboration} {CLAS}),\ }\href@noop {} {\
  (\bibinfo {year} {2013}{\natexlab{b}})},\ \bibinfo {note} {to be
  submitted}\BibitemShut {NoStop}%
\end{thebibliography}%

\end{document}